\documentclass[reqno,12pt]{article}
\input{myPreliminary}
\allowdisplaybreaks
\newcommand{\showName}{1}
\newcommand{\isMain}{1}
\newcommand{\paperTitle}{Anytime-valid goodness-of-fit test by $e$-spacings}
\begin{document}
\if1\isMain{
\title{\bf \paperTitle}
}\else{
\title{\bf Supplementary note to ``\paperTitle''}
}\fi

\if1\showName{
	\author[1,2]{Yunda Hao\thanks{Email: \href{mailto:yundahao@tju.edu.cn}{\nolinkurl{yundahao@tju.edu.cn}}}}
	\author[2]{Kin Wai Chan\thanks{Email: 
				\href{mailto:kinwaichan@cuhk.edu.hk}
				{\nolinkurl{kinwaichan@cuhk.edu.hk}}}}
    \affil[1]{Center for Applied Mathematics, KL-AAGDM, Tianjin University, Tianjin, 300072, China.}
	\affil[2]{Department of Statistics and Data Science, The Chinese University of Hong Kong.}
}\else{
	\author[]{}
}\fi	
\maketitle
\begin{abstract}
To perform an anytime-valid goodness-of-fit test, we propose spacing-based \(e\)-processes that control the type-I error rate under arbitrary stopping rules.
Based on
the maximum spacing of order statistics,
we construct spacing-based \(e\)-processes for
performing nonparametric distribution tests.
The resulting procedures are simple and computationally efficient, and are power-consistent.
We call them $e$-spacing tests.
This approach is especially powerful in detecting localized discrepancy
between the data-generating density and the null density.
Under mild conditions, the logarithm of the
proposed \(e\)-process is asymptotically equivalent to that of the
oracle-optimal \(e\)-process.
The maximal spacing strategy is generalized to
a maximum spacing set strategy in the form of
the \(1/k\) fraction of the largest \(m\)th order spacings,
as a measure of the lower-level set of density.
It boosts power under more global discrepancy from the null.
Furthermore, we propose
a greedy predictable scheme, via argmax or softmax rules,
to combine multiple $e$-spacing tests.
We prove that it optimally selects the best testing strategy.
This approach is readily generalizable to
the multivariate case via
a coordinate-wise Rosenblatt-transform.
Simulation studies demonstrate that the proposed tests
perform favorably relative to the considered anytime-valid competitors as
well as
various types of multiple test-adjusted Kolmogorov--Smirnov tests.
\end{abstract}

\spacingset{1.6}
\newpage

\section{Introduction}
\subsection{Background}

Classical goodness-of-fit (GoF) tests include Pearson's $\chi^2$ test
\citep{pearson1900}, the Cram\'er--von Mises test
\citep{cramer1928,vonMises1928}, the Kolmogorov--Smirnov (KS) test
\citep{an1933sulla,smirnov1948table}, and the Andersen--Darling test
\citep{AndersonDarling1952}. Subsequent work developed alternatives
to KS with greater Bahadur efficiency \citep{berk1979goodness} and
a unified likelihood-ratio framework \citep{zhang2002powerful}.
Multivariate methods include KS extensions
\citep{peacock1983two,lopes2007two,gosset1987three,fasano1987multidimensional}
and characteristic-function tests \citep{fan1997goodness}.
More recent methods use kernelized Stein discrepancy
\citep{liu2016kernelized,chwialkowski2016kernel} or maximum mean
discrepancy \citep{sang2025integral,balasubramanian2021optimality}.

Fixed-sample GoF tests can exceed their nominal type-I error rate
under repeated monitoring. Allocating the error budget across testing
times through $\alpha$-spending can lead to stringent thresholds over
long horizons.
Existing anytime-valid methods have different requirements. The
Monte Carlo procedure of \cite{dombowsky2026besag} generates auxiliary
samples at each update. The watermark tests of \cite{su2026online}
target super-uniform departures of a scalar pivotal statistic.
Sequential Stein tests \citep{zhou2024sequential,martinez2024sequential}
require a choice of kernel and, for continuous models, evaluation of
the gradient of the null log-density. Our $e$-spacing tests use the
specified null CDF to transform the observations and compute the
Bernoulli null probabilities directly from spacing lengths. They
target both localized and broader departures from uniformity and
extend to multivariate settings through the Rosenblatt transform.

\subsection{Setup for anytime-valid goodness-of-fit-tests}
In this article,
we study a sequential goodness-of-fit test problem based on a streaming dataset.
Let $X_1, X_2, \ldots \simIID F^X$ be an independent and identically distributed (i.i.d.)~real-valued stream
from an unknown continuous cumulative distribution function (CDF) $F^X$.
We test $\mathcal{H}_0: F^X = F_0^X$ against $\mathcal{H}_1: F^X \neq F_0^X$,
where $F_0^X$ is a given null continuous CDF.
Denote $\pr$ and $\pr_0$ as the probability measures under $\mathcal{H}_1$ and $\mathcal{H}_0$, respectively.
Then we say that we test
\begin{equation}\label{eq:hypotheses}
    \mathcal{H}_0: \pr = \pr_0
    \qquad \text{against} \qquad
    \mathcal{H}_1: \pr \neq \pr_0,
\end{equation}
Clearly,
$U_i := F^X_0(X_i) \sim \Unif[0, 1]$ if and only if (iff) $X_i \sim F_0^X$.
Hence, testing $\mathcal{H}_0$ on $X_1, \ldots, X_n$
is equivalent to testing standard uniformity on
$U_1, \ldots, U_n$.
Denote the CDFs and densities of $U_i$ as $(F,F_0)$ and $(p,p_0)$
under $\pr$ and $\pr_0$, respectively.

Let \((\Omega,\mathcal{F},\pr)\)
be a probability space supporting the data stream \(X_1,X_2,\ldots\), and let \(\mathcal{F}_t=\sigma(X_1,\ldots,X_t)\) be its natural filtration, with \(\mathcal{F}_0=\{\varnothing,\Omega\}\).
A stopping time may take values in $\mathbb{Z}^+ \cup \{\infty\}$.
For a nonnegative process $\{M_t\}_{t\ge 1}$, define $M_\infty = \lim\inf_{t\rightarrow\infty} M_t$.

The goal of this article is to develop an anytime-valid
goodness-of-fit test for $\cH_0$ against $\cH_1$, i.e.,
a sequence of tests \(\{\psi_t\}_{t\in\mathbb{Z}^+\cup \{\infty\}}\) that reject \(\mathcal{H}_0\) when \(\psi_t = 1\) and satisfy, under \(\mathcal{H}_0\),
\[
\pr_0\left(\sup_{t\in\mathbb{Z}^+\cup \{\infty\}} \psi_t = 1\right) \le \alpha.
\]
or equivalently,
$\pr_0(\psi_{\tau}=1) \leq \alpha$ for any stopping time $\tau \in\mathbb{Z}^+\cup \{\infty\}$ satisfying
$\{\tau\leq t\} \in \mathcal{F}_t$; see, e.g.,
\cite{johari2022always,howard2021time} for more discussions on anytime valid $p$-values.
In addition, we hope that the test
is computationally simple, intuitive, and powerful.

Our approach is motivated by two key concepts: $e$-process and spacings,
which are reviewed as follows:
First,
an $e$-variable $\Lambda_n$
is a nonnegative statistic dependent on the sample
$X_1, \ldots, X_n$
that satisfies $\E_{\pr_0}(\Lambda_n) \leq 1$, whereas
a nonnegative adapted process $\{ \Lambda_t \}_{t \geq 1}$ with
$\Lambda_1 = 1$ is a test martingale if $\E_{\pr_0}(\Lambda_t\mid \mathcal{F}_{t-1}) = \Lambda_{t-1}$, and a test supermartingale if ``$=$'' is
replaced by ``$\le$''.
Both are $e$-processes in the stopping-time sense: $\E_{\pr_0}(\Lambda_\tau) \le 1$ for every almost surely (a.s.) finite
stopping time $\tau$.
The same bound holds for $\tau$ that may be infinite under the preceding
$\liminf$ convention by Fatou’s lemma.
Consequently, by Ville’s inequality gives $\pr_0(\sup_t \Lambda_t \ge 1/\alpha)\le\alpha$; see \cite{ramdas2022testing,GrunwaldHK19}.
One major advantage of $e$‑processes is that they guarantee type-I error control under arbitrary stopping rules; see, e.g.,
\cite{ramdas2025hypothesis, chugg2026values, koning2026anytime, 10.1093/jrsssb/qkag066}.

Second, spacings are the gaps between adjacent order statistics.
They are building blocks for many popular GoF tests.
The use of spacings can be dated back to \cite{preason1902note,cheng1989goodness},
who studied the arithmetic mean and
geometric mean of the spacings.
Spacing-based tests are then developed in,
e.g., \cite{kimball1950asymptotic,darling1953class};
see, also \cite{pyke1965spacings} for a survey.
The aforementioned spacing-based methods
are not sequential in nature
and require an asymptotic argument for approximation.
We introduce spacings here.
Given a sample $X_{1:t} := (X_1, \ldots, X_t)$,
define the order statistics as
$X_{1,t} \leq \cdots \leq X_{t,t}$.
Also denote $X_{0,t}\equiv -\infty$ and $X_{t+1,t}\equiv +\infty$.
Set $U_{0,t} = 0$, $U_{t+1,t} = 1$ and define $U_{i,t} = F_0(X_{i,t})$ for $i=1, \ldots, t$.
Then, at time $t$,
the spacings are
\begin{align}\label{eqt:spacings}
    d_{i,t}
    :=
    F^X_0(X_{i+1, t}) - F^X_0(X_{i, t})
    = U_{i+1,t} - U_{i,t}
\end{align}
for $i=0, \ldots, t$.
More generally, for $m\in\mathbb{Z}^+$, the $m$th order spacings are
\begin{align}\label{eqt:mspacings}
    d_{i,t}^{(m)} :=\left\{
            \begin{array}{ll}
            F^X_0(X_{(i+1)m, t}) - F^X_0(X_{im, t})
    = U_{(i+1)m,t} - U_{im,t} & \text{if $i=0, \ldots, \lfloor \frac{t+1}{m} \rfloor-1$} ; \\
            1 - F_0^X(X_{im, t}) = 1- U_{im,t} & \text{if $i=\lfloor \frac{t+1}{m} \rfloor$ and $\lfloor \frac{t+1}{m} \rfloor \neq \frac{t+1}{m} $}.
            \end{array}
            \right.
\end{align}
Clearly, $d_{i,t}^{(1)} = d_{i,t}$ and $d_{i,t}^{(m)} = \sum_{j=0}^{m-1} d_{im+j,t}$.
We will develop new tests based on $d_{i,t}$ first, and then generalize it to a version based on $d_{i,t}^{(m)}$.

Throughout the article,
we denote the Kullback--Leibler (KL) divergence between $F$ and $F_0$ as
$\KL(F \| F_0)$.
The Bernoulli distribution with success probability $\theta$ is denoted as $\Bern(\theta)$.
The Lebesgue measure is denoted by $\Leb(\cdot)$.

Our main contributions are as follows:
\begin{itemize}
    \item Methodologically, we construct a support-splitting bridge based on spacing from nonparametric tests to
    likelihood ratio tests, producing \(e\)-processes built from the maximum spacing
 \(\max_i d_{i,t}\) over time $t$. This test is denoted as $\Lambda_n^{(1,\infty)}$.
    We call this an $e$-spacing test.
    This test is particularly powerful for detecting localized discrepancy from
    the null density.
    It controls the type-I error rate under arbitrary stopping times, for both finite and infinite sampling horizons.
    Then we generalize it to the maximum spacing set.
    Precisely, we use
    the maximum sum of the largest \(1/k\) fraction of $m$th order spacings over time $t$,
    i.e., $\max_{\mathcal{I}} \sum_{i\in\mathcal{I}} d_{i,t}^{(m)}$, where $|\mathcal{I}| \approx \#(\text{$m$th order spacings})/k$.
    This test is denoted as $\Lambda_n^{(m,k)}$, which gives higher power
    when there is discrepancy over a broader range of support.
    We also introduce a predictable weighting scheme to combine
    tests $\Lambda_n^{(m,k)}$ with multiple values of $k$ to improve power.
    In particular, we prove that, upon using appropriate weighting schemes,
    it automatically selects the best possible candidate $k$.
    It can also automatically switch between the maximum spacing approach
    and the optimally tuned maximum spacing set approach.
    Finally, we extend these methods to the multivariate case.
    \item  Theoretically, we analyze the two proposed $e$-processes
    based on the maximum spacing and on the sum of the largest \(1/k\) fraction of $m$th order spacings.
    With maximum spacing, under mild conditions,
    the $e$-process diverges to infinity under $\cH_1$;
    the conditions approximately hold for the Hölder class on $[0,1]$,
    which includes Lipschitz densities.
    Using the largest \(1/k\) fraction of $m$th order spacings, under \(\mathcal{H}_1\), the logarithm of the \(e\)-process diverges to infinity at a linear rate.
    This behavior is closely related to the asymptotic equipartition property for relative entropy.
    \item Empirically,
    we compare our methods with the unadjusted KS test, several adjusted KS tests, and the kernelized Stein discrepancies (KSD) based $e$‑processes \citep{zhou2024sequential,martinez2024sequential}.
    The simulation results show that our $e$‑spacing tests outperform the adjusted KS tests
    and the KSD‑based $e$‑processes.
    Moreover, Section \ref{sec:realData} demonstrates
    the proposed method in a real-data application
    involving gas sensor shift.
    This analysis provides a practical framework for distribution testing
    after adjusting for numerous recorded confounders.

\end{itemize}

The remainder of this article is organized as follows:
Section~\ref{sec:methodology} introduces the basic principle of the methodology.
Section \ref{sec:maxSpacing} presents the test strategy based on maximum spacing.
Oracle-optimality is derived.
We prove that it can be achieved by our \(e\)-processes based on
a simple minimum density estimator.
Section~\ref{sec:general_aggreation} generalizes
it to maximum spacing set
by using the maximum sum of the largest \(1/k\) fraction of $m$th order spacings.
We show that it tests by assessing the lower-level set of density.
Section~\ref{sec:compare_two_eprocess}
develops a method to
combine multiple \(e\)-processes to improve test power.
An argmax rule and a softmax rule are derived and proven
to be asymptotically optimal
under some conditions.
Section~\ref{sec:multi_rv} extends the results to the multivariate setting.
Section~\ref{sec:simulations} presents simulation studies demonstrating that our proposals are
attractive alternatives to the existing methods.
Section \ref{sec:realData} shows a real-data application on gas sensor shift test.
All technical proofs, additional results and discussions are deferred to a supplement.

\section{Methodology: $e$-spacing likelihood ratio test}\label{sec:methodology}
\subsection{Fundamental principle}
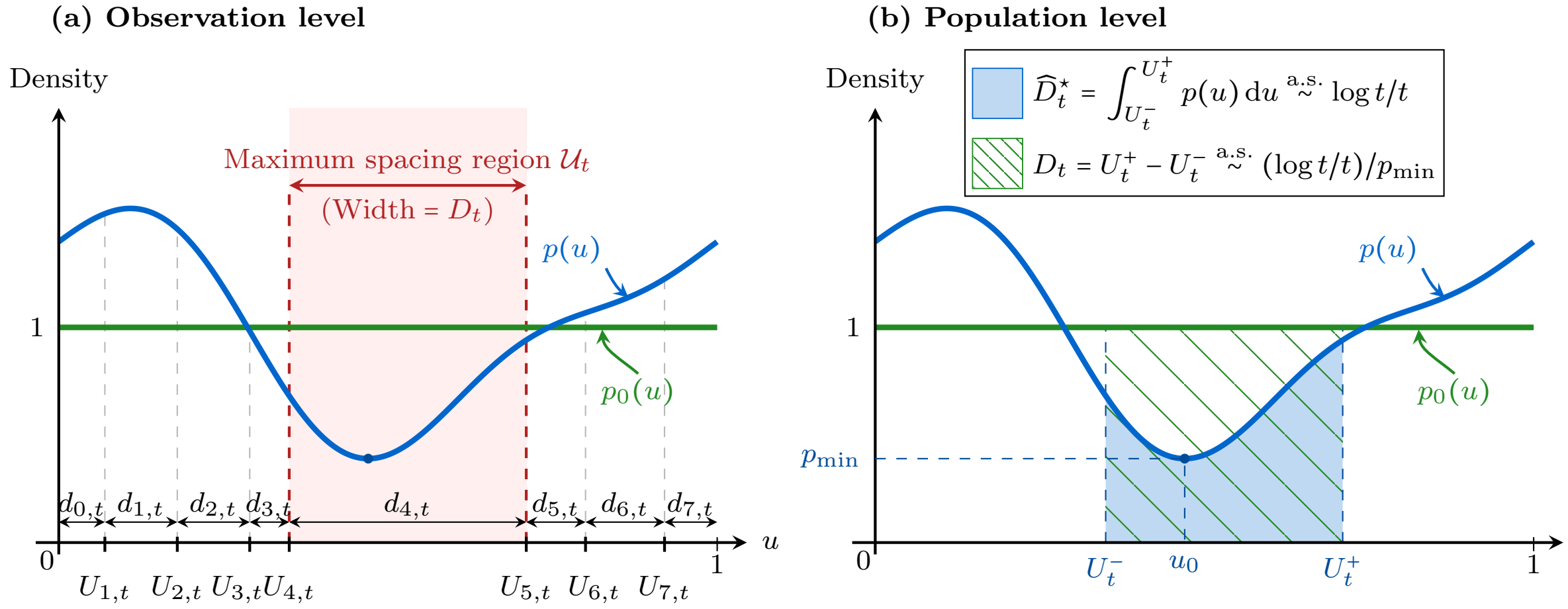
\begin{figure}[t]
\centering
\begin{tikzpicture}[x=0.405\linewidth,y=2.15cm,font=\scriptsize,>=stealth]
\definecolor{profGreen}{RGB}{34,139,34}
\definecolor{profRed}{RGB}{178,34,34}
\definecolor{betterblue}{RGB}{0,102,204}
\colorlet{profblue}{betterblue!75!black}
\def\gapLeft{0.35}
\def\gapRight{0.71}
\def\minimumAt{0.47}
\def\axisTop{2.02}
\pgfmathdeclarefunction{maxgapdensity}{1}{%
  \pgfmathparse{1-.47*cos(360*(#1-.47))-.14*cos(720*(#1-.47))
    -.18*sin(360*(#1-.47))*(1-cos(360*(#1-.47)))}}
\def\densityPath{plot[domain=0:1,samples=131,smooth] (\x,{maxgapdensity(\x)})}
\usetikzlibrary{patterns}
\def\declareGapHatch#1#2{%
  \pgfdeclarepatternformonly{#1}{\pgfpointorigin}
    {\pgfpoint{#2}{#2}}{\pgfpoint{#2}{#2}}
    {\pgfsetlinewidth{0.4pt}\pgfpathmoveto{\pgfpoint{0pt}{#2}}%
     \pgfpathlineto{\pgfpoint{#2}{0pt}}\pgfusepath{stroke}}}
\declareGapHatch{maxGapAreaHatch}{14.14214pt}
\declareGapHatch{maxGapKeyHatch}{4.24264pt}
\def\drawCurvesAndAxes{%
  \draw[profGreen,ultra thick] (0,1) -- (1,1);
  \draw[betterblue,ultra thick] \densityPath;
  \draw[->,thick] (-.035,0) -- (1.045,0) node[right] {$u$};
  \draw[->,thick] (0,-.055) -- (0,\axisTop) node[above] {Density};
  \node[below left,inner sep=1pt] at (0,-.02) {$0$};
  \draw[thin] (1,.025) -- (1,-.04) node[below,inner sep=1pt] {$1$};
  \node[left] at (0,1) {$1$};
}
\def\labelDensities{%
  \node[betterblue,inner sep=1pt] (densP) at (.78,1.36) {$p(u)$};
  \draw[->,betterblue,thick] (densP.south east) to[bend right=12] (.865,{maxgapdensity(.865)});
  \node[profGreen,inner sep=1pt] (densNull) at (.88,.70) {$p_0(u)$};
  \draw[->,profGreen,thick,shorten >=.6pt] (densNull.north) to[out=140,in=-85] (.825,1);
}

\begin{scope}
  \node[anchor=west,font=\scriptsize\bfseries] at (-.035,2.43) {(a) Observation level};
  \fill[pink,opacity=.25] (\gapLeft,0) rectangle (\gapRight,\axisTop);
  \foreach \u in {.07,.18,.29,.80,.92}{
    \draw[thin,dashed,gray!55] (\u,0) -- (\u,{maxgapdensity(\u)});
  }
  \foreach \u in {\gapLeft,\gapRight}{
    \draw[thick,dashed,profRed] (\u,0) -- (\u,1.66);
  }
  \drawCurvesAndAxes
  \fill[profblue] (\minimumAt,{maxgapdensity(\minimumAt)}) circle[radius=1.4pt];
  \foreach \u in {.07,.18,.29,\gapLeft,\gapRight,.80,.92}{
    \draw[thick] (\u,.027) -- (\u,-.043);
  }
  \tikzset{order label/.style={anchor=base,inner sep=0pt,yshift=-1.05em}}
  \foreach \i/\u/\offset in {1/.07/0pt,2/.18/0pt,3/.29/-3.5pt,4/\gapLeft/0pt,5/\gapRight/0pt,6/.80/2.5pt,7/.92/0pt}{
    \node[order label,xshift=\offset] at (\u,-.043) {$U_{\i,t}$};
  }
  \foreach \i/\a/\b in {0/0/.07,1/.07/.18,2/.18/.29,3/.29/\gapLeft,
                        4/\gapLeft/\gapRight,5/\gapRight/.80,6/.80/.92,7/.92/1}{
    \draw[<->,thin] (\a,.095) -- (\b,.095)
      node[midway,above,inner sep=.5pt] {$d_{\i,t}$};
  }
  \draw[<->,thick,profRed] (\gapLeft,1.66) -- (\gapRight,1.66)
    node[midway,above=2pt,inner sep=1pt] {Maximum spacing region \(\mathcal U_t\)}
    node[midway,below=2pt,inner sep=1pt] {(\(\text{Width}=D_t\))};
  \labelDensities
\end{scope}

\begin{scope}[shift={(1.24,0)}]
  \node[anchor=west,font=\scriptsize\bfseries] at (-.035,2.43)
    {(b) Population level};
  \fill[betterblue,opacity=.25] (\gapLeft,0) --
    plot[domain=\gapLeft:\gapRight,samples=61,smooth] (\x,{maxgapdensity(\x)})
    -- (\gapRight,0) -- cycle;
  \fill[pattern=maxGapAreaHatch,pattern color=profGreen]
    (\gapLeft,0) rectangle (\gapRight,1);
  \foreach \u/\side in {\gapLeft/-,\gapRight/+}{
    \draw[profblue,dashed,thin] (\u,0) -- (\u,1);
    \node[profblue,below,inner sep=1pt] at (\u,-.043) {$U_t^{\side}$};
  }
  \drawCurvesAndAxes
  \labelDensities
  \draw[profblue,dashed,thin] (0,.39) -- (\minimumAt,.39) -- (\minimumAt,0);
  \node[left,profblue] at (0,.39) {$p_{\min}$};
  \fill[profblue] (\minimumAt,.39) circle[radius=1.4pt];
  \node[below,profblue,inner sep=1pt] at (\minimumAt,-.043) {$u_0$};
  \matrix[anchor=center,draw,thin,inner sep=.7mm,column sep=.9mm,
          row sep=.8mm,nodes={inner sep=0pt,outer sep=0pt}] at (.50,1.95) {
    \node[minimum size=5mm,draw=betterblue,fill=betterblue,
          fill opacity=.25,line width=.35pt] {}; &
    \node[anchor=west] {$\displaystyle\widehat D_t^{\star}=\int_{U_t^-}^{U_t^+}p(u)\,\mathrm{d}u
          \overset{\text{a.s.}}{\sim}\log t/t$}; \\
    \node[minimum size=5mm,draw=profGreen,pattern=maxGapKeyHatch,
          pattern color=profGreen,line width=.35pt] {}; &
    \node[anchor=west] {$D_t=U_t^+-U_t^-
          \overset{\text{a.s.}}{\sim}(\log t/t)/p_{\min}$}; \\
  };
\end{scope}
\end{tikzpicture}
\caption{
(a) It shows how the maximum spacing $D_t$ is defined. 
Suppose the current sample size is $t=7$, 
a sample $U_{1},\ldots,U_{7}$ is observed.
The order statistics $U_{1,t},\ldots,U_{7,t}$ are indicated on the horizontal axis.
The corresponding spacings $d_{0,t},\ldots,d_{7,t}$ are shown on top of the horizontal axis.
The maximum spacing region $\mathcal{U}_t$ corresponds to the red region, 
whose width is $D_t= \Leb(\mathcal{U}_t)$.
This region will asymptotically cover the minimum point of the density of \(\pr\), which is indicated by a blue dot.
The more $\pr$ deviates from $\pr_0$, the larger the value of $D_t$ will eventually be.
(b) It shows the relationship between $\widehat{D}_t^{\star}$ and $D_t$. 
When a new observation $U_{t+1}$ arrives, it will reduce the 
maximum spacing $D_t$ if it falls inside $\mathcal{U}_t$.
Under $\mathcal{H}_1$, the probability is the blue area, i.e., 
$\widehat{D}_t^{\star}=\pr(D_{t+1}<D_t\mid\mathcal{F}_t)$.
Under $\mathcal{H}_0$, the probability is the green
shaded area. The dot marks $p_{\min}=p(u_0)$.
In particular, 
under Condition \ref{cond:simple_strategy} to be introduced in Section \ref{sec:maxSpacing_mainResult}, 
if $p_{\min}>0$, then 
$\widehat{D}_t^{\star}\sim \log t / t$ and
$D_t \sim (\log t/t)/p_{\min}$ almost surely as $t\rightarrow\infty$.
Since $p_{\min}<1$ under $\mathcal{H}_1$, 
$D_t > \widehat{D}_t^{\star}$ asymptotically almost surely. 
}
\label{fig:H0_faster_covergence}
\end{figure}

At time $t$, we observe the data $X_{1:t} = (X_1, \ldots, X_t)$.
The spacings
$d_{i,t} = U_{i+1,t}- U_{i,t}$
($i=0,\ldots,t$)
defined in (\ref{eqt:spacings})
are the building blocks.
Define the maximum spacing and the indicator for breaking the maximum spacing record at time $t$ as
\begin{equation}\label{eq:max_Gt}
    D_t = \max_{i=0, \ldots, t} d_{i,t}
    \qquad \text{and} \qquad
    I_t = \mathbb{1}(D_t < D_{t-1}).
\end{equation}
Under $\mathcal{H}_0$, we have $U_t\sim\Unif[0,1]$,
which creates the maximum spacing $D_t\inAS 0$ faster than
under any distribution in $\mathcal{H}_1$; see
Figure~\ref{fig:H0_faster_covergence}(a) for a schematic illustration.
So, $D_t$ is a distinguishing signal between $\cH_0$ and $\cH_1$.
It motivates us to build our test on $D_t$.

To define an anytime-valid test,
we use the maximum spacing record-breaking indicator $I_t$,
which satisfies that
$[I_t \mid \mathcal{F}_{t-1}] \sim \Bern\{\E_\pr( I_t \mid \mathcal{F}_{t-1})\}$.
The null-restricted and unrestricted estimators of
the maximum spacing record-breaking probability
$\E_\pr( I_t \mid \mathcal{F}_{t-1})$ are
\begin{align}\label{eqt:D_Dhat}
    D_{t-1} = \E_{\pr_0}( I_t \mid \mathcal{F}_{t-1})
    \qquad \text{and} \qquad
    \widehat{D}_{t-1} = \widehat{\E}_\pr( I_t \mid \mathcal{F}_{t-1}) =: g_{t-1}(U_{1:t-1}) ,
\end{align}
where $\widehat{\E}_\pr(\cdot)$ is an estimator of $\E_{\pr}(\cdot)$
defined via a function
$g_{t-1}: [0, 1]^{t-1} \rightarrow [0, 1]$ to be specified.
Then the proposed $e$-spacing likelihood ratio test (LRT) statistic
at time $n$ is
\begin{align}\label{def:e_process}
    \Lambda_n
        = \prod_{t=1}^n \lambda_t,
    \quad \text{where} \quad
    \lambda_1=1,
    \quad
    \lambda_t
        =
            \left(\frac{\widehat{D}_{t-1}}{D_{t-1}}\right)^{I_t}
            \left(\frac{1-\widehat{D}_{t-1}}{1-D_{t-1}} \right)^{1-I_t} \quad (t=2,3,\ldots) .
\end{align}
Here we define \(\lambda_t = 1\) whenever
\(D_{t-1},\widehat{D}_{t-1} \in \{0,1\}\).
We discuss $\lambda_t$ and $\Lambda_n$ in the sequel:

First, the definition of $\lambda_t$ in (\ref{def:e_process}) is inspired by likelihood ratios, specifically
``likelihood-ratio-type $e$-variables'' for parametric tests; see e.g.,  \cite{hao2024values,hao2024Ksample,grunwald2024optimal,hao2025values} for testing exponential families.
In nonparametric settings, feasible likelihood ratios are harder to construct.
Our approach bridges this by linking $\pr_0$ and $\pr$ to Bernoulli models via spacings.

Second, $D_{t-1}\in\mathcal{F}_{t-1}$ and
$t\mapsto D_{t}$ is non-increasing.
When a new observation $U_t$ arrives,
either
$\{D_t< D_{t-1} \}$ or
its complement $\{D_t= D_{t-1} \}$ occurs.
Under $\mathcal{H}_0$, $U_t \sim \Unif[0,1]$ and is independent of $U_{1:t-1}$, hence,
$\pr_0(D_t< D_{t-1}\mid \mathcal{F}_{t-1}) = D_{t-1}$, i.e.,
            $[I_t \mid \mathcal{F}_{t-1}]\sim \Bern(D_{t-1})$
            under $\mathcal{H}_0$.
            So, we call $D_{t-1}$ the null-restricted estimator
            of $\E_\pr( I_t \mid \mathcal{F}_{t-1})$ in (\ref{eqt:D_Dhat}).
On the other hand,
the user-specified function $g_{t-1}: [0, 1]^{t-1} \rightarrow [0, 1]$
            is selected such that $\widehat{D}_{t-1} = g_{t-1}(U_{1:t-1})$
            is an estimator of $\E_\pr( I_t \mid \mathcal{F}_{t-1})$
            under $\mathcal{H}_0\cup\mathcal{H}_1$
            based on the observed data $U_{1:t-1}$.
So, we similarly interpret $\widehat{D}_{t-1}$ as an estimator of
$\pr(D_t< D_{t-1} \mid \mathcal{F}_{t-1})$, hence,
we call $\widehat{D}_{t-1}$ the unrestricted estimator.
Thus, combining the interpretation of $D_{t-1}$ and $\widehat{D}_{t-1}$,
the statistic $\lambda_t$ is a conditional likelihood ratio
between two Bernoulli distributions for $[I_t \mid \mathcal{F}_{t-1}]$.
            We remark that
            $\widehat{D}_{t-1}$, or equivalently $g_{t-1}$,
            can be chosen arbitrarily and does not affect
            the validity of test; see Proposition \ref{prop:eprocess}.
            For example,
            it can be any estimator of $\E_\pr( I_t \mid \mathcal{F}_{t-1})$,
            or even a constant.
            But, clearly, the choice of $\widehat{D}_{t-1}$ affects the power of the test.
            In Section \ref{sec:general_aggreation},
            we provide a generalized form of $D_t$ based on $m$th order spacing (\ref{eqt:mspacings}).

        Third, the proposed $e$-spacing LRT (\ref{def:e_process}) is
            closely related to the prequential approach \citep{dawid1984present}.
            Here, we sequentially estimate the time-varying
            quantity $\E_\pr( I_t \mid \mathcal{F}_{t-1})$
            for defining the LRT statistic for inferring $\pr$,
            whereas
            the classical prequential approach estimates a fixed unknown
            parameter sequentially and has a stronger favor on prediction.
            Indeed, \cite{dawid1992prequential} connects the prequential approach with Bayesian inference;
            see also \cite{dawid1997prequential,bj/1173707098}.
            Later,
            \cite{grunwald2007minimum} presents the prequential approach from the perspective of minimum description length.

    Fourth, \cite{zhou2024sequential,martinez2024sequential} also developed
    anytime-valid GoF tests using $e$-processes, however, their approaches are based on KSD, which differ fundamentally from ours. KSD is defined via the Stein operator
$
    (T_p h)(x)
    = \sum_{i=1}^d \{ h_i(x) \frac{\partial }{\partial x_i}\log p_0(x)
+ \frac{\partial}{\partial x_i} h_i(x) \}
$
for $h$
in a \(d\)-dimensional reproducing kernel Hilbert space.
The Stein operator requires that \(F_0^X\) has a differentiable density $p_0$.
In contrast, our $e$-spacing method only requires that
$p_0$ is continuous.
The simulation in Section~\ref{sec:simulations} shows that the $e$-spacing method is more powerful than the KSD-based methods in our considered models.

We conclude this subsection with Proposition \ref{prop:eprocess} showing that
the nonnegative process $\{\Lambda_n\}_{n\in\mathbb{Z}^+}$
is an $e$-process, and state explicitly the testing procedure.

\begin{proposition}\label{prop:eprocess}
Suppose $\widehat{D}_{t-1}$ is $\mathcal{F}_{t-1}$-measurable and takes values in $[0, 1]$. Then the stochastic process $\{ \Lambda_n\}_{n\in\mathbb{Z}^+}$ is an $e$-process
with respect to the filtration
$\{ \mathcal{F}_n \}_{n\in\mathbb{Z}^+}$
under $\mathcal{H}_0$,
i.e., for any stopping time $\tau$ with respect to
$\{ \mathcal{F}_n \}_{n\in\mathbb{Z}^+}$,
$\E_{\pr_0} (\Lambda_\tau) \leq 1$.
\end{proposition}

By Proposition \ref{prop:eprocess},
$\Lambda_\tau$
is an $e$-variable for any stopping time $\tau$.
Thus, $\Lambda_n$ is named an $e$-spacing LRT statistic.
Consequently, $\pr_0(\exists \tau < \infty: \Lambda_\tau \ge 1/\alpha) \leq \alpha$
for any $\alpha\in(0,1)$; see \cite{ramdas2020admissible,ramdas2023savi}.
So, an anytime-valid level-$\alpha$ $e$-spacing
LRT at time $n$ is
\begin{align}\label{eqt:test}
    \psi_n = \mathbb{1}\left( \Lambda_n \ge 1/\alpha\right),
\end{align}
i.e., $\mathcal{H}_0$ is rejected at level $\alpha$ at time $n$ if and only if $\psi_n = 1$.
This test maintains level $\alpha$ even if the test is performed sequentially.
The validity is non-asymptotic and works for all $n>1$.

\subsection{General notation}
To facilitate presentation,
we define our proposed $e$-spacing LRT statistic as a function for
future development in the remaining part of this article:

\begin{definition}[$e$-spacing statistic function]\label{def:espacing}
Let $I_t= \mathbb{1}(U_t \in \mathcal{U}_{t-1}) \in\{0,1\}$ be any $\mathcal{F}_t$-measurable record breaking indicator,
where
$U_t = F^X_0(X_t)$ and
$\mathcal{U}_{t-1}\subseteq [0,1]$ is a $\mathcal{F}_{t-1}$-measurable region.
Let $D_t, \widehat{D}_t \in [0,1]$ be any $\mathcal{F}_t$-measurable statistics,
representing
null-restricted and unrestricted estimators of
the record-breaking probability $\E_\pr( I_t \mid \mathcal{F}_{t-1})$.
Generally, we define the $e$-spacing LRT statistic function as
\begin{align}
    \Lambda_n(U_{1:n},D_{1:n-1}, \widehat{D}_{1:n-1}, \mathcal{U}_{1:n-1})
        &= \prod\limits_{t=1}^n\lambda_t(U_t,D_{t-1}, \widehat{D}_{t-1}, \mathcal{U}_{t-1}) ,\quad \text{where} \quad \label{def:GENe_process} \\
    \lambda_t(U_t,D_{t-1}, \widehat{D}_{t-1}, \mathcal{U}_{t-1})
        &=
            \left(\frac{\widehat{D}_{t-1}}{D_{t-1}}\right)^{I_t}
            \left(\frac{1-\widehat{D}_{t-1}}{1-D_{t-1}} \right)^{1-I_t}\quad (t=2,3,\ldots) .
            \label{def:GENe_process_lambda}
\end{align}
Here we define \(\lambda_t(U_t,D_{t-1}, \widehat{D}_{t-1}, \mathcal{U}_{t-1}) = 1\) when
$t=1$ or \(D_{t-1},\widehat{D}_{t-1} \in \{0,1\}\).
\end{definition}

Once the forms of $\mathcal{U}_{t-1}$, $D_{t-1}$ and $\widehat{D}_{t-1}$ are determined,
we can compute the $e$-spacing LRT statistic based on Definition \ref{def:espacing}.
A summary of our proposed
$D_{t-1}$ and $\widehat{D}_{t-1}$ to be developed in this article
is shown in Table~\ref{tab:examples}.
We remark that
Proposition \ref{prop:eprocess}
remains true for any choices of $D_{t-1}$ and $\widehat{D}_{t-1}$
presented in Table~\ref{tab:examples}.

We also remark that, in the maximum‑spacing strategy,
    we have
    \(\mathbb{1}(U_t \in \mathcal{U}_{t-1}) = \mathbb{1}(D_t < D_{t-1})\).
    However, in more general strategies, this equivalence need not hold; see Section~\ref{sec:general_aggreation}.

\begin{table}[t]
\centering
\begin{tabular}{llll}
\toprule
Principle & $I_t$ & $D_{t-1}$ & $\widehat{D}_{t-1}$ \\ \hline
Maximum spacing (\S\ref{sec:maxSpacing})
& (\ref{eqt:record_breaking_maxSpacing})
& $D_{t-1}^{(1,\infty)} = \max_{i=0, \ldots, t-1} d_{i,t-1}$ (\ref{eqt:Dtildestar})
& $\widehat{D}_{t-1}^{\md}= \log t/t$ (\ref{eqt:Dtildestar}) \\
Maximum spacing set (\S\ref{sec:general_aggreation}) & (\ref{eqt:aggregate_I_mathcalI_mathcalU})
& $D_{t-1}^{(m,k)} = \sum_{i\in\mathcal{I}_{t-1}^{(m,k)}} d_{i,t-1}^{(m)}$ (\ref{eq:sum_half_max_spacings})
& $\widehat{D}_{t-1}^{(m,k)}= \lfloor t/k\rfloor/t$ (\ref{eq:sum_half_max_spacings})\\
\bottomrule
\end{tabular}
\caption{A summary of choices of record breaking indicator $I_t$,
null-restricted estimator $D_{t-1}$, and
unrestricted estimator $\widehat{D}_{t-1}$
used in (\ref{def:GENe_process}).
The optimal $k$ can be selected according to Theorem \ref{thm:divergent_mathcalK_optimal_softmax}.
The parameter $m=m_t = \lfloor t^{\phi} \rfloor$ with $\phi\in(0.5, 1)$
ensures testing consistency according to Theorem \ref{thm:sum_spacings_strategy}.
We set $m = \lfloor t^{0.51} \rfloor$ by default.}
\label{tab:examples}
\end{table}

\section{Maximum spacing}\label{sec:maxSpacing}

\subsection{Optimal $\widehat{D}_{t-1}$}\label{sec:oracle_log_optimal}
From (\ref{eq:max_Gt}),
we test $\mathcal{H}_0$ by assessing the likelihood
of breaking the record of maximum spacing $D_{t-1} = \max_{i=0, \ldots, t-1} d_{i,t-1}$.
So, the record-breaking indicator is
\begin{align}\label{eqt:record_breaking_maxSpacing}
	I_{t}
        = \mathbb{1}(D_t < D_{t-1})
		= \mathbb{1}(U_{t}\in\mathcal{U}_{t-1}) ,
    \qquad\text{where} \qquad
    \mathcal{U}_{t-1}
        = [U_{t-1}^-, U_{t-1}^+]
\end{align}
is the interval corresponds to the maximum spacing, i.e.,
\(D_{t-1} = |\mathcal{U}_{t-1}| = U_{t-1}^+ - U_{t-1}^-\).
When we wish to emphasize that the maximum spacing approch is used,
the variables will be denoted with superscript
${}^{(1, \infty)}$,
e.g., $\Lambda_n^{(1, \infty)} = \Lambda_n(U_{1:n},D_{1:n}^{(1, \infty)}, \widehat{D}_{1:n}^{(1, \infty)}, \mathcal{U}_{1:n-1}^{(1,\infty)})$.

We define an optimal test as
a test with the $\widehat{D}_{t-1}$ that
maximizes $\lambda_t$ in (\ref{def:GENe_process_lambda})
for each $t$ under $\cH_1$.
Consequently,
it maximizes the \(e\)-power, defined as \(\E_\pr(\log \Lambda_\tau)\); see \cite{kelly1956new,Shafer:2021,GrunwaldHK19,ramdas2023savi,waudby2024estimating}.
This principle also leads to various results
in gambling, information theory, finance, and machine learning; see, e.g.,  \cite{breimualphanu2011optimal,Cover2006,luenberger1997investment,cesa2006prediction}.

Accordingly, we aim at choosing the sequence
$\widehat{{D}} = (\widehat{D}_1,\widehat{D}_2, \ldots)$ to maximize the
$e$-power $\E_\pr(\log \Lambda_\tau)$.
Let $\mathcal{D}_t$ denote the class of all maps
$\widehat{D}_{t}: [0, 1]^{t} \to [0, 1]$ for $t \in \mathbb{Z}^+$.
Denote $\mathcal{D} = \mathcal{D}_1\times \mathcal{D}_2\times \cdots$.
Formally, we seek to find the best
$\widehat{D}^{\star} = (\widehat{D}_1^{\star}, \widehat{D}_2^{\star}, \ldots)$
defined as
\begin{align}
    \widehat{{D}}^{\star} := \argmax_{\widehat{D}\in \mathcal{D}} \E_\pr(\log \Lambda_\tau)
    &= \argmax_{\widehat{D}\in \mathcal{D}} \sum_{t=1}^\infty \E_{\pr} \E_{\pr}\{\mathbb{1}(t \leq \tau) \log \lambda_t \mid \mathcal{F}_{t-1}\}.\label{eq:optimal_g}
\end{align}
Denote the $\lambda_t$ and $\Lambda_t$
as $\lambda_t^{\star}$ and $\Lambda_t^{\star}$
when $\widehat{{D}} = \widehat{{D}}^{\star}$.
We call $\{\Lambda_t^{\star}\}_{t \geq 1}$
the \emph{oracle log-optimal $e$-process}.
From (\ref{eq:optimal_g}), maximizing $\E_\pr(\log \Lambda_\tau)$ is equivalent to
maximizing $\E_\pr(\log \lambda_t \mid \mathcal{F}_{t-1})$ for each $t\in \{1, \ldots, \tau\}$.
Throughout this paper, the term \emph{oracle log-optimal $e$-process} refers to the $e$-process constructed using $\widehat{D}^{\star}$. It does not imply optimality among all $e$-processes based on the observations.
The solution to (\ref{eq:optimal_g}) is provided in Theorem~\ref{thm:log_optimal}.

\begin{theorem}[Optimality]\label{thm:log_optimal}
Suppose $F$ is continuous and
$0 < D_{t-1} < 1$.
Then
\begin{align}\label{eqt:hatDoptiomal}
    \widehat{D}_{t-1}^{\star} &= \pr\left(D_t < D_{t-1} \mid \mathcal{F}_{t-1} \right),\\
\max_{\widehat{D}_{t-1} \in [0, 1]} \E_\pr\left(\log \lambda_t \mid \mathcal{F}_{t-1}\right)
&=
\E_\pr\left(\log \lambda_t^{\star} \mid \mathcal{F}_{t-1}\right)
= \KL\{ \Bern(\widehat{D}_{t-1}^{\star}) \| \Bern(D_{t-1}) \}, \nonumber
\end{align}
where $\lambda_t^{\star}$ is defined in (\ref{def:GENe_process}) with $\widehat{D}_{t-1} = \widehat{D}_{t-1}^{\star}$.
\end{theorem}

By Theorem \ref{thm:log_optimal}, we have \(\E_\pr\left(\log \lambda_t^{\star} \mid \mathcal{F}_{t-1}\right) = 0\) iff \(\widehat{D}_{t-1}^{\star} = D_{t-1}\).
Moreover, at time \(t\),
\(\E_\pr(\log \Lambda_t^{\star}) = 0\) if and only if \(\pr = \Unif[0,1]\). Equivalently, for any \(t\), if \(\pr \neq \Unif[0,1]\), then \(\E_\pr(\log \Lambda_t^{\star}) > 0\).
The quantity \(\widehat{D}_{t-1}^{\star} = \pr(D_t < D_{t-1} \mid \mathcal{F}_{t-1}) = \E_{\pr}[I_t \mid \mathcal{F}_{t-1}]\) is an oracle choice, i.e.,
unavailable in practice because \(\pr\) is unknown. By Theorem~\ref{thm:log_optimal}, \(\widehat{D}_{t-1}\) is interpreted as an estimator of \(\E_{\pr}( I_t \mid \mathcal{F}_{t-1}) \), as described in Section~\ref{sec:methodology}.

\subsection{Minimum density estimator for $\widehat{D}_{t-1}$}\label{sec:maxSpacing_mainResult}

A good practical choice of \(\widehat{D}_{t-1}\) should approximate
the optimal \(\widehat{D}_{t-1}^{\star} = \pr\left(D_t < D_{t-1} \mid \mathcal{F}_{t-1}\right)\),
as stated in Theorem \ref{thm:log_optimal}.
In this subsection, we derive $\widehat{D}_{t-1}$ based on the asymptotic theory
of the maximum spacing.
It leads us to a neat choice of \(\widehat{D}_{t-1} = \log t/t\).

We begin with the motivation.
Our goal is to find an estimator
\(\widehat{D}_{t-1}\) such that
\begin{equation}\label{eq:hatD_ratio}
    \widehat{D}_{t-1} \sim  \widehat{D}_{t-1}^{\star} \equiv \pr\left(D_t < D_{t-1} \mid \mathcal{F}_{t-1}\right)
\end{equation}
almost surely,
where $a_t \sim b_t$ means $a_t/b_t \to 1$ as $t\to\infty$.
Theorem~3 of~\cite{deheuvels1984strong},
which is copied in Lemma \ref*{lem:Gt_locates_low_den} for reference,
shows that the maximum spacing almost surely occurs near
the minimal density and shrinks to $0$,
provided that $p_{\min} := \min_{u\in [0, 1]} p(u) > 0$.
Hence,
\begin{align}\label{eqt:approx_int}
    \pr\left(D_t < D_{t-1} \mid \mathcal{F}_{t-1}\right)
        = \int_{U_{t-1}^-}^{U_{t-1}^+} p(u) \, \dd u
        \sim (U_{t-1}^+-U_{t-1}^-) p_{\min}
        = D_{t-1} p_{\min}
\end{align}
almost surely.
Next, we note the following properties of $D_{t-1}$.
Under $\mathcal{H}_0$,
$D_{t-1} = \log t/t + O((\log \log t) / t )$ almost surely
by Theorem~3.1 in~\cite{holst1980lengths};
see also \cite{slud1978entropy,devroye1981laws}.
Under $\mathcal{H}_1$,
Theorem~4 of~\cite{deheuvels1984strong} shows that for any continuous distribution $\pr$
with $p_{\min}>0$ such that there exists $u_0\in(0,1)$ satisfying
$p(u)>p(u_0)>0$ for all $u\in(0,u_0)\cup(u_0,1)$
and
$
    \limsup_{h\downarrow0}\{ p(u_0+h)-p(u_0)\}/h  = a \in(0,\infty),
$
we have
\begin{equation}\label{eq:asymt_Dt}
    D_{t-1} \sim \frac{\log t}{ t  } \times \frac{1}{p_{\min}}
\end{equation}
almost surely.
Putting (\ref{eqt:approx_int})
and (\ref{eq:asymt_Dt})
into (\ref{eq:hatD_ratio}), we obtain
$\widehat{D}_{t-1} \sim  \log t /t$ almost surely.
We call it the minimum density estimator of $\widehat{D}_{t-1}^{\star}$.
Then, the resulting $\Lambda_n$, $D_{t-1}$ and $\widehat{D}_t$
are
\begin{align}\label{eqt:Dtildestar}
    \Lambda_n^{\md}
    = \Lambda_n(U_{1:n},D_{1:n-1}^{\md}, \widehat{D}_{1:n-1}^{\md}, \mathcal{U}_{1:n-1}^{(1,\infty)}),
    \quad
    D_{t-1}^{\md} = \max_{i=0, \ldots, t-1} d_{i,t-1},
    \quad
    \widehat{D}_{t-1}^{\md} = \frac{\log t}{t}.
\end{align}
The consistency of the test $\Lambda_n^{\md}$
requires that $\pr$ and $\pr_0$ are sufficiently different from each other.
We measure the difference in terms of the magnitudes of
$D_{t-1} = \E_{\pr_0}(I_t \mid \mathcal{F}_{t-1})$ and
$\widehat{D}_{t-1}^{\star} = \E(I_t \mid \mathcal{F}_{t-1})$, as stated below:

\begin{condition}[Well separation of $\pr$ and $\pr_0$]\label{cond:simple_strategy}
The measures $\pr$ and $\pr_0$ satisfy that
\begin{itemize}
    \item Case 1: $p_{\min}>0$.
        We require
\begin{align}
    D_{t-1} &= c \times \frac{\log t}{t} \{ 1+o_{\AS}(1)\}, \qquad \text{where} \qquad
    c = c_\pr \in(1, \infty),  \label{eqt:cond_D}\\
    \widehat{D}_{t-1}^{\star}
    &= \frac{\log t}{t} \{ 1+o_{\AS}(1)\}
    .\label{eq:cond_Q}
\end{align}
    \item Case 2: $p_{\min}=0$.
    We require $\limsup_{t\rightarrow\infty} D_{t-1}<1$,
\begin{align}
    D_{t-1}&\gg  \frac{\log t}{t}  \qquad \text{almost surely},  \label{eqt:cond_D=0}\\
    \widehat{D}_{t-1}^{\star}
    &= D_{t-1}\times o_{\AS}\left(\frac{1}{\log \delta_t} \right), \qquad \text{where} \qquad \delta_t = \frac{D_{t-1}}{\log t /t}.\label{eq:cond_Q=0}
\end{align}
\end{itemize}
\end{condition}

Clearly, Condition~\ref{cond:simple_strategy} does not hold under \(\cH_0\).
So, Condition~\ref{cond:simple_strategy} puts assumptions on $\mathcal{H}_1$,
which requires that
the maximum spacing must be larger than the null benchmark in both cases.
Essentially, it requires that
$|D_{t-1}-\widehat{D}_{t-1}^{\star}|$ is sufficiently large under $\mathcal{H}_1$.
Consequently, we expect that the test is most powerful
if the localized difference $|p(u)-p_0(u)|$ is large around
the minimizer $u_0$ of the density $p(\cdot)$.
In Case 1,
under \(\mathcal{H}_1\), we have \(p_{\min}<1\). Therefore, requirement~(\ref{eqt:cond_D}) in Condition~\ref{cond:simple_strategy} implies~(\ref{eq:asymt_Dt}), which holds under very mild assumptions on the data-generating mechanism under \(\mathcal{H}_1\). Hence, requirement~(\ref{eqt:cond_D}) is reasonable.
The idea is visualized in Figure \ref{fig:H0_faster_covergence}(b).
Condition~(\ref{eq:cond_Q}) follows directly from Theorem~4 of~\cite{deheuvels1984strong}, i.e., combining~(\ref{eqt:approx_int}) and~(\ref{eq:asymt_Dt}), provided that \(\pr\) satisfies the regularity condition stated above~(\ref{eq:asymt_Dt}). In Case 2,
we require $D_{t-1}$ to be bounded away from 1 almost surely,
which is mild as the maximum spacing $D_{t-1}$ should not approach 1 as $t\rightarrow\infty$. Requirement~(\ref{eqt:cond_D=0}) is also natural when
$p_{\min}=0$, because
the regions where $p(u)\approx 0$ yield sparse samples, making the largest spacing
even larger than in the case that $p_{\min}$
is sufficiently small.
Condition~(\ref{eq:cond_Q=0}) can be interpreted as a strengthened version of (\ref{eqt:approx_int}).
To see it, we observe that
$\widehat{D}_{t-1}^{\star}/D_{t-1} = \int_{\mathcal{U}_{t-1}} p(u) \, \dd u/ |\mathcal{U}_{t-1}|$
is the average density over the maximum-spacing interval $\mathcal{U}_{t-1}$.
Hence, (\ref{eq:cond_Q=0}) means that
$\int_{\mathcal{U}_{t-1}} p(u) \, \dd u/ |\mathcal{U}_{t-1}|\inAS 0$ sufficiently quickly
as $o_{\AS}(1/\log \delta_t)$, where $\delta_t \inAS \infty$.
This is natural when $p_{\min}=0$, because $\mathcal{U}_{t-1}$ eventually lie in a low-density region.
Moreover, $o_{\AS}(1/\log \delta_t)$ tends to zero very slowly, as we allows $\delta_t\inAS\infty$ very slowly according to (\ref{eqt:cond_D=0}).
So, (\ref{eq:cond_Q=0}) is just a mild
strengthening of $\widehat{D}_{t-1}^{\star}/D_{t-1} \inAS 0$.
For example, if $U_i \simIID \Beta(2,1)$, then $p(u)=2u\mathbb{1}(0\leq u\leq 1)$.
In this case, $D_{t-1} \asymp t^{-1/2}$ and $\widehat{D}_{t-1}^{\star}/D_{t-1} \asymp (\log t)/t^{1/2} = o_{\AS}(1/\log \delta_t)$, so both (\ref{eqt:cond_D=0})--(\ref{eq:cond_Q=0}) hold.
Generally, simple sufficient conditions to ensure
Condition \ref{cond:simple_strategy} hold are
\begin{itemize}
    \item Case 1: $D_{t-1} \sim c \log t/t$
            and
            $\widehat{D}_{t-1}^{\star} \sim \log t/t$ for some $c\in(1,\infty)$.
    \item Case 2: $t^{-\beta}\lesssim D_{t-1} \lesssim 1$ and
    $\widehat{D}_{t-1}^{\star} \lesssim \log t/t$
    for some $\beta\in(0,1)$.
\end{itemize}
We remark that the conditions for cases 1--2
are different because the asymptotic behaviors of $D_{t-1}$ are usually different in scales:
$D_{t-1}\asymp \log t/t$ and $D_{t-1}\gg \log t/t$, respectively.
Also,
we argue that Condition (\ref{eq:cond_Q=0}) is mild and
easily verifiable; see the example below:
\begin{example}\label{example:delta_t_to_infty}
If $F$ is atomless and
$\delta_t \inAS \infty$,
then,
\begin{align*}
    0\leq \frac{\widehat D_{t-1}^{\star}}{D_{t-1}}
    = \frac{F(U_{t-1}^{+})-F(U_{t-1}^{-})}{D_{t-1}}
    &\leq \frac{\max_{0\leq i< t} \{F(U_{i+1,t-1})- F(U_{i,t-1}\} )}{D_{t-1}}\\
    &\overset{(a)}{=}\frac{O_{\AS}(\log t/t)}{D_{t-1}}
    = O_{\AS}\left(\frac{1}{\delta_t}\right)
    = o_{\AS}\left( \frac{1}{\log\delta_t} \right)
    \to0 .
\end{align*}
where $(a)$ holds by (\ref{eq:asymt_Dt}) with $p_{\min} =1$ since $F(U_i)\sim \Unif[0,1]$.
Consequently, (\ref{eq:cond_Q=0}) holds.
Some specific examples are provided in Section \ref*{proof:examples}
of the supplement.
\qed
\end{example}

Based on Condition \ref{cond:simple_strategy},
the consistency of the proposed test can be proved:
\begin{theorem}[Consistency]\label{thm:maxSpacing_strategy}
Let $(D_t, \widehat{D}_t) = (D_t^{\md}, \widehat{D}_t^{\md})$ for all $t$,
according to (\ref{eqt:Dtildestar}).
Under $\mathcal{H}_1$,
if Condition~\ref{cond:simple_strategy} holds, then as $n \rightarrow \infty$
\begin{enumerate}
    \item in Case 1, $\log \Lambda_n^{\md} /L_n \inAS c_{\pr}-1-\log c_{\pr} >0$,
        where $L_n = \sum_{t=2}^n \log t/t \asymp (\log n)^2$,
    \item in Case 2, $\log \Lambda_n^{\md} \geq  S_n/8 + O_{\AS}(1)$,
        where $S_n = \sum_{t=2}^n D_{t-1}  \gg (\log n)^2$, and
    \item in either case, $\Lambda_n^{\md} \inAS \infty$ and $\E(\Lambda_n^{\md}) \rightarrow \infty$.
\end{enumerate}
\end{theorem}

Theorem~\ref{thm:maxSpacing_strategy} shows that the proposed test
is consistent, i.e.,
$\pr(\sup_{t\in\mathbb{Z}^+} \psi_{t} = 1) = 1$ under $\cH_1$,
i.e., the proposed anytime-valid test $\psi$ in (\ref{eqt:test})
rejects $\mathcal{H}_0$ under any distribution in
$\mathcal{H}_1$ with probability one, provided that
Condition~\ref{cond:simple_strategy} is satisfied.
In the following proposition,
we further show
that the $e$-process based on \(\widehat{D}_{t-1} =\widehat{D}_{t-1}^{\md} \) is asymptotically equivalent,
in terms of the logarithmic growth of the $e$-process,
to the ``oracle'' $e$-process using
\(\widehat{D}_{t-1} =\widehat{D}^{\star}_{t-1} =: \widehat{D}^{\star\md}_{t-1}\)
as defined in (\ref{eqt:hatDoptiomal}).
Denote
\(\Lambda_n^{\star\md} = \Lambda_n(U_{1:n},D_{1:n-1}^{\md}, \widehat{D}_{1:n-1}^{\star\md}, \mathcal{U}_{1:n-1}^{(1,\infty)})\).

\begin{proposition}[Asymptotic equivalence]\label{prop:power_ratio_with_oracle}
Let $D_t = D_t^{\md}$ for all $t$, according to (\ref{eqt:Dtildestar}).
Under $\mathcal{H}_1$, if Condition~\ref{cond:simple_strategy} holds,
then
$\log \Lambda_n^{\star\md} / \log \Lambda_n^{\md} \inAS 1$.
\end{proposition}

Proposition \ref{prop:power_ratio_with_oracle} shows that
our practical test $\log\Lambda_n^{\md}\inAS\infty$
at the same speed as the oracle test $\log\Lambda_n^{\star\md}\inAS\infty$.
This is verified in our simulation studies; see Section \ref{sec:simulations}.

\section{Maximum spacing set}\label{sec:general_aggreation}
\subsection{A general framework of aggregation strategies}\label{sec:general_framework}
The maximum spacing strategy only utilizes the information
contained in one spacing interval, i.e.,
$I_t = \mathbb{1}(D_t<D_{t-1}) = \mathbb{1}(U_t \in [U_{t-1}^-, U_{t-1}^+])$
in (\ref{eqt:record_breaking_maxSpacing}),
for detecting the localized discrepancy at the minimum density.
This may reduce efficiency.
To improve it, we aggregate information from multiple intervals.
Let the selection rule $\mathcal{I}_t \subseteq \{0, \ldots, t\}$
be $\mathcal{F}_t$-measurable.
Then we generalize $I_t$ in (\ref{eqt:record_breaking_maxSpacing}) to
\begin{equation}\label{eq:general_Gt}
    I_t^\agg = \mathbb{1}(U_t \in \mathcal{U}_{t-1}^\agg) ,
    \qquad \text{where} \qquad
    \mathcal{U}_{t-1}^\agg = \bigcup_{i\in\mathcal{I}_{t-1}} [U_{i,t-1}, \; U_{i+1,t-1}].
\end{equation}
Correspondingly,
define the null-restricted and unrestricted estimators of
$\E_\pr( I_t^\agg \mid \mathcal{F}_{t-1})$ as
\begin{align}\label{eq:D_and_Dhat}
    D_{t-1}^\agg
    = \E_{\pr_0}( I_t^\agg \mid \mathcal{F}_{t-1})
    = \sum_{i\in\mathcal{I}_{t-1}} d_{i,t-1}
    \qquad \text{and} \qquad
    \widehat{D}_{t-1}^\agg
    = \widehat{\E}_\pr( I_t^\agg \mid \mathcal{F}_{t-1})
    =: g_{t-1}(U_{1:t-1}),
\end{align}
where $g_{t-1}: [0,1]^{t-1}\rightarrow[0,1]$ is as defined before.
Then define the generalized $\Lambda_n$ and $\lambda_t$ as in (\ref{def:GENe_process}).
We show that
$\{\Lambda_n^{\agg}\}_{n\in\mathbb{Z}^+}$ is an $e$-process
and derive the optimal $\widehat{D}_{t-1}^\agg$:

\begin{proposition}[Generalized Proposition~\ref{prop:eprocess}]\label{prop:gen_eprocess}
Define $\Lambda_n^{\agg}$ according to (\ref{eq:general_Gt}) and (\ref{def:GENe_process_lambda}).
The stochastic process $\{ \Lambda_n^{\agg} \}_{n\in\mathbb{Z}^+}$ is an $e$-process
with respect to the filtration
$\{ \mathcal{F}_n \}_{n\in\mathbb{Z}^+}$
under $\mathcal{H}_0$,
i.e., for any stopping time $\tau$ with respect to
$\{ \mathcal{F}_n \}_{n\in\mathbb{Z}^+}$,
$\E_{\pr_0} (\Lambda_\tau^{\agg}) \leq 1$.
\end{proposition}

\begin{theorem}[Generalized  Theorem~\ref{thm:log_optimal}]\label{thm:general_log_optimal}
Define $\Lambda_n^\agg$ and $\lambda_t^\agg$ according to (\ref{eq:general_Gt})--(\ref{eq:D_and_Dhat}).
Suppose $F$ is continuous
and $0<D_{t-1}^\agg<1$.
Then for any $t$,
\begin{align}
\widehat{D}_{t-1}^{\agg \star} &= \pr\left(U_t \in \mathcal{U}_{t-1}^\agg \mid \mathcal{F}_{t-1} \right),  \label{eq:hat_D_optimal} \\
\max_{\widehat{D}_{t-1}^\agg \in [0, 1]} \E_\pr\left(\log \lambda_t^\agg \mid \mathcal{F}_{t-1}\right)
&=
\E_\pr\left(\log \lambda_t^{\agg \star} \mid \mathcal{F}_{t-1}\right)
= \KL\{ \Bern(\widehat{D}_{t-1}^{\agg \star}) \| \Bern(D_{t-1}^{\agg})\}, \label{eq:general_optimal_epower}
\end{align}
where $\lambda_t^{\agg \star}$ is defined with
$\widehat{D}_{t-1}^\agg = \widehat{D}_{t-1}^{\agg \star}$.
\end{theorem}

Consequently, define $\widehat{D}^{\agg \star} := (\widehat{D}^{\agg \star}_1,\ldots, \widehat{D}^{\agg \star}_\tau)$.
We have
$
\widehat{D}^{\agg \star} = \argmax_{\widehat{D}^{\agg}\in \mathcal{D}^{\agg}} \E_\pr(\log \Lambda_\tau^{\agg}),
$
where $\mathcal{D}^{\agg} = \mathcal{D}_1^{\agg}\times \mathcal{D}_2^{\agg}\times \cdots$ and $\mathcal{D}_t^{\agg}$ denote the class of all maps
$\widehat{D}_{t}^{\agg}: [0, 1]^{t} \to [0, 1]$.
Similar to Theorem~\ref{thm:log_optimal}, we also obtain that,
for any \(t\), if $\mathcal{H}_1$ is true, then \(\E_{\pr}(\log \Lambda_t^{\agg \star}) > 0\).
The proposed approach is general in the sense that
            the aggregation strategy $\mathcal{I}_{t-1}$ and
            the proximal function $g_{t-1}$
            can be set flexibly at time $t$ by the users.
            The flexibility in choosing $\mathcal{I}_{t-1}$
            facilitates a balance of computational efficiency and statistical efficiency, as we argued in the next subsections.
            On the other hand, the flexibility of $g_{t-1}$
            allows users to incorporate various
            estimation strategy, prior belief,
            preferred parametric model, and
            non-parametric model.
            Some examples are shown in Section \ref*{sec:additionalExample_hatD}
            of the supplement.

\subsection{Maximum $1/k$ fraction $m$th order $e$-spacing LRT}\label{sec:max_sum_half_spacings}
In this subsection,
we propose a data-dependent aggregation strategy.
At time $t-1$,
let $m = m_{t-1} \in \mathbb{Z}^+$ be the order of spacing in (\ref{eqt:mspacings}),
which may depend on $t-1$.
We construct the new version of $D_{t-1}$, denoted as $D_{t-1}^{(m,k)}$,
in two steps.
First,
we partition the ordered sample $\{ U_{i,t-1} \}_{i=0}^{t-1}$
into
\begin{itemize}
    \item $K_{t-1}$ non-overlapping blocks each of size $m$, where $K_{t-1} := \lfloor t/m \rfloor$, and
    \item one incomplete block of size $t-K_{t-1}m = t - \lfloor t/m \rfloor m$,
which exists only when $t/m\not\in\mathbb{Z}$.
\end{itemize}
So, we can compute the $m$th order spacings $d_{i,t-1}^{(m)}$ as defined in (\ref{eqt:mspacings}), i.e.,
\begin{align}\label{eqt:d_new_definition}
    d_{i,t-1}^{(m)}
    = \left\{
    \begin{array}{ll}
        U_{(i+1)m,t-1} - U_{im,t-1}, & \text{if $i = 0, \ldots, K_{t-1}-1$},\\
        1-U_{K_{t-1} m,t-1}, & \text{if $i = K_{t-1}$  and $t/m \not\in\mathbb{Z}$}.
    \end{array}\right.
\end{align}
Second, we
record the indices $i$ of the $\lfloor (t/m)/k \rfloor$ largest spacings as
\[
    \mathcal{I}_{t-1}^{(m,k)}
    = \left\{
        \mathcal{I} \subseteq \{0, \ldots, K_{t-1}\} :
        |\mathcal{I}|= \lfloor (t/m)/k \rfloor \;\;\text{and}\;\;
        d_{i,t-1}^{(m)} > d_{i',t-1}^{(m)} \,\,\text{for}\,\, i'\not\in\mathcal{I}
    \right\},
\]
which takes approximately $1/k$ fraction of all spacings in (\ref{eqt:d_new_definition}).
Then define
\begin{align}\label{eqt:aggregate_I_mathcalI_mathcalU}
	I_{t}^{(m,k)}
		= \mathbb{1}(U_{t}\in\mathcal{U}_{t-1}^{(m,k)})
    \qquad\text{and} \qquad
    \mathcal{U}_{t-1}^{(m,k)} = \bigcup_{i\in\mathcal{I}_{t-1}^{(m,k)}} [U_{im,t-1}, \; U_{(i+1)m,t-1}],
\end{align}
where $[U_{im,t-1}, \; U_{(i+1)m,t-1}] := [U_{im,t-1}, 1]$ when $i>t/m -1$,
which happens when $t/m\not\in\mathbb{Z}$.
Motivated by Glivenko--Cantelli theorem,
we choose $\widehat{D}_{t-1}$ as
$\int_{\mathcal{U}_{t-1}^{(m,k)}} \dd \widehat{F}_{t-1}
\sim m |\mathcal{I}_{t-1}^{(m,k)}|/t \sim \lfloor t/k \rfloor / t$,
where $\widehat{F}_{t-1}$ is the empirical CDF of $U_{1:(t-1)}$;
see Section~\ref*{sec:additionalExample_hatD} and Example~\ref*{sec:Glivenko-Cantelli}
in the supplement for further discussion.
Then the aggregated statistics in (\ref{eq:D_and_Dhat}) are
\begin{equation}\label{eq:sum_half_max_spacings}
    D_{t-1}^{(m,k)} = \sum_{i\in\mathcal{I}_{t-1}^{(m,k)}}  d_{i,t-1}^{(m)}
    \qquad \text{and} \qquad
    \widehat{D}_{t-1}^{(m,k)} = \frac{\lfloor t/k \rfloor}{t}.
\end{equation}
Set $D_{t-1}^{(m,k)} = 0$ if $t < \max(m,k)$.
Consequently, our proposed $e$-process is given by
\begin{equation}\label{eq:def_mk_eprocess}
    \Lambda_n^{(m,k)}
    = \Lambda_n(U_{1:n}, D_{1:n-1}^{(m,k)}, \widehat{D}_{1:n-1}^{(m,k)},\mathcal{U}_{1:n-1}^{(m,k)})
    \;\;\text{and} \;\;
    \lambda_t^{(m,k)}
    = \lambda_t(U_t, D_{t-1}^{(m,k)}, \widehat{D}_{t-1}^{(m,k)},\mathcal{U}_{t-1}^{(m,k)}).
\end{equation}
In a nutshell,
this aggregating strategy uses the sum of the maximum $1/k$ fraction $m$th order spacings.
Before presenting the main result, we state two lemmas.

\begin{lemma}\label{lem:Gt_to_1half}
Let $U_1, U_2, \ldots \simIID \Unif[0, 1]$ and $m = \lfloor t^\phi \rfloor$, where
$\phi \in (0.5,1)$.
Define $D_{t-1}^{(m,k)}$ as in (\ref{eq:sum_half_max_spacings}).
For any fixed $k\in (1,\infty)$, we have that $D_{t-1}^{(m,k)} \inAS 1/k$.
\end{lemma}

\begin{lemma}\label{lem:sum_spacings_strategy}
Let $U_1, U_2, \ldots\simIID F \neq\Unif[0,1]$,
where $F$ has a continuous density $p$.
Let $m = \lfloor t^\phi \rfloor$ for some $\phi \in (0.5,1)$.
Let $\mathcal{U}^{(k)} = \{ u : p(u)\leq p_k\}$
be the $(1/k)$-lower level set, i.e.,
$\int_{\mathcal{U}^{(k)}} p(u) \, \dd u = 1/k$,
for some constant $p_k$.
If $\Leb\{u\in[0,1]:p(u)=p_k\}=0$,
then $\Leb(\mathcal{U}_{t-1}^{(m,k)}\triangle\mathcal{U}^{(k)}) \inAS 0$,
\[
    D_{t-1}^{(m,k)} \inAS \Leb(\mathcal{U}^{(k)}) =: \pi_k \in(1/k,1), \qquad \text{and} \qquad
    \pr(U_t \in \mathcal{U}_{t-1}^{(m,k)} \mid \mathcal{F}_{t-1}) \inAS \frac{1}{k},
\]
where $\triangle$ is symmetric difference.
\end{lemma}

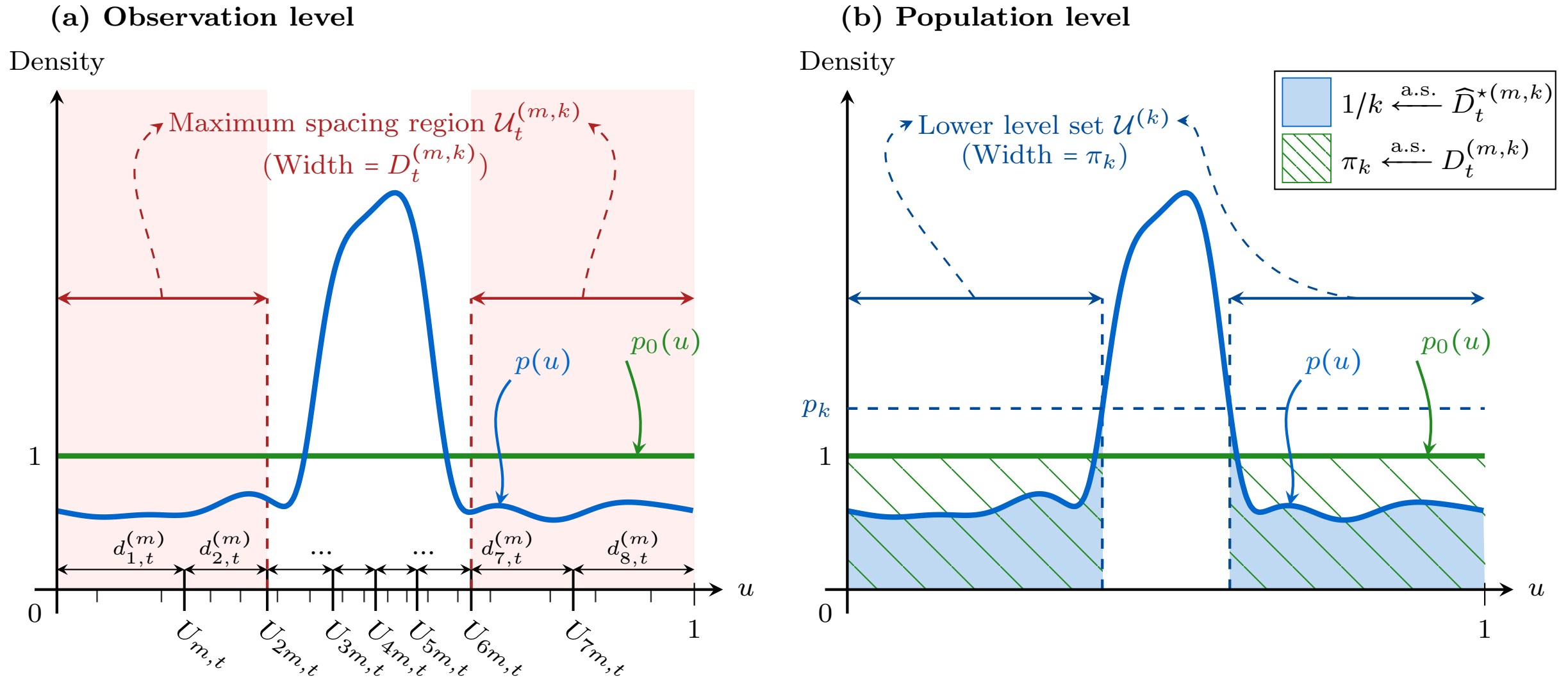
\begin{figure}[t]
    \centering
\begin{tikzpicture}[x=0.405\linewidth,y=1.38cm,font=\scriptsize] 
\tikzset{>=stealth, every path/.style={thick}} 
\def\ybase{1} 
\def\arrowy{0.15}
\definecolor{profGreen}{RGB}{34,139,34}    
\definecolor{profRed}{RGB}{178,34,34}      
\definecolor{betterblue}{RGB}{0,102,204}
\colorlet{profblue}{betterblue!75!black}
\definecolor{paleblue}{RGB}{165, 195, 220}
\definecolor{lightGreen}{RGB}{144,238,144} 
\def\leftCut{0.40}
\def\rightCut{0.60}
\def\leftObserved{0.33}
\def\rightObserved{0.65}
\def\plotTop{3.74}
\def\spanHeight{2.18}
\pgfmathdeclarefunction{spacingdensity}{1}{%
    \pgfmathparse{exp(-(((#1-0.5)/0.10)^4))}%
    \let\spacingcore\pgfmathresult
    \pgfmathparse{cos(360*(#1-0.5))}%
    \let\spacingwave\pgfmathresult
    \pgfmathparse{(#1-0.5)/0.10}%
    \let\peakoffset\pgfmathresult
    \pgfmathparse{0.589073512440+2.278118747811*(\spacingcore
        +0.055*(1-\spacingcore)*cos(180*(#1-0.5))^2*cos(1800*(#1-0.5)))
        +0.045*sin(360*(#1-0.5))*(\spacingwave-cos(36))
        *(3*\spacingwave+2-cos(36))
        +0.46*\peakoffset*(max(0,1-(\peakoffset)^2))^2}%
}
\pgfmathsetmacro{\cutoff}{spacingdensity(\leftCut)}
\def\spacingDensityPath{plot[domain=0:1,samples=151,smooth]
    (\x,{spacingdensity(\x)})}
\usetikzlibrary{patterns}
\tikzset{label arrow/.style={->,semithick,dashed},
         density arrow/.style={->,>=stealth,thick,out=229,in=78}}
\def\declareSpacingHatch#1#2{%
  \pgfdeclarepatternformonly{#1}{\pgfpointorigin}
    {\pgfpoint{#2}{#2}}{\pgfpoint{#2}{#2}}
    {\pgfsetlinewidth{0.4pt}\pgfpathmoveto{\pgfpoint{0pt}{#2}}
     \pgfpathlineto{\pgfpoint{#2}{0pt}}\pgfusepath{stroke}}}
\declareSpacingHatch{spacingNullHatch}{14.14214pt}
\declareSpacingHatch{spacingLegendHatch}{4.24264pt}

\def\drawDensityAxes{%
    \draw[profGreen,ultra thick] (0,\ybase) -- (1,\ybase);
    \draw[betterblue,ultra thick] \spacingDensityPath;
    \draw[->,thick] (-0.035,0) -- (1.045,0) node[right] {$u$};
    \draw[->,thick] (0,-0.18) -- (0,\plotTop) node[above] {Density};
    \node[below left] at (0,0) {$0$};
    \draw[thin] (1,0.05) -- (1,-0.12);
    \node[below] at (1,-0.12) {$1$};
    \node[left] at (0,\ybase) {$1$};
}
\def\labelDensities{%
    \node[betterblue,inner sep=1pt] (LP) at (0.764,{\cutoff+0.35}) {$p(u)$};
    \draw[density arrow,betterblue] (LP.south west) to (0.694,{spacingdensity(0.694)});
    \node[profGreen,inner sep=1pt,anchor=east] (LPzero) at (1.02,1.85) {$p_0(u)$};
    \draw[density arrow,profGreen] (LPzero.south west) to [out=290,in=80] (0.91,\ybase);
}
\def\shadeInterval#1#2#3{%
    \begin{scope}[every path/.style={}]
        \clip (#1,0) rectangle (#2,\plotTop);
        \fill[#3,opacity=0.25] \spacingDensityPath -- (1,0) -- (0,0) -- cycle;
    \end{scope}
}
\begin{scope}
\node[anchor=west,font=\scriptsize\bfseries] at (-0.035,4.27) {(a) Observation level};
\foreach \a/\b in {0/\leftObserved,\rightObserved/1}{
    \fill[pink,opacity=0.25] (\a,0) rectangle (\b,\plotTop);
}
\coordinate (edge0) at (0,\arrowy);
\foreach \blockrank/\obs [count=\i] in {
    m/0.20,2m/\leftObserved,3m/0.433,4m/0.50,
    5m/0.565,6m/\rightObserved,7m/0.81,{}/1}{
    \coordinate (edge\i) at (\obs,\arrowy);
    \draw[<->,thin] (edge\the\numexpr\i-1\relax) -- (edge\i);
    \ifnum\i<8
        \draw[black,line width=0.75pt] (\obs,\arrowy) -- (\obs,-0.16);
        \node[rotate=-35,anchor=north west,inner sep=0pt] at (\obs,-0.20)
            {$U_{\blockrank,t}$};
    \fi
}
\foreach \i/\pos in {1/.65,2/.5,7/.36,8/.5}{
    \path (edge\the\numexpr\i-1\relax) -- (edge\i)
        node[pos=\pos,above,inner sep=0.4pt,font=\fontsize{6}{7}\selectfont]
        {$d_{\i,t}^{(m)}$};
}
\foreach \a/\b in {2/4,4/6}{
    \path (edge\a) -- (edge\b) node[midway,above=2.5pt,inner sep=0pt] {$\cdots$};
}
\foreach \obs in {
    0.063,0.164,
    0.241,0.288,
    0.346,0.397,
    0.448,0.482,
    0.519,0.549,
    0.582,0.629,
    0.680,0.774,
    0.845,0.932}{
    \draw[black!80,line width=0.45pt] (\obs,0) -- (\obs,-0.095);
}

\foreach \obs in {\leftObserved,\rightObserved}{
    \draw[dashed,profRed,thick] (\obs,0) -- (\obs,\spanHeight);
}
\foreach \a/\b/\region in {0/\leftObserved/leftRegion,\rightObserved/1/rightRegion}{
    \draw[<->,thick,profRed] (\a,\spanHeight) -- (\b,\spanHeight)
        coordinate[midway] (\region);
}
\drawDensityAxes
\labelDensities
\node[text=profRed,align=center,inner sep=1pt] (observedRegion)
    at (0.5,3.35)
    {Maximum spacing region \(\mathcal U_t^{(m,k)}\)\\(Width \(=D_t^{(m,k)}\))};
\draw[profRed,label arrow] (leftRegion) to[out=96,in=215] ([yshift=.48em]observedRegion.west);
\draw[profRed,label arrow] (rightRegion) to[out=84,in=-35] ([yshift=.48em]observedRegion.east);
\end{scope}

\begin{scope}[shift={(1.24,0)}]
\node[anchor=west,font=\scriptsize\bfseries] at (-0.035,4.27) {(b) Population level};
\foreach \a/\b/\region in {0/\leftCut/leftLevel,\rightCut/1/rightLevel}{
    \shadeInterval{\a}{\b}{betterblue}
    \fill[pattern=spacingNullHatch,pattern color=profGreen]
        (\a,0) rectangle (\b,\ybase);
    \draw[<->,thick,profblue] (\a,\spanHeight) -- (\b,\spanHeight)
        coordinate[midway] (\region);
}
\foreach \cut in {\leftCut,\rightCut}{
    \draw[dashed,profblue,thick] (\cut,0) -- (\cut,\spanHeight);
}
\drawDensityAxes
\draw[dashed,profblue,thick] (0,\cutoff) -- (1,\cutoff);
\node[left,text=profblue] at (0,\cutoff) {$p_k$};
\labelDensities
\node[text=profblue,inner sep=1pt] (regionWidth) at (0.31,3.24)
    {(Width \(=\pi_k\))};
\node[text=profblue,anchor=south,inner sep=1pt] (regionTitle) at (regionWidth.north)
    {Lower level set \(\mathcal{U}^{(k)}\)};
\draw[profblue,label arrow] (leftLevel)
    to[out=126,in=205] (regionTitle.west);
\draw[profblue,label arrow] (rightLevel)
    .. controls (.54,2.33) and ([xshift=4mm]regionTitle.east) .. (regionTitle.east);

\begin{scope}[shift={(1.02,3.88)},x=1cm,y=1cm]
    \draw[black,thin] (-2.30,-1.22) rectangle (.6,0);
    \node[minimum size=5mm,inner sep=0pt,outer sep=0pt,
          draw=betterblue,fill=betterblue,fill opacity=0.25,line width=0.35pt] at (-1.97,-.32) {};
    \node[anchor=west,inner sep=0pt] at (-1.63,-.32)
        {$1/k \xleftarrow{\text{a.s.}} \widehat{D}_t^{\star(m,k)}$};
    \node[minimum size=5mm,inner sep=0pt,outer sep=0pt,
          draw=profGreen,pattern=spacingLegendHatch,pattern color=profGreen,
          line width=0.35pt] at (-1.97,-.90) {};
    \node[anchor=west,inner sep=0pt] at (-1.63,-.90)
        {$\pi_k \xleftarrow{\text{a.s.}} D_t^{(m,k)}$};
\end{scope}
\end{scope}
\end{tikzpicture}
\caption{Set $(m,k)=(3,2)$. 
(a) It shows how maximum spacing region $\mathcal{U}_{t}$ is computed. 
Suppose $t=23$, so $U_1, \ldots, U_{23}$ are observed. The order statistics $U_{m,t}, \ldots, U_{7m,t}$ are shown.
Their gaps define the $m$th order spacings $d_{1,t}^{(m)}, \ldots, d_{8,t}^{(m)}$.
The largest $1/k=50\%$ of them forms the maximum spacing region $\mathcal{U}_{t}^{(m,k)}$, whose total width is $D_t^{(m,k)} = \Leb(\mathcal{U}_{t}^{(m,k)})$. 
(b) It shows the comparison between $\pi_k$ and $1/k$.
The population cutoff $p_k$ defines the lower level set 
$\mathcal{U}^{(k)}$ 
so that the area under $p$ on this region is $1/k$. 
The area under $p_0$ on this region is $\pi_k$, 
which must be greater than $1/k$ under $\mathcal{H}_1$.}
\label{fig:1k_fraction}
\end{figure}

The proofs of Lemmas~\ref{lem:Gt_to_1half}--\ref{lem:sum_spacings_strategy}
are given in Section~\ref*{appendix:lemmas} of the supplement.
Lemmas~\ref{lem:Gt_to_1half} and
\ref{lem:sum_spacings_strategy} concern the limiting value of $D_{t-1}^{(m,k)}$ under $\cH_0$ and $\cH_1$, respectively.
It is not hard to see that if $V_t\sim\Unif[0,1]$ is independent on $U_{1:n}$, then
\[
    \pr( V_t \in \mathcal{U}_{t-1}^{(m,k)} \mid \mathcal{F}_{t-1}) = D_{t-1}^{(m,k)}
    \qquad \text{and} \qquad
    \pr( U_t \in \mathcal{U}_{t-1}^{(m,k)} \mid \mathcal{F}_{t-1}) = \widehat{D}_{t-1}^{\star(m,k)},
\]
which are simply the null-restricted and optimal unrestricted
record-breaking probabilities.
Hence, Lemma~\ref{lem:sum_spacings_strategy} shows that the null-coupled
probability is no smaller than its unrestricted counterpart in the limit.
Specifically,
$\pr(V_t \in \mathcal{U}_{t-1}^{(m,k)} \mid \mathcal{F}_{t-1})
\inAS \pi_k > 1/k$ and $\pr(U_t \in \mathcal{U}_{t-1}^{(m,k)} \mid \mathcal{F}_{t-1})
\inAS 1/k$.
Figure~\ref{fig:1k_fraction} illustrates this property.
For any $\pr$ supported on $[0,1]$, Lemma~\ref{lem:sum_spacings_strategy} also gives
$\mathcal{U}_{t-1}^{(m,k)} \inAS \mathcal{U}^{(k)}$,
where $\mathcal{U}^{(k)}$ is a fixed region within the
$(1/k)$-lower level set of the density.
The maximum spacing set strategy thus targets discrepancies
$|p(u)-p_0(u)|$ over a broader region than the single maximum
spacing strategy. Based on Lemma~\ref{lem:sum_spacings_strategy}, we can characterize the limiting behavior of \(\Lambda_n^{(m,k)}\)
as follows:

\begin{theorem}[Almost surely-limit of $\Lambda_n^{(m,k)}$]\label{thm:sum_spacings_strategy}
Let
$m = \lfloor t^\phi \rfloor$, where $\phi \in (0.5,1)$.
Define $\pi_k$ and $p_k$ as in Lemma~\ref{lem:sum_spacings_strategy}.
If $\Leb\{u\in[0,1]:p(u)=p_k\}=0$,
then
\begin{align}
    \frac{1}{n}\log \Lambda_n^{(m,k)}
    \inAS \KL_k := \KL\{ \Bern(1/k) \| \Bern(\pi_k)\}\label{eq:k_fraction_power}.
\end{align}
\end{theorem}
By Lemma~\ref{lem:sum_spacings_strategy}, \(\mathcal{U}_{t-1}^{(m,k)}\)
converges almost surely to a fixed region,
so
\(I_t^{(m,k)}=\mathbb{1}(U_t \in \mathcal{U}_{t-1}^{(m,k)})\)
in \eqref{eq:general_Gt}
follow
\(\Bern(1/k)\) under \(\mathcal{H}_1\)
and \(\Bern(\pi_k)\) under \(\mathcal{H}_0\).
Consequently, \(\Lambda_n^{(m,k)}\) is asymptotically the likelihood ratio
of i.i.d.~Bernoulli sample.
Then, the asymptotic equipartition property for relative entropy \citep{Cover2006} yields
\eqref{eq:k_fraction_power} in probability.
For theoretical completeness,
we establish this result almost surely
via a different argument.
Also, Theorem~\ref{thm:sum_spacings_strategy} implies that the almost sure limit of
\( n^{-1}\log \Lambda_n^{(m,k)}\) is non-negative,
and is zero if and only if $\mathcal{H}_0$ is true, in which case \(\pi_k = 1/k\) by Lemma~\ref{lem:Gt_to_1half}.

A statistical interpretation of (\ref{eq:k_fraction_power}) is that
the user-specified constant $k$ determines
how the discrepancy between $\pr$ and $\pr_0$ is measured.
Specifically, $\Lambda_n^{(m,k)}$ quantifies
evidence against $\mathcal{H}_0$ by evaluating
the $(1/k)$-lower level set of $p(\cdot)$ and
comparing its support size with its value under the null.
Thus, $\Lambda_n^{(m,k)}$ is expected to achieve high power when
the departure of $\pr$ from $\pr_0$ is concentrated in this lower level set.
Selection of $k$ is, therefore, equivalent to selection of discrepancy measure.
The optimal selection of $k$ will be discussed in Section \ref{sec:compare_two_eprocess}.

\section{Optimal tuning via combining multiple strategies}\label{sec:compare_two_eprocess}

\subsection{Predictable combinations of candidate strategies}
We note two facts:
(i) Theorem~\ref{thm:sum_spacings_strategy} shows that a larger
$\KL_k$ yields a more powerful test, and
(ii) a predictable linear combination of $e$-processes remains an $e$-process \citep{ramdas2023savi}.
These motivate us to tune $k$ by combining
$\Lambda_t^{(m,k_1)}, \Lambda_t^{(m,k_2)}, \ldots$ using predictable weights
to greedily target the largest $\Lambda_t^{(m,k)}$.

At time $t$,
let $\mathcal{K}_{t-1} \subseteq(1, \infty)$
be a finite set of candidate values of $k$, where $0<|\mathcal{K}_{t-1}|<\infty$.
Assume that
$\emptyset\neq\mathcal{K}_0 \subseteq\mathcal{K}_1 \subseteq\cdots \subseteq(1, \infty)$
such that
$|\mathcal{K}_t|\rightarrow \infty$ as $t\rightarrow\infty$.
We consider predictable weights
$\{ w_{t,k} : k\in\mathcal{K}_{t-1} \} \in \mathcal{F}_{t-1}$
such that $w_{t,k}\geq 0$ and $\sum_{k\in\mathcal{K}_{t-1}} w_{t,k} = 1$.
Then, at time $t$, define
\begin{align}\label{eq:cond_e_R}
    \Lambda_n^{(m,\mathcal{K})}
        = \prod_{t=1}^n \lambda_t^{(m,\mathcal{K})},
    \quad \text{where} \quad
    \lambda_t^{(m,\mathcal{K})}
        = \sum_{k\in\mathcal{K}_{t-1}} w_{t,k} \lambda_t^{(m, k)}.
\end{align}
The basic properties of $\Lambda_n^{(m,\mathcal{K})}$ are shown below:

\begin{proposition}\label{prop:weitghedSum_validity}
Let $m = \lfloor t^\phi \rfloor$, where $\phi \in (0.5,1)$.
Assume that
(\ref{eq:k_fraction_power}) holds for each $k\in\mathcal{K}_{\infty}$.
Then the following results holds:
(i) $\{\Lambda_n^{(m,\mathcal{K})}\}_{n\in\mathbb{Z}^+_0}$ is an
$e$-process under $\mathcal{H}_0$; and
(ii) if there is a sequence
$\widehat{k}_{t-1} \in \mathcal{K}_{t-1}$
and a fixed $k^{\circ} \in \mathcal{K}_{\infty}$
such that
\begin{align}\label{eqt:condition_validy_weightedVer}
    \frac{1}{n}\log \prod_{t=1}^n \lambda_t^{(m, \widehat{k}_{t-1})} \inAS \KL_{k^{\circ}}
    \qquad \text{and} \qquad
    \frac{1}{n}\log \prod_{t=1}^n w_{t, \widehat{k}_{t-1}} \inAS 0.
\end{align}
then $\liminf_{n\rightarrow\infty} n^{-1}\log \Lambda_n^{(m, \mathcal{K})} \geq \KL_{k^{\circ}}$ almost surely under $\mathcal{H}_1$.
\end{proposition}

Proposition~\ref{prop:weitghedSum_validity}(i) ensures the validity of $\Lambda_n^{(m,\mathcal{K})}$
for any predictable weights $\{ w_{t,k} : k \in \mathcal{K}_{t-1} \}$,
whereas part (ii) ensures that $\Lambda_n^{(m,\mathcal{K})}$
is a consistent test, i.e., $\Lambda_n^{(m, \mathcal{K})}\inAS \infty$ under $\mathcal{H}_1$,
provided that $\KL_{k^{\circ}} > 0$.
The first condition in \eqref{eqt:condition_validy_weightedVer}
requires that $\mathcal{K}_{t-1}$ contains
$\widehat{k}_{t-1}$ such that
the test based on $\widehat{k}_{t-1}$ is asymptotically equivalent to that based on $k^{\circ}$.
The second condition in \eqref{eqt:condition_validy_weightedVer}
requires that the weight assigned to $\widehat{k}_{t-1}$ is not too small
so that $\lambda_t^{(m, \widehat{k}_{t-1})}$ remains active eventually.
In other words,
the weighted strategy $\Lambda_n^{(m,\mathcal{K})}$
performs at least as well as
$\Lambda_n^{(m,k^{\circ})}$,
where the limiting $k^{\circ}$ is determined by
the dominating candidates $\widehat{k}_{t-1}$,
which are controlled by the forms of the weights $w_{t,k}$.
Note that $\limsup_{n\rightarrow\infty} n^{-1}\log\Lambda_n^{(m, \mathcal{K})}$
may be strictly greater than $\KL_{k^{\circ}}$ because other components in the weighted sum $\lambda_t^{(m,\mathcal{K})}$
may contribute additional exponential growth.

In the following Section \ref{sec:finite_mathcalK},
we will provide concrete choices of
$\mathcal{K}_t$ (see Examples \ref{eg:finite_K}--\ref{eg:infinite_K})
and $w_{t,k}$ (see Definition \ref{def:argmax_softmax})
so that the limiting $k^{\circ}$ performs well.

\subsection{Optimal combinations}\label{sec:finite_mathcalK}
An increasingly richer set $\mathcal{K}_t$, i.e., $|\mathcal{K}_t|\rightarrow\infty$,
allows us to consider more candidate values of $k$,
however, too many choices of $k$ may blur the weighting scheme $w_{t,k}$.
Consequently, we need to control $|\mathcal{K}_t|$
and design appropriate weights $w_{t,k}$ to ensure that the limit $k^{\circ}$
performs well.
We study this trade-off in this section.

In view of Theorem~\ref{thm:sum_spacings_strategy}, we define the best possible $k$ as
\begin{align}\label{eq:def_k*}
    k^{\star} = \argmax_{k\in\mathcal{K}_{\infty}} \KL_k,
    \quad \text{where} \quad
    \mathcal{K}_{\infty} = \bigcup_{t=0}^{\infty} \mathcal{K}_t .
\end{align}
The goal is to select $k^{\star}$ asymptotically.
We propose two forms of $w_{t,k}$:

\begin{definition}\label{def:argmax_softmax}
The weights $w_{t,k}$'s are in the
\emph{argmax form} and \emph{softmax form} if
\begin{align}\label{eqt:argmax_softmax}
    w_{t,k} = \mathbb{1}(k = \widehat{K}_{t-1})
    \qquad \text{and} \qquad
    w_{t,k} = \frac{\exp\{ \gamma\Lambda_{t-1}^{(m,k)}\}}{\sum_{k'\in\mathcal{K}_{t-1}} \exp\{ \gamma\Lambda_{t-1}^{(m,k')}\}} ,
\end{align}
where
$\widehat{K}_{t-1} = \min \{ \argmax_{k\in\mathcal{K}_{t-1}} \Lambda_{t-1}^{(m,k)} \}$
and
$\gamma>0$ is a learning rate.
Set $\gamma=1$ by default.
\end{definition}

\begin{condition}
The argmax condition (\ref{eqt:argmax_cond})
and softmax conditions (\ref{eqt:softmax_cond})
are
\begin{subequations}\label{eq:adaptive_conditions}
\begin{gather}
    \lim_{\varepsilon\downarrow 0} \limsup_{n\rightarrow\infty} \frac{1}{n} \sum_{t=1}^n \sup_{\substack{k\in\mathcal{K}_{t-1}\\ |k-k^{\star}|\leq\varepsilon}} \left| \log \lambda_t^{(m,k)}-\log\lambda_t^{(m,k^{\star})}\right|= 0 .\label{eqt:argmax_cond}\\
    \frac{1}{t} \log |\mathcal{K}_{t}| \rightarrow 0
    \qquad \text{and} \qquad
    \lim_{\varepsilon\downarrow 0} \limsup_{n\rightarrow\infty} \frac{1}{n} \sum_{t=1}^n \log\bigg\{ 1 + \sum_{\substack{k\in\mathcal{K}_{t-1}\\ 0<|k-k^{\star}|\leq\varepsilon}} w_{t,k} / w_{t,k^{\star}} \bigg\}= 0 .\label{eqt:softmax_cond}
\end{gather}
\end{subequations}
\end{condition}

For the argmax form,
$\widehat{K}_{t-1}$ is
the minimum value among all maximizers.
It is possible to replace it by
a randomly selected maximizer.
The asymptotic properties will not be affected.
The condition (\ref{eqt:argmax_cond})
requires that all candidates $k\in\mathcal{K}_{t-1}$ that are close to
$k^{\star}$ have similar asymptotic value of $\log \lambda_{t}^{(m,k)}$ on average.
It holds if
$|\log \lambda_t^{(m, k)}-\log \lambda_t^{(m, k^{\star})}| \leq C_t |k-k^{\star}|$,
where $C_t>0$ are constants satisfying
$\sum_{t=1}^n C_t/n = O(1)$.
We need (\ref{eqt:argmax_cond}) to ensure that
\[
    \frac{1}{n}\sum_{t=1}^n \log \lambda_t^{(m, \widehat{k}_{t-1})} - \log \lambda_t^{(m, k^{\star})}\inAS 0
\]
for any $\widehat{k}_{t-1}\inAS k^{\star}$.
For the softmax form,
the first condition in (\ref{eqt:softmax_cond})
requires $|\mathcal{K}_t|$ to be
subexponential, i.e., $|\mathcal{K}_t| = e^{o(t)}$.
It controls the size of the candidate set $\mathcal{K}_{t-1}$.
The second condition is needed for handling
a dense grid $\mathcal{K}_{t-1}$.
It says that the total relative weight $w_{t,k}/w_{t,k^{\star}}$
of candidates that are close to $k^{\star}$
is negligible on average.
By Ces\`{a}ro summability, the second condition in (\ref{eqt:softmax_cond}) holds if
$w_{t,k^{\star}} \inAS 1$.

We prove that, under regularity conditions,
both forms lead to tests that are at least as
powerful as the oracle test $\Lambda_n^{(m,k^{\star})}$:

\begin{theorem}[Optimal tuning]\label{thm:divergent_mathcalK_optimal_softmax}
Let $m = \lfloor t^\phi \rfloor$, where $\phi \in (0.5,1)$.
Suppose that
$w_{t,k}$ is in the argmax form
with (\ref{eqt:argmax_cond}), or
in the softmax form
with (\ref{eqt:softmax_cond}).
If
(\ref{eq:k_fraction_power}) holds for each $k\in\mathcal{K}_{\infty}$,
the maximizer $k^{\star}$ is unique, and for any $\varepsilon>0$
\begin{align}\label{eqt:condition_validy_weightedVer_softmax}
    \Delta_{\varepsilon} := \KL_{k^{\star}} - \sup_{\substack{k\in\mathcal{K}_{\infty}\\ |k-k^{\star}|>\varepsilon}} \KL_{k} >0
    \qquad \text{and} \qquad
    \eta_n := \max_{k\in\mathcal{K}_n} \left\vert \frac{1}{n}\log \Lambda_n^{(m,k)} - \KL_k \right\vert \inAS 0,
\end{align}
then $\liminf_{n\rightarrow\infty} n^{-1}\Lambda_n^{(m, \mathcal{K})} \geq \KL_{k^{\star}}$ almost surely under $\mathcal{H}_1$.
\end{theorem}

Theorem \ref{thm:divergent_mathcalK_optimal_softmax}
shows that both strategies can adaptively choose the best possible $k^{\star}$ and
perform as well as the oracle $e$-process asymptotically.
The first condition in (\ref{eqt:condition_validy_weightedVer_softmax})
requires that
$\KL_{k^{\star}}$
is well separated,
by a fixed positive gap $\Delta_{\varepsilon}$,
from $\KL_k$ for any competing
$k \in \mathcal{K}_{\infty}$ that is at least $\varepsilon$ away from $k^{\star}$.
This is a usual condition in M-estimation; see \cite{vandervaart1998}.
The second condition in (\ref{eqt:condition_validy_weightedVer_softmax})
is used to handle a divergent number of candidates $k$.
This is a uniform strong law for $n^{-1} \log \Lambda_n^{(m,k)}$,
and is stronger than the first condition in (\ref{eqt:condition_validy_weightedVer}).

We remark that
$\mathcal{K}_{\infty}$ is allowed to be
dense in $(1, \infty)$
for both argmax and softmax rules.
If $\mathcal{K}_{\infty}$
is restricted to a finite set,
then the regularity conditions (\ref{eqt:argmax_cond})--(\ref{eqt:softmax_cond})
can be removed, and the optimality result can be proved straightforwardly;
see Section \ref*{sec:armax} of the supplement.
Two practical choices of $\{\mathcal{K}_t\}$ are given below:

\begin{example}[Finite set of strategies]\label{eg:finite_K}
Let $\mathcal{K}_t \equiv \mathcal{K}$
for all $t\geq 0$ for a fixed finite set $\mathcal{K}$.
According to the simulation studies in Section \ref{sec:simulations},
$\mathcal{K}=\{2,20\}$ performs well.
\end{example}

\begin{example}[Divergent set of strategies]\label{eg:infinite_K}
Let $\mathcal{K}_t$ be defined as dyadic grids, e.g.,
\[
    \mathcal{K}_t = \left\{  1+ \frac{j}{2^{r_t}} : j = 1, \ldots, \lfloor (t+1)^{1/4}\rfloor 2^{r_t}\right\},
    \qquad \text{where} \qquad
    r_t = \left\lfloor \frac{1}{4} \log_2(t+1) \right\rfloor.
\]
This ensures that
\(|\mathcal K_t|=\lfloor(t+1)^{1/4}\rfloor 2^{r_t}=O(t^{1/2})\).
For example,
$\mathcal{K}_t = \{2\}$ for $0\leq t\leq 14$;
$\mathcal{K}_t = \{1.5, 2, 2.5, 3\}$ for $15\leq t\leq 79$; and
$\mathcal{K}_t = \{1.5, 2, 2.5, 3, 3.5, 4\}$ for $80\leq t\leq 254$.
\end{example}

\begin{example}[Automatic switch between $\lambda_t^{(1,\infty)}$ and $\lambda_t^{(m,k)}$]\label{eg:switch}
We can further combine
(i) the maximum spacing strategy $\lambda_t^{(1,\infty)}$ and
(ii) any optimally tuned maximum spacing set strategy $\lambda_t^{(m, \mathcal{K})}$.
It is permitted by
Proposition \ref{prop:weitghedSum_validity}
with some trivial modification.
We call it an automatic switch between methods in Sections
\ref{sec:maxSpacing} and \ref{sec:general_aggreation}.
With the argmax rule for switching, we have
$\Lambda_n^{\text{switch}}= \prod_{t=1}^n \{w_t \lambda_t^{(1,\infty)} + (1-w_t) \lambda_t^{(m,\mathcal{K}_{t-1})}\}$,
where $w_t = \mathbb{1}(\Lambda_{t-1}^{(1,\infty)} > \Lambda_{t-1}^{(m,\mathcal{K}_{t-1})})$.
\end{example}

\subsection{Suboptimality of simple average}
In (\ref{eq:cond_e_R}),
one simple way to combine multiple strategies $\mathcal{K}_{t-1}$
is simple averaging:
\begin{definition}\label{def:simpleAvg}
If $w_{t,k} \equiv 1/|\mathcal{K}_{t-1}|$ for $k\in\mathcal{K}_{t-1}$,
$w_{t,k}$'s are in the \emph{simple average form}.
\end{definition}

For fixed finite $\mathcal{K}_{t-1}\equiv\mathcal{K}$, the simple average
$\lambda_t^{(m,\mathcal{K})}$ dominates every symmetric combining
function \cite[Proposition~3.1]{VovkWang2021}.
To our knowledge, the case $|\mathcal{K}_{n-1}|\inAS\infty$ remains
unexplored. Here, Proposition~\ref{prop:weitghedSum_validity} does not
apply because both conditions in
(\ref{eqt:condition_validy_weightedVer}) fail. Simple averaging does
not eventually select a fixed $k^{\circ}$, and
$(\log \prod_{t=1}^n w_{t,\widehat{k}_{t-1}})/n=-(\sum_{t=1}^n \log |\mathcal{K}_{t-1}|)/n
\inAS -\infty$,
We derive the property in this case below:

\begin{proposition}[Average rule]\label{prop:average_rule}
Let $m=\lfloor t^\phi \rfloor$, $\phi\in(0.5,1)$.
Suppose that $w_{t,k}$ is in the simple average form.
Let $\mathcal{K}_t$ be
non-empty, deterministic and equally spaced such that
\begin{equation}\label{cond:Kt}
    \mathcal{K}_{t-1}= \{1+j\Delta_t: \{j: 1\le j\le |\mathcal{K}_{t-1}|\}\cap \{j: j\in \mathbb{N}\}\},\quad
    |\mathcal{K}_{t-1}|\asymp \frac{t^{\nu}}{\Delta_t},
    \quad\text{and}\quad
    \log\left(\frac{1}{\Delta_t}\right) = O(\log t),
\end{equation}
for some $\nu>0$ and $\Delta_t\in(0,1]$.
If $p$ is continuous on $[0,1]$ and $\Leb\{u:p(u)=0\}=0$,
then,
under $\cH_1$,
$
\log\Lambda_n^{(m,\mathcal{K})} /n\inAS 0.
$
\end{proposition}

Proposition \ref{prop:average_rule} shows that
the logarithm of the $e$-process under the simple average rule
grows sublinearly.
Thus, its asymptotic logarithmic growth rate is lower than those of the argmax and softmax rules,
provided that their growth rates are strictly positive.
We remark that
the conditions in (\ref{cond:Kt}) are satisfied by, e.g.,
the dyadic grids in
Example~\ref{eg:infinite_K}.

\section{Multivariate distribution test}\label{sec:multi_rv}

\subsection{General multivariate aggregation strategy}\label{sec:definition}
Our framework extends naturally to the multivariate setting.
We
present the multivariate versions of
maximum‑spacing and the combined strategies in Sections~\ref{sec:multi_variate_max_spacing}--\ref{sec:robust_combine}, respectively.

We consider the GoF test for
an $\mathbb{R}^d$-valued data stream
$X_1,X_2,\ldots \simIID F^X$, where  the support of $F^X$ is $\mathbb{R}^d$
Write \(X_i = (X_i^{[1]},\ldots,X_i^{[d]})^\T\) for each $i$.
Let
$F^X_0(x^{[1]},\ldots,x^{[d]}) = \pr_0\left( X^{[1]} \le x^{[1]}, \ldots, X^{[d]} \le x^{[d]}\right)$.
Denote the conditional distribution function by
\begin{align}\label{eqt:condtionCDF}
    F^X_0(x^{[j]} \mid x^{[1]},\ldots,x^{[j-1]})
    &=  \pr_0\left( X_1^{[j]} \le x^{[j]} \mid X_1^{[1]} = x^{[1]}, \ldots, X_1^{[j-1]}= x^{[j-1]}\right),
\end{align}
and its density function by
$f^X_0(x^{[j]} \mid x^{[1]},\ldots,x^{[j-1]}) = \dd F^X_0(x^{[j]} \mid x^{[1]},\ldots,x^{[j-1]})/\dd x^{[j]}$
for $j=1, \ldots, d$, where the conditioning set is empty for $j=1$.
We construct
\begin{align}\label{eqt:def_U_multi}
U_i^{[j]} = F^X_0\left(X_i^{[j]} \mid X_i^{[1]},\ldots,X_i^{[j-1]}\right),
\quad j=1,\ldots,d,
\end{align}
whose null property is shown in Lemma~\ref{lemma:uniform_trans} below;
see also
\cite{rosenblatt1952remarks}.

\begin{lemma}[Theorem~1 of \cite{justel1997multivariate}]\label{lemma:uniform_trans}
Define $U_i^{[1]}, \ldots, U_i^{[d]}$ as in (\ref{eqt:condtionCDF})
for a fixed $i$.
Then $U_i^{[1]}, \ldots, U_i^{[d]} \simIID \Unif[0, 1]$ if and only if $F^X_0 = F^X$.
\end{lemma}

The multivariate spacings
are defined as follows:
At time \(t\), for the \(j\)th coordinate, we sort
\(U_{1}^{[j]},\ldots,U_{t}^{[j]}\) in an ascending order as
$
    0 \equiv U_{0,t}^{[j]} \le U_{1,t}^{[j]} \le \cdots \le U_{t,t}^{[j]} \le U_{t+1,t}^{([j]} \equiv 1,
$
and define spacings \(d_{i,t}^{[j]} = U_{i+1,t}^{[j]} - U_{i,t}^{[j]}\) for \(i=0,\ldots,t\).
For a given index set
\(\mathcal{I}_t^{[j]} \subseteq \{0,1,\ldots,t\}\),  define
\begin{equation}\label{eq:spacings_each_dimension}
    D_t^{[j]} = \sum_{i \in \mathcal{I}_t^{[j]}} d_{i,t}^{[j]}
    \qquad \text{and} \qquad
    \mathcal{U}_t^{[j]} = \bigcup_{i \in \mathcal{I}_t^{[j]}} [U_{i,t}^{[j]},\, U_{i+1,t}^{[j]}].
\end{equation}
Let \(j^{\star}_t = \min(\argmax_{j =1, \ldots, d} D_t^{[j]})\)
be the smallest index that maximizes \(j \mapsto D_t^{[j]}\).
Define
the multivariate version of $D_t$ and $\widehat{D}_t$ as
\begin{equation}\label{eq:multi_IGn}
    D_t =D_t^{[j_t^{\star}]} = \max_{j=1,\ldots, d} D_t^{[j]}
    \qquad \text{and} \qquad
    \widehat{D}_{t-1} = g_{t-1}(U_{1:t-1}) ,
\end{equation}
where \(g_{t-1} : [0,1]^{t-1} \to [0,1]\)
is defined as before.
In analogy with (\ref{eq:general_Gt}), we set
$
    I_t = \mathbb{1}\{U_t^{[j^{\star}_{t-1}]} \in \mathcal{U}_{t-1}^{[j^{\star}_{t-1}]}\}.
$
As in (\ref{eq:D_and_Dhat}),
we call \(D_{t-1}\) and \(\widehat{D}_{t-1}\)
the null-restricted and unrestricted estimators of \(\E_\pr(I_t \mid \mathcal{F}_{t-1})\), respectively.
The $e$-process \(\Lambda_n\) is then given by
$\Lambda_n = \Lambda_n(U_{1:n},D_{1:n-1}, \widehat{D}_{1:n-1},\mathcal{U}_{1:n-1})$.
By arguments analogous to those in Proposition~\ref{prop:gen_eprocess} and Theorem~\ref{thm:general_log_optimal}, we immediately obtain the corresponding
results in the multivariate setting.

\begin{proposition}\label{prop:multi_eprocess}
Let $d\geq1$.
Let $D_t$ and $\widehat{D}_{t-1}$ be defined as (\ref{eq:multi_IGn}).
The stochastic process $\{ \Lambda_n\}_{n\in\mathbb{Z}^+}$ is an $e$-process
with respect to the filtration
$\{ \mathcal{F}_n \}_{n\in\mathbb{Z}^+}$
under $\mathcal{H}_0$,
i.e., for any stopping time $\tau$ with respect to
$\{ \mathcal{F}_n \}_{n\in\mathbb{Z}^+}$,
$\E (\Lambda_\tau) \leq 1$.
\end{proposition}

\begin{theorem}\label{thm:full_general_optimal}
Let $d\geq 1$.
Let $D_t$ be defined as (\ref{eq:multi_IGn}).
Suppose $F$ is continuous.
Denote
$\Lambda_n^{\star} = \Lambda_n(U_{1:n}, D_{1:n-1}, \widehat{D}_{1:n-1}^{\star},\mathcal{U}_{1:n-1}^{[j^{\star}]})$ and
$\lambda_t^{\star} = \lambda_t(U_t,D_{t-1}, \widehat{D}_{t-1}^{\star},\mathcal{U}_{t-1}^{[j_{t-1}^{\star}]})$, where
\begin{align}\label{eqt:multi_Dhat_optimal}
    \widehat{D}_{t-1}^{\star} := \pr\left(U_t^{[j_{t-1}^{\star}]} \in \mathcal{U}_{t-1}^{[j_{t-1}^{\star}]} \mid \mathcal{F}_{t-1} \right)
    \quad \text{and} \quad
    \mathcal{U}_{1:n-1}^{[j^{\star}]} := (\mathcal{U}_1^{[j_1^{\star}]},\ldots, \mathcal{U}_{n-1}^{[j_{n-1}^{\star}]}).
\end{align}
Then (i)
$
\max_{\widehat{D}_{t-1} \in [0, 1]} \E_\pr\left(\log \lambda_t \mid \mathcal{F}_{t-1}\right)
=
\E_\pr\left(\log \lambda_t^{\star} \mid \mathcal{F}_{t-1}\right)
= \KL\{ \Bern(\widehat{D}_{t-1}^{\star}) \| \Bern(D_{t-1})\} ;
$
and (ii)
$\widehat{D}^{\star} := (\widehat{D}^{\star}_1,\widehat{D}^{\star}_2, \ldots)
= \argmax_{\widehat{D}\in \mathcal{D}} \E_\pr(\log \Lambda_\tau)$.
\end{theorem}

Hence, we obtain a level-$\alpha$ anytime-valid goodness-of-fit test:
$
\pr_0(\exists \tau < \infty: \Lambda_\tau \ge 1/\alpha) \leq \alpha.
$
The next two subsections present two concrete constructions of \(\mathcal{I}_t^{[j]}\).

\begin{remark}
Note that \(\Lambda_n\) is constructed via Lemma~\ref{lemma:uniform_trans}, so the transformed vector \((U^{[1]},\ldots,U^{[d]})\) depends on the ordering of \((X^{[1]},\ldots,X^{[d]})\). Consequently, permuting the coordinates of the original sample \(X_1, X_2,\ldots\) can change the value of \(\Lambda_n\). To address this potential bias,
we can build an $e$‑process
for each ordering of \((X^{[1]},\ldots,X^{[d]})\), and then aggregate the resulting \(d!\) $e$‑processes using the method in Section \ref{sec:compare_two_eprocess}.
\end{remark}

\subsection{Multivariate maximum $e$-spacing LRT}\label{sec:multi_variate_max_spacing}
Let \(\mathcal{I}_t^{[j]}\)
be the singleton containing \(i_t^{\star} := \min( \arg\max_{i=1,\ldots,t} d_{i,t}^{[j]})\).
Then $D_t$ and $\mathcal{U}_t^{[j]}$ become
\begin{align}\label{eqt:Dt_multi}
    D_t^{(1,\infty)} = \max_{j =1, \ldots, d} \max_{i =1, \ldots, t} d_{i,t}^{[j]}
    \qquad \text{and} \qquad
    \mathcal{U}_t^{\md[j]} = [U_{i_t^{\star},t}^{[j]},\, U_{i_t^{\star}+1,t}^{[j]}] .
\end{align}
This means that
\(D_t^{(1, \infty)}\) may occur at different coordinates
$j=1, \ldots, d$ at different times $t=1, 2, \ldots$.
Consequently, \(D_t^{(1, \infty)} / D_t^{[j]} \ge 1\) for all $t$ and $j$.
Together with $\widehat{D}_{t-1}^{(1, \infty)} = \log t /t$,
the proposed multivariate maximum $e$-spacing LRT is
$\Lambda_n^{\md} = \Lambda_n(U_{1:n},D_{1:n-1}^{\md}, \widehat{D}_{1:n-1}^{\md},\mathcal{U}_{1:n-1}^{\md [j^{\star}]})$.
Consequently, we have the following result:

\begin{theorem}[Consistency]\label{thm:multi_simple}
Let $d\geq 1$
and \(U_i = (U_i^{[1]},\ldots,U_i^{[d]})^\T, i\in\{1,\ldots,n\}\). Suppose that there exists $j\in\{1,\ldots,d\}$ such that $U_i^{[j]} \not\sim \Unif[0,1]$.
If Condition~\ref{cond:simple_strategy} holds for (\ref{eqt:Dt_multi}),
then as $n \rightarrow \infty$
\begin{enumerate}
    \item in Case 1, $\log \Lambda_n^{\md} /L_n \inAS c_{\pr}-1-\log c_{\pr} >0$,
        where $L_n = \sum_{t=2}^n \log t/t \asymp (\log n)^2$,
    \item in Case 2, $\log \Lambda_n^{\md} \geq  S_n/8 + O_{\AS}(1)$,
        where $S_n = \sum_{t=2}^n D_{t-1}  \gg (\log n)^2$, and
    \item in either case, $\Lambda_n^{\md} \inAS \infty$ and $\E(\Lambda_n^{\md}) \rightarrow \infty$.
\end{enumerate}
\end{theorem}

Theorem~\ref{thm:multi_simple} shows that the proposed test is consistent whenever at least one marginal distribution is non-uniform, that is, $U_i^{[j]} \not\sim \Unif[0,1]$ for some $j$ and Condition~\ref{cond:simple_strategy} holds for (\ref{eqt:Dt_multi}), then
$\pr(\sup_{t\in\mathbb{Z}^+} \psi_t = 1) = 1$ under $\mathcal{H}_1$. In other words, under these conditions, the anytime-valid test $\psi$ in (\ref{eqt:test}) rejects $\mathcal{H}_0$ with probability one under $\mathcal{H}_1$.
If all transformed components are $\Unif[0,1]$ marginally, but are dependent,
then Theorem \ref{thm:multi_simple} does not apply.
Further research directions and discussion
are provided in Section~\ref*{sec:discussion} of the supplement.

\subsection{Multivariate $1/k$ fraction $m$th order $e$-spacing LRT}\label{sec:robust_combine}
We define the multivariate versions for
Sections~\ref{sec:general_aggreation}--\ref{sec:compare_two_eprocess}
by using
the sum of the \(\lfloor t/(mk) \rfloor\) largest $m$th order spacings in each dimension.
At time $t-1$,
let the $m$th order spacings in the $j$th coordinate be
\(d_{i,t-1}^{(m)[j]} = U_{(i+1)m,t-1}^{[j]} - U_{im,t-1}^{[j]}\), if $i=0, \ldots, \lfloor t/m \rfloor-1$; $d_{i,t-1}^{(m)[j]} =1-U_{\lfloor t/m \rfloor m,t-1}$, if $i = \lfloor t/m \rfloor$ and $t/m \not\in\mathbb{Z}$.
Define
\begin{gather}\label{eqt:Dhatmk_multi}
D_{t-1}^{(m,k)}
    = \max_{j=1, \ldots, d} D_{t-1}^{(m,k)[j]}
    = D_{t-1}^{(m,k)[j^{\star}_{t-1}]},\ \text{where}\\
D_{t-1}^{(m,k)[j]} =\sum_{i\in\mathcal{I}_{t-1}^{(m,k)[j]}} d_{i,t-1}^{(m)[j]} ,\qquad
j^{\star}_{t-1} = \min\left( \argmax_{j =1, \ldots, d} D_{t-1}^{(m,k)[j]} \right)\ \text{and}\nonumber\\
\mathcal{I}_{t-1}^{(m,k)[j]}
= \left\{
    \mathcal{I} \subseteq \{0, \ldots, K_{t-1}\} :
    |\mathcal{I}|= \lfloor (t/m)/k \rfloor \;\;\text{and}\;\;
    d_{i,t-1}^{(m)[j]} > d_{i',t-1}^{(m)[j]} \,\,\text{for}\,\, i'\not\in\mathcal{I}
\right\}.\nonumber
\end{gather}
Similar to the definitions of $\widehat{D}_{t-1}^{(m,k)}$ and $I_t^{(m,k)}$ in
the univariate case in (\ref{eqt:aggregate_I_mathcalI_mathcalU})--(\ref{eq:sum_half_max_spacings}), we  use
\begin{align}\label{eq:multi_Dhat_mk}
    \widehat{D}_{t-1}^{(m,k)} = \lfloor t/k \rfloor / t
    \qquad \text{and} \qquad
    I_{t}^{(m,k)}
		= \mathbb{1}\left(U_{t}^{[j_{t-1}^{\star}]}\in\mathcal{U}_{t-1}^{(m,k)[j_{t-1}^{\star}]}\right),
\end{align}
where
\begin{gather*}
    \mathcal{U}_{t-1}^{(m,k)[j_{t-1}^{\star}]} = \bigcup_{i\in\mathcal{I}_{t-1}^{(m,k)[j_{t-1}^{\star}]}} \left[U_{im,t-1}^{[j_{t-1}^{\star}]}, \; U_{(i+1)m,t-1}^{[j_{t-1}^{\star}]}\right],
\end{gather*}
$[U_{im,t-1}, \; U_{(i+1)m,t-1}] := [U_{im,t-1}, 1]$ when $i>t/m -1$, which happens when $t/m\not\in\mathbb{Z}$.
Then the multivariate $1/k$ fraction $m$th order $e$-spacing test statistic is
$$
\Lambda_n^{(m,k)} = \Lambda_n(U_{1:n}, D_{1:n-1}^{(m,k)}, \widehat{D}_{1:n-1}^{(m,k)}, \mathcal{U}_{1:n-1}^{(m,k)[j^{\star}]}).
$$

By the definitions of
\(D_{t-1}^{(m,k)}\) and \(D_{t-1}^{(m,k)[j]}\), we have \(D_{t-1}^{(m,k)} / D_{t-1}^{(m,k)[j]} \ge 1\) for all \(t\geq 1\) and all $j\in \{1,\ldots, d\}$.
Under \(\mathcal{H}_0\), Lemmas~\ref{lemma:uniform_trans} and \ref{lem:Gt_to_1half} imply that
\(D_{t-1}^{(m,k)[j]} \inAS 1/k\).
Under \(\mathcal{H}_1\), for each nonuniform transformed marginal,
following Lemma~\ref{lem:sum_spacings_strategy},
we have $D_{t-1}^{(m,k)[j]} \inAS \pi_k^{[j]} \in (1/k,\, 1)$
and $\pr\left(U_{t}^{[j]}\in\mathcal{U}_{t-1}^{(m,k)[j]} \mid \mathcal{F}_{t-1}\right) \inAS 1/k$.
Then the analogues of Theorem~\ref{thm:sum_spacings_strategy}
follows:
\begin{theorem}\label{thm:multi_k_spacings}
Let $d\geq 1$.
Let $m = \lfloor t^\phi \rfloor$, where $\phi \in (0.5,1)$.
Assume that there exists a unique \(j^{\star}\in\{1,\ldots,d\}\) such that $j^{\star} := \argmax_{j\in\{1,\ldots,d\}} \pi_k^{[j]}$. Under the marginal conditions stated above,
then the property in \eqref{eq:k_fraction_power} holds almost surely, i.e.,
\begin{equation}\label{eq:asymp_KL_multi}
    \frac{1}{n}\log \Lambda_n^{(m,k)}
    \inAS \KL_k := \KL \{ \Bern(1/k) \| \Bern(\pi_k^{[j^{\star}]}) \}.
\end{equation}
\end{theorem}

Theorem \ref{thm:multi_k_spacings} shows that the properties
of the multivariate version are similar to the univariate case.
Following Section~\ref{sec:compare_two_eprocess},
we can combine $\Lambda_n^{(m,k)}$ across different values of $k>1$.
The resulting $e$-process achieves the same result as in
Theorem~\ref{thm:divergent_mathcalK_optimal_softmax}.
The proof is a trivial modification; thus, it is omitted.

\section{Monte Carlo experiments}\label{sec:simulations}
\subsection{$e$-spacing distribution test with different strategies}\label{sec:powerMCexp}
We investigate the finite-sample performance of the proposed test
in this section.
We sequentially test $\mathcal{H}_0$ against $\mathcal{H}_1$ at level $\alpha$
until reaching the maximum sample size $n_{\max}$.
In other words,
we reject $\mathcal{H}_0$ at time $\tau$ if $\Lambda_\tau \ge 1/\alpha$ and $\tau \le n_{\max}$.
We set $\alpha = 5\%$ and $n_{\max} = 10000$.
We study the proposed $\Lambda_n$ under the
following six possible strategies:
\begin{itemize}
    \item Maximum spacing strategy $\Lambda_n^{(1,\infty)}$;
        see (\ref{eqt:Dtildestar}).
    \item Maximum spacing set strategy $\Lambda_n^{(m, \mathcal{K})}$ in (\ref{eq:cond_e_R}) with
    $m=\lfloor t^{0.51}\rfloor$, where
    $\{\mathcal{K}_t\}$ is selected as
    (i) a fixed finite $\mathcal{K}=\{2,20\}$ under argmax rule,
    and (ii) a dyadic set $\mathcal{K}_t$ in Example \ref{eg:infinite_K} under
    argmax, softmax, and simple average rules;
    see Definitions \ref{def:argmax_softmax}--\ref{def:simpleAvg}.
    \item Automatic switch strategy $\Lambda_n^{\text{switch}}$,
            which uses argmax rule to switch
            between $\Lambda_n^{(1,\infty)}$ and
            dyadic argmax strategy; see Example \ref{eg:switch}.
\end{itemize}
Their power is computed under four different data generating processes (DGPs) of $\pr$:
\begin{itemize}
    \item[(a)] Let \(X_1, X_2, \ldots \simIID \Normal(\mu, \sigma^2)\).
    We test
\(\mathcal{H}_0: (\mu, \sigma) = (0,1)\)
against
\(\mathcal{H}_1: (\mu, \sigma) \neq (0,1)\).
    \item[(b)] Let \(X_1, X_2, \ldots \simIID \Beta(\alpha, \beta)\).
    We test
\(\mathcal{H}_0: (\alpha, \beta) = (1,1)\)
against
\(\mathcal{H}_1: (\alpha, \beta) \neq (1,1)\).
    \item[(c)]
            Let $X_1, X_2, \ldots \simIID \mathrm{Notch}(\theta_1,\theta_2)$,
            which denotes a two-notch distribution with
            density $p(x)
                = 0.15c\mathbb{1}(x\in G_1 \cup G_2) + c\mathbb{1}(x\in [0,1]\setminus G_1 \cup G_2)$,
                where
                $c=\{1-0.085(|\theta_1|+|\theta_2|)\}^{-1}$,
                $G_1 = [0.25+0.1\min\{0,\theta_2\}, 0.25+0.1\max\{0,\theta_2\}]$, and
                $G_2=[0.75+0.1\min\{0,\theta_1\}, 0.75+0.1\max\{0,\theta_1\}]$.
            Note that $|\theta_1|$ and $|\theta_2|$
            control the notch widths.
            Positive (negative) values of $\theta_1$ and $\theta_2$
            place the notches to the right (left) of 0.25 and 0.75, respectively.
            Note also that
            $\mathrm{Notch}(\theta_1,\theta_2) = \Unif[0,1]$
            iff $\theta_1=\theta_2=0$.
            This model creates localized drops, i.e., notches, in $p(u)$.
            We test $\mathcal{H}_0:(\theta_1,\theta_2)=(0,0)$
            against  $\mathcal{H}_1:(\theta_1,\theta_2)\neq(0,0)$.
    \item[(d)] Let $X_1, X_2, \ldots \simIID
    \kappa t_{\nu}$,
        where $\kappa>0$ is a scale parameter and $t_{\nu}$
        denotes Student's $t$ distribution with $\nu$ degrees of freedom.
        Note that $t_{\infty} = \Normal(0,1)$.
        We test $\mathcal{H}_0:(\nu,\kappa)=(\infty,1)$
            against  $\mathcal{H}_1:(\nu,\kappa)\neq(\infty,1)$.
\end{itemize}

\begin{figure}[!tb]
	\centering
    \includegraphics[width=.92\textwidth]{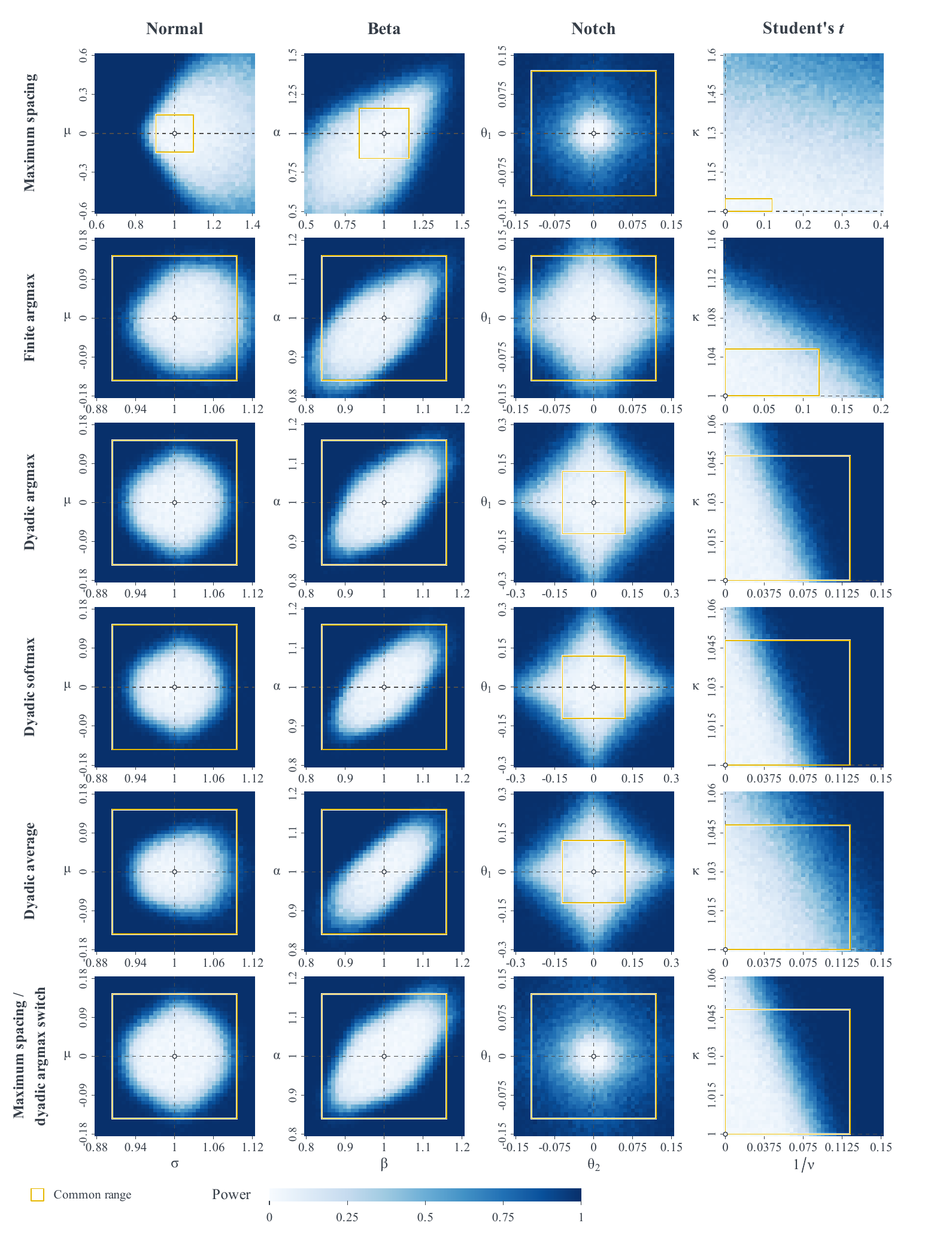}
	\caption{
    The power of the proposed test $\Lambda_n$
    with six strategies under four DGPs
    in Section \ref{sec:powerMCexp}.
    The null case is indicated by a small circle.
    A darker cell means higher power.
    Rectangles indicate the common range of parameters within each DGP.
    Each experiment is replicated
2,000 times under $\mathcal{H}_0$ and 200 times under each $\mathcal{H}_1$.
    }
	\label{fig:heatmaps}
\end{figure}

The results are reported in
Figure~\ref{fig:heatmaps}.
First, we note that neither the maximum spacing strategy
nor the maximum spacing set strategies dominate the other.
The maximum spacing strategy is more capable of detecting localized gaps
between $p(u)$ and $p_0(u)$ in, e.g., the notch case.
This type of subtle change is difficult to detect,
especially based on standard tests that rely on
consistent density estimator.
Hence, this is the most appealing feature of our proposal.
On the other hand, the maximum spacing set strategy
helps retrieve power in a more global difference of $\mid p(u)-p_0(u)\mid$, e.g., difference in dispersion or heavy-tailedness
in the Student's $t$ case.
Finally, the automatic switch strategy $\Lambda_n^{\text{switch}}$
takes the strength of both approaches, and it shows the
highest power or nearly the highest power in all cases.
We remark that even though $p$ is not continuous in $\mathrm{Notch}(\theta_1,\theta_2)$,
our proposed $\Lambda_n$'s
are still valid and powerful in the simulation experiments.
Due to space constraints, we defer the experiments
for the multivariate case to
Section \ref*{sec:multi_MC}
of the supplement.

\subsection{Power comparison with other methods}\label{sec:comparison_other_methods}
\begin{figure}
    \centering
    \includegraphics[width=.95\linewidth]{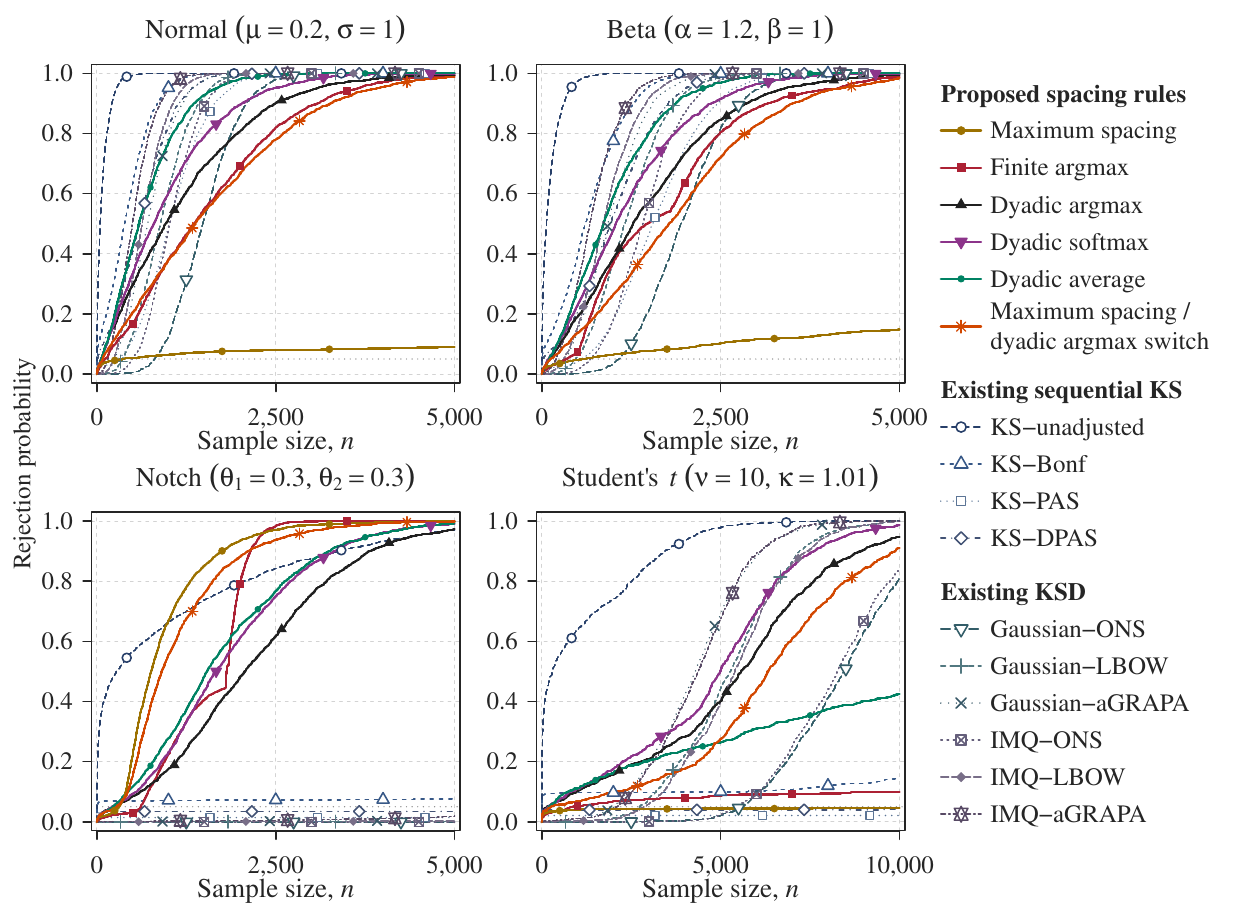}
    \caption{The null rejection probabilities of different tests
    are plotted against the sample size $n$
     under each model.
     The nominal level is $\alpha = 0.05$.
     Each case is replicated 2,000 times.
    }
    \label{fig:power_vs_n}
\end{figure}

We compare the following tests:
\begin{itemize}
   \item The KS distribution test is considered.
            We reject $\cH_0$ at all times \( \geq t\) at level $\alpha$
            if the \(p\)-value based on the first \(t\) observations is
            below \( \alpha_t \).
            Four types of $\alpha_t$ are considered:
            \begin{itemize}
                \item (\textsf{KS-unadjusted}) The unadjusted KS test uses $\alpha_t \equiv \alpha$.
                \item (\textsf{KS-Bonf}) The Bonferroni‑corrected KS test uses $\alpha_t = \alpha/t$.
                \item (\textsf{KS-PAS}) The KS test with polynomial alpha spending
            uses $\alpha_t = 6\alpha(\pi t)^{-2}$.
                \item (\textsf{KS-DPAS}) The KS test with delayed polynomial alpha spending,
            i.e., testing every 10 observations,
            uses $\alpha_t = 6\alpha(\pi t/10 )^{-2} \mathbb{1}(t/10\in\mathbb{Z})$.
            \end{itemize}
            Note that \textsf{KS-unadjusted} and even \textsf{KS-Bonf} are not
            anytime valid because the tests may continue indefinitely
            and the maximum number of tests is not fixed in advance.

    \item \cite{zhou2024sequential,martinez2024sequential} developed KSD-based $e$‑processes.
They use the same type of $e$‑process but different kernels and algorithms for choosing ``betting fraction''.
\cite{martinez2024sequential} use the inverse
multiquadratic (IMQ) kernel with the ``aGRAPA'' and ``LBOW'' algorithms,
while \cite{zhou2024sequential} use the Gaussian kernel with the ``ONS'' algorithm.
We evaluate all six kernel–algorithm combinations,
denoted as $\mathbb{K}$-$\mathbb{A}$,
where $\mathbb{K}\in\{\textsf{IMQ}, \textsf{Gaussian}\}$ and
$\mathbb{A}\in\{ \textsf{aGRAPA},\textsf{LBOW},\textsf{ONS} \}$.

    \item Our proposed $e$-spacing test $\Lambda_n$ with the six strategies
            defined in Section \ref{sec:powerMCexp}.
\end{itemize}
The four DGPs in Section \ref{sec:powerMCexp} are considered
when $(\mu, \sigma)=(0.2, 1)$, $(\alpha, \beta)=(1.2,1)$,
$(\theta_1, \theta_2)=(0.3, 0.3)$, and $(\nu, \kappa)=(10,1.01)$, respectively.
The type-I error probabilities are reported in Table~\ref{tab:null-rejection-probabilities}, whereas
the power curves against $n$ are shown in Figure \ref{fig:power_vs_n}.

\begin{table}[!t]
\centering
\caption{The type-I error probability (\%) under each DGP and $n$
        at the nominal level of $5\%$, based on 2,000
            replications each.
  The type-I error probabilities for other KSD-based tests
    are at most $0.1\%$; see Table \ref*{tab:null-rejection-probabilities_full}
    in the supplement.
  }
\label{tab:null-rejection-probabilities}
\small
\setlength{\tabcolsep}{4pt}
\begin{tabular}{ll rr rr rr}
\toprule
& & \multicolumn{2}{c}{Normal/ Student's $t$}
& \multicolumn{2}{c}{Beta}
& \multicolumn{2}{c}{Notch}
\\
\cmidrule(lr){3-4}\cmidrule(lr){5-6}
\cmidrule(lr){7-8}
Type & Method
& \(n=2500\) & \(n=5000\)
& \(n=2500\) & \(n=5000\)
& \(n=2500\) & \(n=5000\)
\\
\midrule
$e$-spacing & Maximum spacing & 2.30 & 2.35 & 2.10 & 2.10 & 2.45 & 2.50 \\
(proposal) &Finite argmax   & 3.30 & 3.65 & 2.80 & 3.30 & 2.95 & 3.40 \\
&Dyadic argmax   & 3.15 & 3.20 & 3.05 & 3.10 & 4.00 & 4.00 \\
&Dyadic softmax  & 3.60 & 3.70 & 3.05 & 3.10 & 4.15 & 4.15 \\
&Dyadic average  & 3.95 & 4.05 & 3.35 & 3.35 & 4.65 & 4.75 \\
&Switch          & 2.35 & 2.45 & 3.10 & 3.15 & 3.65 & 3.65 \\
\cmidrule(lr){1-8}
KS-based &\textsf{KS-unadjusted} & 61.40 & 65.95 & 60.35 & 64.30 & 62.55 & 65.80 \\
&\textsf{KS-Bonf}       &  8.20 &  8.25 &  7.35 &  7.35 &  7.05 &  7.05 \\
&\textsf{KS-PAS}        &  1.50 &  1.50 &  1.50 &  1.50 &  1.40 &  1.40 \\
&\textsf{KS-DPAS}       &  4.00 &  4.00 &  3.95 &  3.95 &  3.55 &  3.55 \\
\cmidrule(lr){1-8}
KSD-based &\textsf{Gaussian-aGRAPA} & 0.10 & 0.15 & 0.10 & 0.20 & 0.20 & 0.20 \\
&\textsf{IMQ-aGRAPA}      & 0.20 & 0.25 & 0.30 & 0.35 & 0.25 & 0.25 \\
\bottomrule
\end{tabular}
\end{table}

Under $\mathcal{H}_0$,
Table~\ref{tab:null-rejection-probabilities} implies that
\textsf{KS-unadjusted} and \textsf{KS-Bonf}
over-reject $\mathcal{H}_0$, thus, they are not anytime-valid tests.
All other methods can control the level at 5\%.
While \textsf{KS-PAS} and \textsf{KS-DPAS}
successfully control the null rejection rate under
$\mathcal{H}_0$, their power is low.
By decreasing the testing frequency, \textsf{KS-DPAS}
manages to recover some power, but the power remains unsatisfactorily low.
The KSD-based tests show good power in the normal and beta cases,
however, the performance varies across the choice of $\mathbb{K}$ and $\mathbb{A}$.
Most notably, their power drops sharply for the notch and Student’s $t$ cases. In the Notch distribution, their power is nearly zero. By contrast, our test shows strong power for localized spikes, such as the notch distribution. In the normal and beta cases, its power is comparable to other anytime-valid tests. No test uniformly dominates the others. In the beta and notch cases, the finite argmax strategy shows twists in power because it switches to another candidate $k$ after a certain sample size, which boosts up the power.

\subsection{Null rejection time comparison with other methods}\label{sec:different_existing_method}
We compare different tests in terms of the null rejection times, i.e.,
the stopping times $\tau$.
Under $\mathcal{H}_1$, $\tau$ measures the delay of rejecting the false $\mathcal{H}_0$.
So, the shorter the delay, the better is the test.

We test $\mathcal{H}_0: X_1, X_2, \ldots, X_{2000} \simIID t_1$
and follow the same experimental setting
as stated in Figure 2 of \cite{zhou2024sequential}.
We generate data from the $t_1$, $t_5$, $t_{10}$, and $t_{\infty}$.
So, the first one is under $\mathcal{H}_0$, while the last three are under $\mathcal{H}_1$.
The stopping time $\tau$ capped at 2,000 is computed.

\begin{figure}[t]
        \centering
        \includegraphics[width=0.95\textwidth]{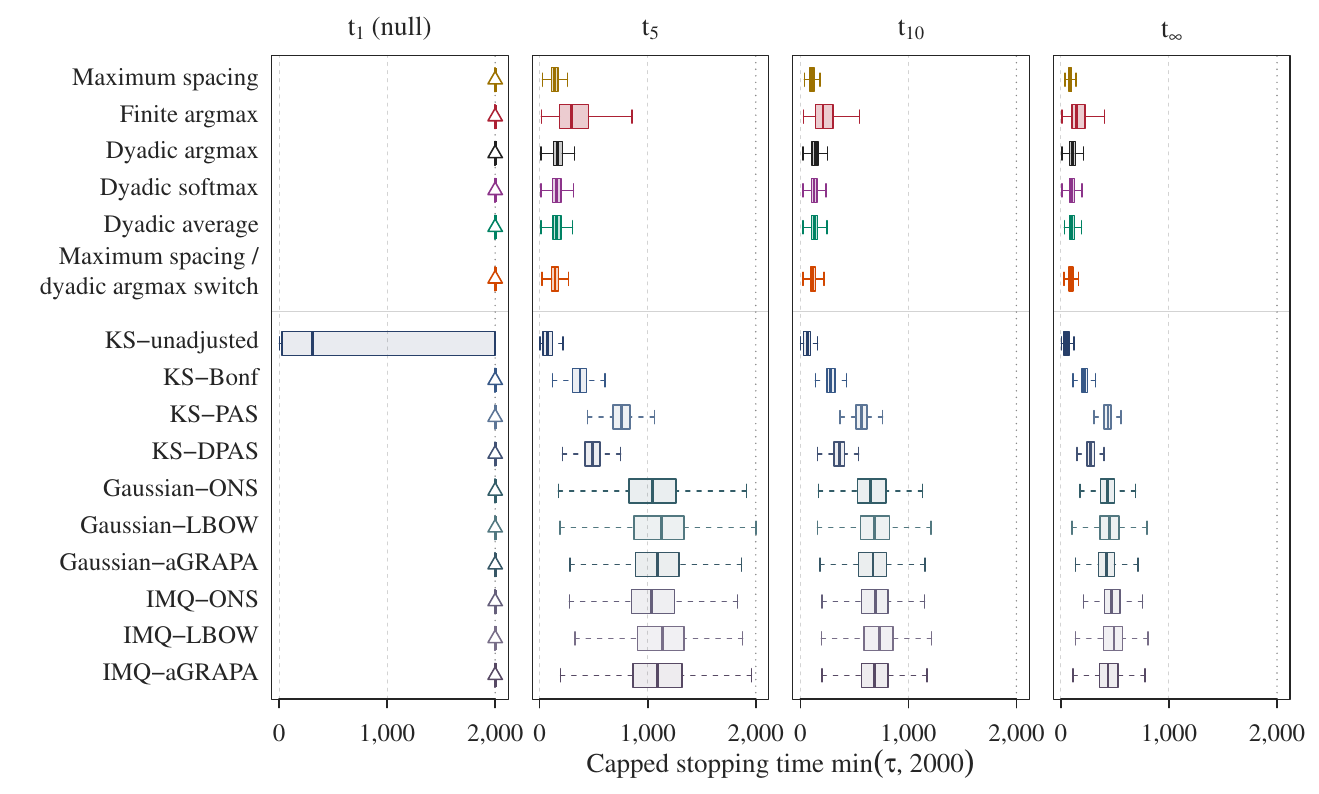}
    \caption{The box plots of the stopping time $\tau$ (capped at 2,000)
    of different tests under the DGPs $t_1$, $t_5$, $t_{10}$, and $t_{\infty}$, respectively.
    Triangles indicate that $\tau\geq 2000$.
    Each experiment is repeated 500 times.
    }
	\label{fig:relative_size}
\end{figure}

The results are shown in Figure~\ref{fig:relative_size}.
Under $\mathcal{H}_1$,
all of our proposed $e$-spacing-based tests reject
$\mathcal{H}_0$ substantially earlier than the KS-based and kernel-based tests.
The only exception is the \textsf{KS-unadjusted} method,
which rejects $\mathcal{H}_0$ very early,
no matter under $\mathcal{H}_0$ or $\mathcal{H}_1$,
due to its invalid type-I error control,
as demonstrated in the first plot of Figure~\ref{fig:relative_size}.
It is interesting to observe that the KSD-based tests
experience more significant delay when the data have heavier tails,
whereas our $e$-spacing based tests can quickly reject $\mathcal{H}_0$
in all cases of $\mathcal{H}_1$.

Overall, our proposed test achieves high power and
is well suited to sequential testing.
In the settings considered,
it rejects a false $\mathcal{H}_0$ substantially earlier
than adjusted KS tests
and other recently developed kernel-based methods.

\section{Real-data application: Gas-sensor array drift test}\label{sec:realData}
A multisensor measurement system is frequently used
in environmental monitoring and quality control,
where repeated measurements are collected
under changing operating conditions.
However, the sensitivity of sensors may degrade
during prolonged operation because of
sensor-surface contamination or changes in background conditions.
So, detecting sensor drift is of interest \citep{Rudnitskaya2018}.
In what follows, a sensor drift means
a change in the conditional distribution of the sensor response relative
to a stable reference period,
after controlling for recorded confounding variables.

We study the gas sensor dataset in \cite{Vergara2012},
which contains $n=13,910$ measurements from 16 chemical sensors exposed to
6 gases: Ethanol, Ethylene, Ammonia, Acetaldehyde, Acetone, and Toluene.
The gases were measured at different concentration levels
together with 128 real-valued sensor response features.
The dataset was collected in ten chronological batches
with unequal sizes $n_1, \ldots, n_{10}$.
So, $n = n_1 + \cdots + n_{10}$.
Note that there is no chronological relationship among observations within the same batch.

In the $i$th observation,
denote $X_{ij}\in\mathbb{R}$ as the $j$th feature ($j=1, \ldots, 128$),
and $C_i\in[0,\infty)$, $G_i\in\{1,\ldots,6\}$, and  $B_i \in \{1, \ldots, 10\}$
as the gas concentration, gas label, and the batch index, respectively.
Let $\mathbb{I}_b = \{N_{b-1}+1, \ldots, N_{b}\}$ be the $b$th batch of observation indices, where $N_0=0$ and $N_b = n_1 + \cdots + n_b$.
So, $|\mathbb{I}_b| = n_b$.
We partition $\mathbb{I}_1\cup\cdots\cup \mathbb{I}_{10}$ into four parts:
\[
    \underbrace{\left( \mathbb{I}_1 \cup \mathbb{I}_2 \cup \mathbb{I}_3 \right)}_{\text{randomized to}\;\mathbb{I}_{\text{fit}},\mathbb{I}_{\text{cal}},\mathbb{I}_{\text{con}}} \cup \underbrace{\left( \mathbb{I}_4 \cup \cdots \cup \mathbb{I}_{10} \right) }_{\mathbb{I}_{\text{test}}}
    \qquad \text{such that} \qquad
    |\mathbb{I}_{\text{fit}}| : |\mathbb{I}_{\text{cal}}| : |\mathbb{I}_{\text{con}}| \approx 50:25:25
\]
and the gas type proportions in
$\mathbb{I}_{\text{fit}}, \mathbb{I}_{\text{cal}}, \mathbb{I}_{\text{con}}$
remain the same as $\mathbb{I}_1 \cup \mathbb{I}_2 \cup \mathbb{I}_3$,
up to rounding.
The sets $\mathbb{I}_{\text{fit}},\mathbb{I}_{\text{cal}},\mathbb{I}_{\text{con}},\mathbb{I}_{\text{test}}$
are used for
model fitting, null-score calibration,
null-control analysis, and testing, respectively.
We perform the analysis in three steps:

First,
based on $\mathbb{I}_{\text{fit}}$,
we perform principal component (PC) analysis on the 128 features
to reduce the dimension.
Denote the first eight PC scores of the $i$th observation as
$P_{i,1}, \ldots, P_{i,8}$ for $i=1, \ldots, n$.
Based on $\mathbb{I}_{\text{fit}}$ again,
a linear regression model is fitted for each $j=1,\ldots, 8$:
\begin{align*}
    \E(P_{i,j} \mid G_i , L_i) = \alpha_{j, G_i} + \beta_{j,G_i}L_i + \gamma_{j, G_i} L_i^2,
    \qquad \text{where} \qquad
    L_i = \log (1+C_i).
\end{align*}
So, it includes
the interaction between $G_i$ with $(1,L_i, L_i^2)$.
The log-plus-one transformation is used for $C_i \geq 0$
to account for nonlinear gas-sensor responses \citep{Hirobayashi2003}
and to prevent large values of $C_i$ from dominating the regression fit.
Let the fitted regression function be
$\widehat{P}_j(g, \ell) = \widehat{\alpha}_{j, g} + \widehat{\beta}_{j,g}\ell + \widehat{\gamma}_{j, g} \ell^2$.
We measure the sensor response in the $j$th PC direction
after adjusting for gas type and concentration by the studentized residuals:
\begin{align}\label{eqt:studentized_residuals}
    z_{ij} = Z_{ij}/\widehat{\sigma}_{i,G_i},
    \qquad \text{where} \qquad
    Z_{i,j} = P_{i,j}-\widehat{P}_j(G_i, L_i)
\end{align}
and $\widehat{\sigma}_{j,g}^2$ is the sample variance
of $\{Z_{ij}: G_i=g, i\in\mathbb{I}_{\text{fit}}\}$.
In (\ref{eqt:studentized_residuals}), $Z_{ij}$ is studented by gas type
because of the heterogeneity across $g$.
Residuals are commonly used as scores for quality control; see, e.g., \cite{JacksonMudholkar1979}.
Then we define one global score and eight local scores as
\[
    S_{i0} = \frac{1}{8}\sum_{j=1}^8 |z_{ij}|
    \qquad \text{and} \qquad
    S_{ij} = |z_{ij}|,\quad j=1,\ldots,8.
\]

Second, for score channel $j\in\{0, 1, \ldots, 8\}$,
define the reference null score distribution by
$F_{0j}(s \mid g)
        = \pr(S_{i,j}\leq s \mid \mathcal{F}, G_i=g)$,
where
$i \in \mathbb{I}_{\text{cal}}$ and
$\mathcal{F}$ denote all information generated from observations
with indices $\mathbb{I}_{\text{fit}}$.
We aim at testing
\[
    \mathcal{H}_0: \pr(S_{ij}\leq \cdot \mid \mathcal{F}, G_i=g) = F_{0j}(\cdot \mid g),
    \qquad i\in\mathbb{I}_{\text{cal}}\cup\mathbb{I}_{\text{test}}.
\]
for all $g$.
To do it, we use the randomized rank probability-integral transform (PIT).
Let $n_{b,g} = |\{ i\in\mathbb{I}_b : G_i = g\}|$.
For each batch index $b=4, \ldots, 10$ and each gas type $g=1,\ldots, 6$,
we randomly select $\min(3, n_{b,g})$ observations in $\mathbb{I}_{\text{test}}$
from the $(b,g)$-cell.
The selected observations are arranged in ascending batch index,
and the orders are randomized within each batch.
Denote them as $S_{i''_1,j}, \ldots, S_{i''_{T}, j}$,
where $T = \sum_{b=4}^{10}\sum_{g=1}^6 \min(3,n_{b,g}) = 120$
in this dataset, because gas 6 is absent from batches 4--5.
For each $i''_t$, we randomly select $i'_t \in \mathbb{I}_{\text{cal}}$
such that $G_{i'_t}= G_{i''_t}$ without replacement.
Similar to \cite{HenziLaw2024},
for the $t$th pair $(S_{i'_t,j}, S_{i''_t,j})$, we define
\begin{align}\label{eqt:def_V_realData}
    U_{t,j} = \frac{\mathbb{1}( S_{i'_t,j} <S_{i''_t,j} )+V_{t,j}}{2},
    \qquad\text{where} \qquad
    V_{t,j}\sim\Unif[0,1].
\end{align}
Under $\mathcal{H}_0$,
$\pr(S_{i'_t,j}<S_{i''_t,j}\mid\mathcal{F},G_t)=1/2$.
Consequently, $[U_{t,j}\mid\mathcal{F},G_t]\sim\Unif[0,1]$.
So, testing $\mathcal{H}_0$ can be done by
testing (conditional) uniformity on $U_{1,j},\ldots, U_{120,j}$.

\begin{figure}
    \centering
    \includegraphics[width=\linewidth]
    {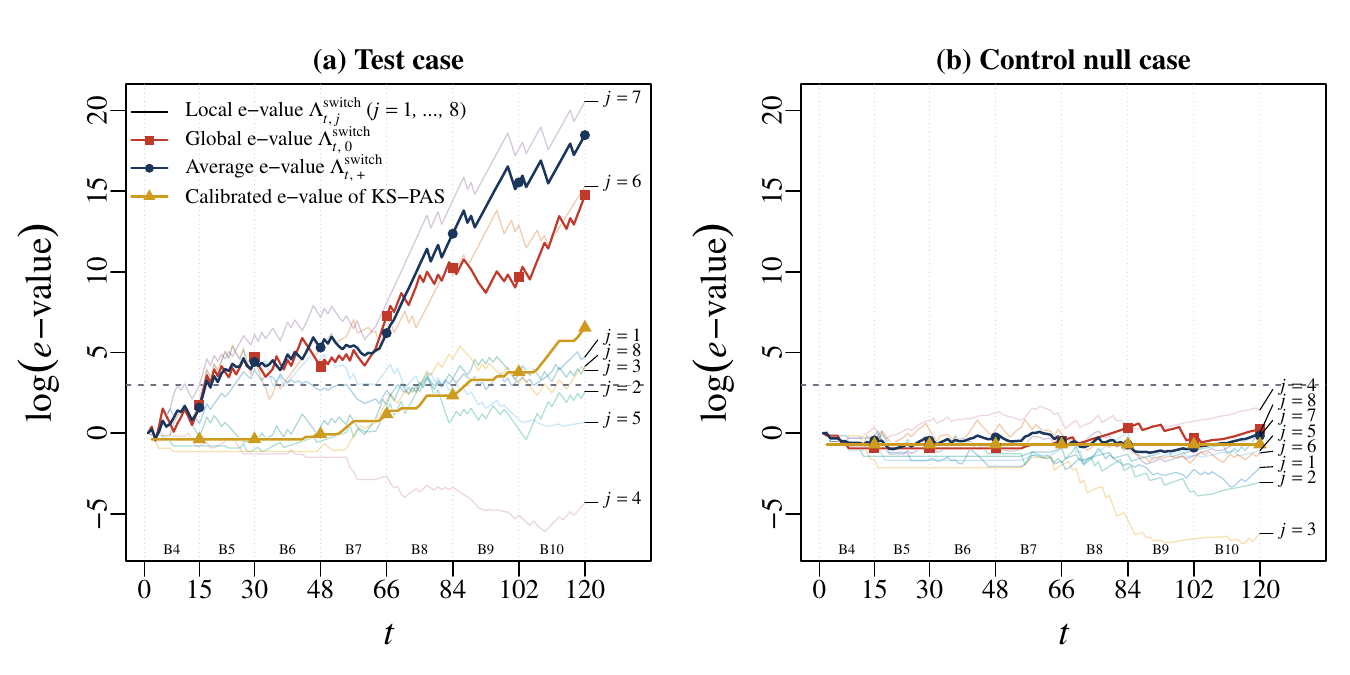}
    \caption{The log $e$-value paths
    in the test case (left panel) and the control null case (right panel)
    for the global test $\Lambda_{t,0}^{\text{switch}}$,
    local tests ($\Lambda_{t,1}^{\text{switch}}, \ldots, \Lambda_{t,8}^{\text{switch}}$),
    the average test ($\Lambda_{t,+}^{\text{switch}}$),
    and the projected \textsf{KS-PAS} test.
    The vertical dashed lines partition the batches.
    The horizontal dashed line is $\log(1/5\%)$.
    }
    \label{fig:A05_realData}
\end{figure}

Third, for each score channel $j=0, \ldots, 8$,
we apply the proposed test
with the automatic switch strategy defined in Section \ref{sec:powerMCexp}
to $U_{1,j},\ldots, U_{t,j}$ sequentially.
For each $j=0, \ldots, 8$, denote it as $\Lambda_{t,j}^{\text{switch}}$.
For reference, we also compute
(i) the averaged $e$-value across eight PC directions,
i.e., $\Lambda_{t,+}^{\text{switch}} = \sum_{j=1}^8\Lambda_{t,j}^{\text{switch}}/8$,
which is also an $e$-value, and
(ii) the \textsf{KS-PAS} method defined in Section \ref{sec:comparison_other_methods} with
a $e$-value computed based on the $p$-to-$e$ calibrator $e(p)=1/(2p^{1/2})$
\citep{VovkWang2021}.
The log $e$-values of all tests are reported in Figure \ref{fig:A05_realData} (a).
The first reject times based on
$\Lambda_{t,0}^{\text{switch}}$, $\Lambda_{t,+}^{\text{switch}}$, and \textsf{KS-PAS}
are 17 (batch 5), 17 (batch 5), and 88 (batch 9).
Hence, the proposed methods can detect shift earlier.
From the local test, we observe that the 6th and 7th PC directions
show the largest $e$-values, thus,
the changes reported in
the global and average measures are mainly driven by them.
Practitioner can, therefore, investigate this specific direction of the shift.
However, we remark that this analysis is an example of
a one-sample test against a reference null sample
and is different from online change point
or anomaly detection, as studied in \cite{gosmann2021cp,leung2024nonparametric,leung2025onlinegeneralizedmethodmoments}.
For reference, the above PIT procedure is repeated
with $i''\in\mathbb{I}_{\text{con}}$ instead of
$i''\in\mathbb{I}_{\text{test}}$, but the
control observations are not restricted by batch index.
By exchangeability, this control stream satisfies $\mathcal{H}_0$;
see Figure \ref{fig:A05_realData} (b).
All tests do not report shit as expected.
Since the above analysis depends on the random partition and generation of $V_{t,j}$ in (\ref{eqt:def_V_realData}),
we replicate the above analysis 1,000 times as a sensitivity analysis.
The probabilities of rejecting $\mathcal{H}_0$ at $t\leq 120$
based on $\Lambda_{t,0}^{\text{switch}}$, $\Lambda_{t,+}^{\text{switch}}$, and \textsf{KS-PAS}
are 91.4\%, 94.0\%, and 76\%, respectively.
The restricted mean rejection time estimates are
69.4, 64.9, and 98.1, respectively,
where the restriction time is set at the total sample size 120.
It confirms that $\Lambda_{t,0}^{\text{switch}}$ and $\Lambda_{t,+}^{\text{switch}}$
are powerful across replications;
see Figure \ref*{fig:A05_realData_rejTime} of the supplement for the
rejection probability curves based on the Kaplan--Meier estimator.

\section*{Acknowledgements}
Chan is the corresponding author.
Hao completed a part of this work while he was a postdoctoral fellow at The Chinese University of Hong Kong; he has since completed his fellowship and moved to Tianjin University.
The authors would like to thank
Peter Grünwald for his comments on the selection of $\widehat{D}_{t-1}^{(1,\infty)}$.
Although we did not adopt his original suggestion,
it inspired us to derive our choice of $\widehat{D}_{t-1}^{(1,\infty)} = (\log t)/t$.
The authors also thank Zhengyu Zhou and Weiwei Liu for generously sharing the code
from their paper, and Alexander Henzi, Limin Li and Qianxin Yi for helpful discussions.

\section*{Statement on AI use}
The authors used ChatGPT to assist with proofreading, language editing, figure generation, code development, numerical experiments, and drafting and checking mathematical proofs. The authors thoroughly reviewed and verified all material retained in the article and take full responsibility for the final work.

\appendix
\setcounter{equation}{0}
\setcounter{table}{0}
\setcounter{figure}{0}
\renewcommand{\theequation}{\thesection.\arabic{equation}}
\renewcommand{\thetable}{\thesection.\arabic{table}}
\renewcommand{\thefigure}{\thesection.\arabic{figure}}

\section{Proofs of main results}
\subsection{Proofs of Propositions \ref{prop:eprocess} and \ref{prop:gen_eprocess}}\label{sec:proof_prop:eprocess}
Proposition~\ref{prop:eprocess} is a special case of
Proposition \ref{prop:gen_eprocess}.
So, we only prove Proposition \ref{prop:gen_eprocess}.

Recall that,
by definition, $\lambda_1 = 1$.
At time $n\geq 1$, we have
\begin{align*}
\E_{\pr_0}\left(\lambda_n \mid \mathcal{F}_{n-1} \right)
=& \frac{\widehat{D}_{n-1}}{D_{n-1}} \cdot \pr_0\left(U_n \in \mathcal{U}_{n-1} \mid \mathcal{F}_{n-1} \right)
+ \frac{1- \widehat{D}_{n-1}}{1-D_{n-1}} \cdot \pr_0\left(U_n \notin \mathcal{U}_{n-1} \mid \mathcal{F}_{n-1}\right)\\
\overset{(a)}{=}&
\frac{\widehat{D}_{n-1}}{D_{n-1}} \cdot D_{n-1}
+ \frac{1 - \widehat{D}_{n-1}}{1-D_{n-1}} \cdot (1-D_{n-1})
= 1,
\end{align*}
where $(a)$ follows from $U_1, \ldots U_n \simIID \Unif[0, 1]$ under $\cH_0$. We get
\begin{equation}
\E_{\pr_0}\left(\Lambda_n \mid \mathcal{F}_{n-1} \right)
= \Lambda_{n-1} \cdot \E_{\pr_0}\left(\lambda_n \mid \mathcal{F}_{n-1} \right)
= \Lambda_{n-1}.
\end{equation}
By definition, $\{\Lambda_n\}_{n \geq 1}$ are non-negative random variables,
hence, $\E_{\pr_0}\left( |\Lambda_n | \right) = \E_{\pr_0}\left(\Lambda_n \right) < \infty$.
Therefore, $\{\Lambda_n\}_{n \geq 1}$ is a non-negative martingale with initial condition $\Lambda_1 = 1$, which is also called a test martingale \citep{shafer2011test}.
\cite{ramdas2020admissible} shows that a test martingale
$\{\Lambda_n\}_{n \geq 1}$ is an $e$-process under simple null $\pr_0$, i.e.,
$\E_{\pr_0}(\Lambda_\tau) \leq 1$
holds for any stopping time $\tau$ with respect to the filtration $\{\mathcal{F}_n\}_{n \geq 1}$.

\subsection{Proofs of Theorems  \ref{thm:log_optimal} and \ref{thm:general_log_optimal}}\label{proof:thm42}
Theorem~\ref{thm:log_optimal} is a special case
of Theorem \ref{thm:general_log_optimal}.
So, we only prove Theorem \ref{thm:general_log_optimal}.

Recall that $\lambda_t$ is the test statistic defined in~(\ref{def:GENe_process_lambda}). We get
\begin{align}
&\E_\pr\left(\log \lambda_t \mid \mathcal{F}_{t-1}\right)\nonumber\\
&\qquad= \E_\pr\left[\log \left\{\frac{\widehat{D}_{t-1}}{D_{t-1}} \mathbb{1}(U_t \in \mathcal{U}_{t-1}) + \frac{1 - \widehat{D}_{t-1}}{1-D_{t-1}} \mathbb{1}(U_t\notin \mathcal{U}_{t-1}) \right\} \Bigg\mid \mathcal{F}_{t-1} \right]\nonumber\\
&\qquad\overset{(a)}{=}
\pr\left(U_t \in \mathcal{U}_{t-1} \mid \mathcal{F}_{t-1} \right) \cdot \log\frac{\widehat{D}_{t-1}}{D_{t-1}}
+ \pr\left(U_t \notin \mathcal{U}_{t-1} \mid \mathcal{F}_{t-1} \right) \cdot \log\frac{1 - \widehat{D}_{t-1}}{1-D_{t-1}}\nonumber\\
&\qquad=
\widehat{D}_{t-1}^{\star}\cdot \log \widehat{D}_{t-1}
+ (1-\widehat{D}_{t-1}^{\star})\cdot \log(1 - \widehat{D}_{t-1})
- \widehat{D}_{t-1}^{\star} \cdot \log D_{t-1}
- (1 - \widehat{D}_{t-1}^{\star}) \cdot \log (1-D_{t-1})\nonumber\\
&\qquad\overset{(b)}{\leq}
\widehat{D}_{t-1}^{\star}\cdot \log \widehat{D}_{t-1}^{\star}
+ (1-\widehat{D}_{t-1}^{\star})\cdot \log(1 - \widehat{D}_{t-1}^{\star})
- \widehat{D}_{t-1}^{\star} \cdot \log D_{t-1}
- (1 - \widehat{D}_{t-1}^{\star}) \cdot \log (1-D_{t-1})\nonumber\\
&\qquad=
\pr\left(U_t \in \mathcal{U}_{t-1} \mid \mathcal{F}_{t-1} \right) \cdot \log\frac{\widehat{D}_{t-1}^{\star}}{D_{t-1}}
+ \pr\left(U_t \notin \mathcal{U}_{t-1} \mid \mathcal{F}_{t-1} \right) \cdot \log\frac{1 - \widehat{D}_{t-1}^{\star}}{1-D_{t-1}}\nonumber\\
&\qquad=
\E_\pr\left( \log \lambda_t^{\star} \mid \mathcal{F}_{t-1}\right)
=\KL(B_\pr \,\|\, B_{\pr_0}),\label{align:proof_thm42}
\end{align}
where (a) holds since $U_{1:t-1}$ is independent of $U_t$, (b) follows from Gibbs' inequality with equality holds iff $\widehat{D}_{t-1} = \widehat{D}_{t-1}^{\star}$. Then (\ref{eq:general_optimal_epower}) follows for any fixed $\pr$.

\subsection{Proof of Theorem \ref{thm:maxSpacing_strategy}}\label{proof:thm32}
\subsubsection{Preliminary and outline}
For notational simplicity,
throughout this proof we write \(D_t\), \(\widehat{D}_t\), and \(\Lambda_n\)
instead of \(D_t^{\md}\), \(\widehat{D}_t^{\md}\), and \(\Lambda_n^{\md}\).
Recall that $\log \lambda_t = I_t A_{t-1} + (1-I_t) B_{t-1}$,
where $A_{t-1} = \log\{\widehat{D}_{t-1}/D_{t-1}\}$
and  $B_{t-1} = \log\{(1-\widehat{D}_{t-1})/(1-D_{t-1})\}$.
We decompose $\Lambda_n = \prod_{t=2}^n \lambda_t$ into two parts:
\begin{align}
    \log \Lambda_n
        &= \left( \sum_{t=2}^n \mu_t \right)
            +  \sum_{t=2}^n (\log\lambda_t-\mu_t) ,
    \quad \text{where} \quad \label{eqt:decomposition_logLambda} \\
    \mu_t &:= \E(\log\lambda_t\mid\mathcal{F}_{t-1})
        = \widehat{D}_{t-1}^{\star} A_{t-1} + (1-\widehat{D}_{t-1}^{\star}) B_{t-1}
    \quad \text{and}  \quad
    \widehat{D}_{t-1}^{\star} = \E(I_t \mid \mathcal{F}_{t-1}).
        \label{eqt:mu_definition}
\end{align}
Note that $\{\log \lambda_t - \mu_t\}_{t=2}^{\infty}$ is a martingale difference sequence
with respect to $\mathcal{F}_t$.
We proceed with the proof of $\Lambda_n \inAS \infty$
in two cases;
see Sections \ref{sec:case1_consistency} and \ref{sec:case2_consistency}, respectively.
In both cases, we will derive the limit of the two terms on the right-hand side of (\ref{eqt:decomposition_logLambda}) one by one:
\begin{itemize}
    \item $\sum_{t=2}^n \mu_t$ is handled by
        using the conditions made in the theorem, whereas
    \item $\sum_{t=2}^n (\log\lambda_t-\mu_t)$
        is handled by the martingale strong law of large numbers
        in Theorem 2.18 of \cite{hallHeyde1980}.
\end{itemize}
Then, in Section \ref{sec:case12_Epower}, we prove $\E(\Lambda_n)\rightarrow\infty$.

\subsubsection{Case 1: $p_{\min} > 0$}\label{sec:case1_consistency}
Note that Condition~\ref{cond:simple_strategy} implies
\begin{align}\label{eqt:order_D_Dhat}
    D_{t-1}   &= \frac{c \log t}{t} + o_{\AS}\left( \frac{\log t}{t} \right)
    \quad \text{and} \quad
    \widehat{D}_{t-1}^{\star} = \frac{\log t}{t} + o_{\AS}\left( \frac{\log t}{t} \right).
\end{align}
Using the first relation in (\ref{eqt:order_D_Dhat}),
the definition $\widehat{D}_{t-1} = \log t/t$, and Taylor's theorem, we have
\begin{align}
    A_{t-1}   &= -\log c + o_{\AS}(1), \label{eqt:A_expansion}\\
    B_{t-1} &= \log(1-\widehat{D}_{t-1}) - \log(1-D_{t-1})
            = -\widehat{D}_{t-1}
                + D_{t-1} + o_{\AS}\left( \frac{\log t}{t}+ D_{t-1} \right)
                \nonumber \\
            &= (c-1) \frac{\log t}{t} + o_{\AS}\left( \frac{\log t}{t} \right).\label{eqt:B_expansion}
\end{align}
Putting (\ref{eqt:order_D_Dhat}), (\ref{eqt:A_expansion}), and (\ref{eqt:B_expansion}) into the definition of $\mu_t$ in (\ref{eqt:mu_definition}), we have
\[
    \mu_t = (c-1-\log c) \frac{\log t}{t} + o_{\AS}\left( \frac{\log t}{t} \right).
\]
By the convergence property of Ces\`{a}ro means,
\begin{align}\label{eqt:sumOfMu_limit}
    \frac{1}{L_n} \sum_{t=2}^n \mu_t \inAS c-1-\log c,
    \quad \text{where} \quad
    L_n = \sum_{t=2}^n \frac{\log t}{t}.
\end{align}
Now, note that
\begin{align}
    \E\{(\log\lambda_t-\mu_t)^2\mid \mathcal{F}_{t-1}\}
    &= \Var(\log\lambda_t \mid \mathcal{F}_{t-1})
    = (A_{t-1}-B_{t-1})^2\widehat{D}_{t-1}^{\star} (1-\widehat{D}_{t-1}^{\star}) \label{eqt:condVarloglambda}\\
    &\leq (A_{t-1}-B_{t-1})^2 \widehat{D}_{t-1}^{\star}. \nonumber
\end{align}
Together with (\ref{eqt:A_expansion}), (\ref{eqt:B_expansion}), and
the property that $L_t \asymp (\log t)^2$ as $t\rightarrow\infty$,
we have
\[
    \sum_{t=2}^{\infty}\E\left\{ \left( \frac{\log\lambda_t-\mu_t}{L_t} \right)^2 \mid \mathcal{F}_{t-1} \right\}
    \lesssim \sum_{t=2}^{\infty} \frac{\log t/t}{(\log t)^4}
    = \sum_{t=2}^{\infty} \frac{1}{t(\log t)^3}
    <\infty\qquad \text{almost surely}.
\]
By the martingale strong law of large numbers (see Theorem 2.18 of \cite{hallHeyde1980}), $\sum_{t=2}^n (\log\lambda_t-\mu_t) / L_n \inAS 0$.
Together with (\ref{eqt:sumOfMu_limit}),
we obtain
\begin{align}\label{eqt:consistency_result_case1_proof}
    \log\Lambda_n /L_n \inAS c-1-\log c>0.
\end{align}
Thus, the desired result follows in case 1.

\subsubsection{Case 2: $p_{\min}=0$}\label{sec:case2_consistency}
Condition~\ref{cond:simple_strategy} implies that
$\widehat{D}_{t-1}^{\star} \inAS 0$.
So, the three convergence in Condition~\ref{cond:simple_strategy},
the convergence $\widehat{D}_{t-1}^{\star} \inAS 0$, and
the fact that $\log t/t \rightarrow 0$ imply that
there is an almost surely finite $T$ such that for all $t\geq T$,
\begin{align}\label{eqt:goodCondition_bounds}
    \frac{D_{t-1}}{\log t/t} \geq 4, \quad\;
    0\leq D_{t-1} < 1-\varepsilon, \quad \;
    \frac{\widehat{D}_{t-1}^{\star} \log\left( \frac{D_{t-1}}{\log t/t}\right)}{D_{t-1}} \leq \frac{1}{8}, \quad\;
    \widehat{D}_{t-1}^{\star} \leq \frac{1}{4} , \quad \;
    \frac{\log t}{t} \leq \frac{1}{2},
\end{align}
where $\varepsilon>0$ is a constant.
So, for $t\geq T$, we can use (\ref{eqt:goodCondition_bounds}) to obtain
\begin{align}
    A_{t-1} &= \frac{-D_{t-1}}{\widehat{D}_{t-1}^{\star}}\frac{\widehat{D}_{t-1}^{\star}\log\left(\frac{D_{t-1}}{\log t/t}\right)}{D_{t-1}}
        \geq  \frac{-D_{t-1}}{8\widehat{D}_{t-1}^{\star}}, \label{eqt:bound_A_case2} \\
    B_{t-1} &= -\log(1-D_{t-1}) + \log\left(1- \frac{\log t}{t}\right)
    \geq D_{t-1}- \frac{2\log t}{t} \label{eqt:bound_log(1-x)}
    \geq \frac{D_{t-1}}{2}, \\
    |A_{t-1}-B_{t-1}|
    	&\leq \left\vert \log \left( \frac{D_{t-1}}{\log t/t} \right)\right\vert
	    		+ \left\vert \log \left( \frac{1-D_{t-1}}{1-\log t/t} \right)\right\vert
		\leq C_{\varepsilon} + \log \left( \frac{D_{t-1}}{\log t/t} \right) , \label{eqt:bound_|A-B|_case2}
\end{align}
where $C_{\varepsilon}>0$ is finite almost surely;
and (\ref{eqt:bound_log(1-x)})--(\ref{eqt:bound_|A-B|_case2}) are
obtained by using the following inequalities:
\begin{itemize}
	\item (\ref{eqt:bound_log(1-x)}) uses
		$-\log(1-x) \geq x$ for $x\in[0,1)$, and
		$-\log(1-x) \leq 2x $ for $x\in[0,1/2]$; and
	\item (\ref{eqt:bound_|A-B|_case2}) uses
		$0<\varepsilon<1-D_{t-1}\leq 1$ and
		$1/2 \leq 1-\log t/t \leq 1$.
\end{itemize}
Putting
(\ref{eqt:goodCondition_bounds}),
(\ref{eqt:bound_A_case2}), and (\ref{eqt:bound_log(1-x)})
into the definition of $\mu_t$ in (\ref{eqt:mu_definition}), we have
for $t\geq T$,
\begin{align}\label{eqt:bound_mut_case2}
    \mu_t \geq - \frac{D_{t-1}}{8}  + \left( 1- \frac{1}{4}\right) \frac{D_{t-1}}{2}
    \geq \frac{D_{t-1}}{4}
\end{align}
Next, similar to (\ref{eqt:condVarloglambda}), we use
$\widehat{D}_{t-1}^{\star} \geq 0$ and (\ref{eqt:bound_|A-B|_case2}) to obtain
\begin{align}
    \E\{(\log\lambda_t-\mu_t)^2\mid \mathcal{F}_{t-1}\}
    &\leq (A_{t-1}-B_{t-1})^2 \widehat{D}_{t-1}^{\star} \nonumber\\
    &\leq \left\{ C_{\varepsilon} + \log \left( \frac{D_{t-1}}{\log t/t} \right) \right\}^2 \widehat{D}_{t-1}^{\star} \nonumber\\
    &= \left[ \frac{\left\{ C_{\varepsilon} + \log \left( \frac{D_{t-1}}{\log t/t} \right) \right\}^2}{\log \left( \frac{D_{t-1}}{\log t/t} \right)} \right]
    \widehat{D}_{t-1}^{\star} \log \left( \frac{D_{t-1}}{\log t/t} \right) \nonumber\\
    &\leq \left[ \left(1+\frac{C_{\varepsilon}}{\log 4}\right)^2 \log \left( \frac{D_{t-1}}{\log t/t} \right) \right]
    \widehat{D}_{t-1}^{\star} \log \left( \frac{D_{t-1}}{\log t/t} \right) \label{eqt:use_ineq_(C+x)^2/x} \\
    &\leq \left[ \left(1+\frac{C_{\varepsilon}}{\log 4}\right)^2 \log t \right]\times
    \frac{\widehat{D}_{t-1}^{\star} \log \left( \frac{D_{t-1}}{\log t/t} \right) }{D_{t-1}} \times D_{t-1}
    \label{eqt:use_ineq_D/(logt/t)} \\
    &\leq \left[ \left(1+\frac{C_{\varepsilon}}{\log 4}\right)^2 \log t \right]\times
    \frac{1}{8} \times D_{t-1}.\label{eqt:bound_2ndMoment_case2}\\
    &= C_{\varepsilon}' D_{t-1} \log t ,\label{eqt:bound_2ndMoment_case2_final}
\end{align}
where $C_{\varepsilon}'>0$ is finite almost surely,
and (\ref{eqt:use_ineq_(C+x)^2/x})--(\ref{eqt:bound_2ndMoment_case2})
are obtained as follows:
\begin{itemize}
    \item (\ref{eqt:use_ineq_(C+x)^2/x}) uses
the inequality that $(C_{\varepsilon}+x)^2/x \leq (1+C_{\varepsilon}/\log 4)^2 x$ for all $x\geq \log 4$
    and puts $x=\log\{D_{t-1}/(\log t /t)\}$ into this inequality;
    \item (\ref{eqt:use_ineq_D/(logt/t)}) follows from
$D_{t-1}/(\log t/t)\leq 1/(\log t/t) \leq t$ for all $t\geq 2$; and
    \item (\ref{eqt:bound_2ndMoment_case2}) follows from the third inequality in (\ref{eqt:goodCondition_bounds}).
\end{itemize}
Let $S_n = \sum_{t=2}^n D_{t-1}$, which is $\mathcal{F}_{n-1}$-measurable. By the first inequality in (\ref{eqt:goodCondition_bounds}) and
$\sum_{t=1}^n \log t /t \asymp (\log n)^2$,
we know that there is an almost surely finite $T'\geq T$ such that
for all $t\geq T'$,
$S_n \geq c_{T'} (\log n)^2$ or
$\log n \leq c_{T'}^{-1/2}S_n^{1/2}$
for some almost surely finite $c_{T'}>0$.
Together with (\ref{eqt:bound_2ndMoment_case2_final}), we have
for $t\geq T'$,
\begin{align}
    \frac{\E\{(\log\lambda_t-\mu_t)^2\mid \mathcal{F}_{t-1}\}}{S_t^2}
        &\leq C_{\varepsilon}'c_{T'}^{-1/2} \frac{D_{t-1}}{S_t^{3/2}}
        = C_{\varepsilon}'c_{T'}^{-1/2} \frac{S_{t}-S_{t-1}}{S_t^{3/2}} \nonumber\\
        &\leq C_{\varepsilon}'c_{T'}^{-1/2}\int_{S_{t-1}}^{S_t} u^{-3/2} \, \dd u. \label{eqt:integralBound_L2_condition_case2}
\end{align}
Hence, there is an almost surely finite $c'_{T'}$,
such that we can write the sum of (\ref{eqt:integralBound_L2_condition_case2}) as
\begin{align*}
    \sum_{t=2}^{\infty} \frac{\E\{(\log\lambda_t-\mu_t)^2\mid \mathcal{F}_{t-1}\}}{S_t^2}
    &\leq \left[ \sum_{t=2}^{T'-1} \frac{\E\{(\log\lambda_t-\mu_t)^2\mid \mathcal{F}_{t-1}\}}{S_t^2} \right] + \left[ \sum_{t=T'}^{\infty} \frac{\E\{(\log\lambda_t-\mu_t)^2\mid \mathcal{F}_{t-1}\}}{S_t^2} \right]\\
    &\leq c'_{T'} +C_{\varepsilon}c_{T'}^{-1/2}\sum_{t=T'}^{\infty}\int_{S_{t-1}}^{S_t} u^{-3/2}du\\
    &= c'_{T'} +C_{\varepsilon}c_{T'}^{-1/2} \int_{S_{T'-1}}^{\infty} u^{-3/2}du\\
    &= c'_{T'} + 2C_{\varepsilon}c_{T'}^{-1/2} S_{T'-1}^{-1/2} \\&<\infty \qquad \text{almost surely}.
\end{align*}
By the martingale strong law of large numbers (see Theorem 2.18 of \cite{hallHeyde1980}),
\begin{align}\label{eqt:martingaleTerm/S_case2}
    \frac{1}{S_n}\sum_{t=2}^n (\log\lambda_t-\mu_t) \inAS 0.
\end{align}
Hence, there is an almost surely finite $N\geq n$ so that
for all $n\geq N$
\begin{align}\label{eqt:bound_forL2condition_smallVersion}
    \sum_{t=2}^n (\log\lambda_t-\mu_t) \geq \frac{-S_n}{8} .
\end{align}
By (\ref{eqt:bound_mut_case2}) and (\ref{eqt:bound_forL2condition_smallVersion}),
we have, for $n\geq N$,
\begin{align}
    \log \Lambda_n
    &\geq \left(\sum_{t=2}^{T'-1}(\mu_t- D_{t-1}/4) \right)
            +\frac{1}{4}\sum_{t=2}^n D_{t-1} - \frac{S_n}{8} \nonumber\\
    &= \left(\sum_{t=2}^{T'-1}(\mu_t- D_{t-1}/4) \right)
        +\frac{S_n}{8}
    \gg (\log n)^2
    \rightarrow \infty\qquad \text{almost surely}. \label{eqt:logLambda_case2_limit}
\end{align}
Thus, the desired result follows in case 2.

\subsubsection{Limit of $\E(\Lambda_n)$}\label{sec:case12_Epower}
In both cases, $\Lambda_n \inAS \infty$.
For any fixed $K$, we have $\min(\Lambda_n, K) \inAS K$.
By Fatou's lemma,  $\liminf_{n\rightarrow\infty} \E(\Lambda_n) \geq K$.
Since $K$ is arbitrary, we have $\E(\Lambda_n)\rightarrow\infty$.

\subsection{Proof of Proposition~\ref{prop:power_ratio_with_oracle}}
For notational simplicity,
throughout this proof we write \(D_t\), \(\widehat{D}_t\), \(\widehat{D}_t^{\star}\), \(\Lambda_n\), and $\Lambda^{\star}_n$ instead of
\(D_t^{\md}\), \(\widehat{D}_t^{\md}\), \(\widehat{D}_t^{\star\md}\), \(\Lambda_n^{\md}\) and \(\Lambda_n^{\star\md}\).

\subsubsection{Case 1: $p_{\min}>0$}
The proof for case 1 is similar to the proof of Theorem~\ref{thm:maxSpacing_strategy},
as explained below.
From (\ref{eqt:order_D_Dhat}), we have
\[
    \widehat{D}_{t-1}^{\star}
    = \E(I_t\mid \mathcal{F}_{t-1})
    = \log t /t + o_{\AS}(\log t/t).
\]
Upon replacing $\widehat{D}_{t-1} = \log t/t$
by $\widehat{D}_{t-1}^{\star}$ in
the proof of Theorem \ref{thm:maxSpacing_strategy},
we obtain analogously that
$\log\Lambda_n^{\star} /L_n \inAS c-1-\log c$.
Therefore, in view of (\ref{eqt:consistency_result_case1_proof}),
we obtain
\[
    \frac{\log\Lambda_n^{\star}}{\log\Lambda_n}
    \inAS 1.
\]

\subsubsection{Case 2: $p_{\min}=0$}
Note that $\log \lambda_t^{\star} = I_t A_{t-1}^{\star} + (1-I_t) B_{t-1}^{\star}$, where
$A_{t-1}^{\star} = \log\{\widehat{D}^{\star}_{t-1}/D_{t-1}\}$
and
$B_{t-1}^{\star} = \log\{(1-\widehat{D}^{\star}_{t-1})/(1-D_{t-1})\}$.
Similar to (\ref{eqt:mu_definition}),
denote the drift of $\log\lambda_t^{\star}$ by
$\mu_t^{\star} := \E(\log\lambda_t^{\star}\mid\mathcal{F}_{t-1}) = \widehat{D}_{t-1}^{\star} A_{t-1}^{\star} + (1-\widehat{D}_{t-1}^{\star}) B_{t-1}^{\star}$.
So, we can decompose the difference
$\log \Lambda_n - \log \Lambda_n$ as
\begin{align}
    \log \Lambda_n^{\star} - \log \Lambda_n
        &= \left\{ \sum_{t=2}^n (\mu_t^{\star}-\mu_t) \right\}
            +
            \left\{
            \sum_{t=2}^n (\log\lambda_t^{\star}-\mu_t^{\star})
            \right\}
            -
            \left\{
            \sum_{t=2}^n (\log\lambda_t-\mu_t)
            \right\}\nonumber\\
        &= \left\{ \sum_{t=2}^n (\mu_t^{\star}-\mu_t) \right\}
            +
            \left\{
            \sum_{t=2}^n (\log\lambda_t^{\star}-\mu_t^{\star})
            \right\}+ o_{\AS}(S_n),
            \label{eqt:decompose_difference_logLambda}
\end{align}
where the last line uses the result in (\ref{eqt:martingaleTerm/S_case2}).
Next, we are going to prove that the first two terms in (\ref{eqt:decompose_difference_logLambda})
are also $o_{\AS}(S_n)$.
\begin{itemize}
    \item
For the first term in (\ref{eqt:decompose_difference_logLambda}),
by Gibbs' inequality, we know that
\begin{align}
    \mu_t^{\star}-\mu_t
        &= \underbrace{\widehat{D}_{t-1}^{\star}\log\frac{\widehat{D}_{t-1}^{\star}}{\log t/t}}_{Y_t'}
         + \underbrace{(1-\widehat{D}_{t-1}^{\star})\log\frac{1-\widehat{D}_{t-1}^{\star}}{1-\log t/t}}_{Y_t''}
         \geq 0. \label{eqt:diff_mu_case2}
\end{align}
Now, we find upper bounds for $Y_t'$ and $Y_t''$.
By Condition~\ref{cond:simple_strategy},
we know that
\begin{align}\label{eqt:bound_for_3_D_case2}
    \widehat{D}_{t-1}^{\star} = o_{\AS}(D_{t-1}),\qquad
    \log t /t = o_{\AS}(D_{t-1}).
\end{align}
So, eventually,
\begin{align}\label{eqt:bound_3D}
    \widehat{D}_{t-1}^{\star} \leq D_{t-1}, \qquad
    \widehat{D}_{t-1}^{\star} \leq 1/2 , \qquad
    \log t/t \leq 1/2.
\end{align}
Using the first inequality in (\ref{eqt:bound_3D}) and Condition~\ref{cond:simple_strategy}, we eventually have that
\begin{align}\label{eqt:Y'}
    Y_t' \leq D_{t-1}\times o_{a.s.}\left(\frac{1}{\log\frac{D_{t-1}}{\log t/t}}\right) \log\frac{D_{t-1}}{\log t/t}
    = o_{\AS}(D_{t-1}).
\end{align}
Using $\widehat{D}_{t-1}^{\star}\in[0,1]$, triangle inequality, and
$|\log(1-x)| \leq 2x$ for $0\leq x\leq 1/2$ one by one,
we eventually have that
\begin{align}
    Y_t''
    &\leq \left|\log\frac{1-\widehat{D}_{t-1}^{\star}}{1-\log t/t}\right|
    \leq |\log(1-\widehat{D}_{t-1}^{\star})| + |\log(1-\log t/t)|
    \leq 2\widehat{D}_{t-1}^{\star} + 2 \log t/t \nonumber\\
    &= o_{\AS}(D_{t-1}),\label{eqt:Y''}
\end{align}
where the last equality follows from (\ref{eqt:bound_for_3_D_case2}).
So, (\ref{eqt:diff_mu_case2}), (\ref{eqt:Y'}), and (\ref{eqt:Y''})
imply that
$\mu_t^{\star}-\mu_t = o_{\AS}(D_{t-1})$.
By the convergence property of Ces\`{a}ro means, we have
\begin{align}
    \sum_{t=2}^n (\mu_t^{\star}-\mu_t) = o_{\AS}(S_n). \label{eqt:diff_mu_case2_result}
\end{align}
\item
Next, we handle the second term in (\ref{eqt:decompose_difference_logLambda}).
Since $\log \{\widehat{D}_{t-1}^{\star}/D_{t-1}\} \inAS -\infty$,
we know that
there is an almost surely finite $T$ such that for $t\geq T$,
\[
    |A_{t-1}^{\star}-B_{t-1}^{\star}|
        \leq \left| \log\frac{\widehat{D}_{t-1}^{\star}}{D_{t-1}} \right| +
            \left| \log\frac{1-\widehat{D}_{t-1}^{\star}}{1-D_{t-1}} \right|
        \leq - \log\frac{\widehat{D}_{t-1}^{\star}}{D_{t-1}}  + C_T'
        \leq C_T''\left( 1- \log\frac{\widehat{D}_{t-1}^{\star}}{D_{t-1}}\right),
\]
where $C_T',C_T''>0$ are almost surely finite.
So, for $t\geq T$
\begin{align}
    \E\{ (\log\lambda_t^{\star}-\mu_t^{\star})^2 \mid \mathcal{F}_{t-1}\}
        &= (A_{t-1}^{\star}-B_{t-1}^{\star})^2 \widehat{D}_{t-1}^{\star} (1-\widehat{D}_{t-1}^{\star})  \nonumber\\
        &\leq (C_T'')^2 \left\{\frac{\widehat{D}_{t-1}^{\star}}{D_{t-1}}\left( 1- \log\frac{\widehat{D}_{t-1}^{\star}}{D_{t-1}} \right)^2 \right\} \times D_{t-1}\times 1 \nonumber \\
        &\leq C_T D_{t-1},
\end{align}
where the last line follows from the fact that $\sup_{0<u\leq 1} u(1-\log u)^2 <\infty$, and
$C_T>0$ is almost surely finite.
So
\begin{align*}
    \sum_{t=2}^{\infty} \frac{\E\{ (\log\lambda_t^{\star}-\mu_t^{\star})^2 \mid \mathcal{F}_{t-1}\}}{S_t^2}
    &\lesssim C_T' + C_T\sum_{t=T}^{\infty} \frac{D_{t-1}}{S_t^2}
    = C_T' + C_T\sum_{t=T}^{\infty} \frac{S_{t}-S_{t-1}}{S_t^2} \\
    &\leq C_T' + C_T\sum_{t=T}^{\infty} \int_{S_{t-1}}^{S_t} u^{-2} \, \dd u
    = C_T' + C_T\int_{S_{T-1}}^{\infty} u^{-2} \, \dd u \\
    &<\infty \qquad \text{almost surely},
\end{align*}
where $C_T'>0$ is almost surely finite.
By the martingale strong law of large numbers (see Theorem 2.18 of \cite{hallHeyde1980}), we have
\begin{align}\label{eqt:MD_term_case2}
    \sum_{t=2}^n (\log\lambda_t^{\star}-\mu_t^{\star})  = o_{\AS}(S_n).
\end{align}
\end{itemize}

Putting (\ref{eqt:diff_mu_case2_result}) and (\ref{eqt:MD_term_case2}) into (\ref{eqt:decompose_difference_logLambda}),
we have $\log \Lambda_n^{\star} - \log \Lambda_n = o_{\AS}(S_n)$.
From (\ref{eqt:logLambda_case2_limit}), we know that $S_n/\log \Lambda_n =O_{\AS}(1)$.
So, $\log \Lambda_n^{\star} / \log \Lambda_n \inAS 1$.

\subsection{Proof of Theorem \ref{thm:sum_spacings_strategy}}\label{proof:thm_45}
For notational simplicity,
throughout this proof we write \(D_t\), \(\widehat{D}_t\), and \(\Lambda_n\)
instead of \(D_t^{(m,k)}\), \(\widehat{D}_t^{(m,k)}\), and \(\Lambda_n^{(m,k)}\).
By definition, we can expand $\log \Lambda_n /n$ as:
\begin{align*}
	\frac{1}{n}\log \Lambda_n
	= \frac{1}{n}\sum_{t=2}^n \left\{ I_t A_{t-1} + \left(1-I_t\right) B_{t-1} \right\},
    \quad\text{where} \quad
    A_{t-1} := \log \frac{\widehat{D}_{t-1}}{D_{t-1}} ,\quad
    B_{t-1} := \log \frac{1-\widehat{D}_{t-1}}{1-D_{t-1}}  ,
\end{align*}
and $I_t$ is defined in (\ref{eqt:aggregate_I_mathcalI_mathcalU}).
Note that
$D_{t-1} \inAS \pi_k \in (1/k, 1)$ according to Lemma~\ref{lem:sum_spacings_strategy},
and $\widehat{D}_{t-1} \rightarrow 1/k$.
Because both limits are bounded away from $0$ and $1$, we also have
\begin{equation}\label{eq:L1_L0}
    A_{t-1} \inAS \log \frac{1/k}{\pi_k} =: A
    \qquad \text{and} \qquad
    B_{t-1} \inAS \log \frac{1-1/k}{1-\pi_k} =: B.
\end{equation}
Let $M_t = I_t - \E_\pr(I_t \mid \mathcal{F}_{t-1})$.
Since $|M_t|\leq 1$, $\E(M_t\mathcal{F}_{t-1})=0$, and $M_t$ is $\mathcal{F}_{t}$ measurable,
we know that $\{M_t\}_{t=2}^n$ is a bounded martingale difference sequence
with respect to $\mathcal{F}_{t}$.
By the strong law of large numbers for martingale differences,
we have
\begin{align}\label{eqt:asConv_Mt}
    \frac{1}{n} \sum_{t=2}^n M_t \inAS 0.
\end{align}
Since $\E_\pr(I_t \mid \mathcal{F}_{t-1})= \pr(U_t \in \mathcal{U}_{t-1} \mid \mathcal{F}_{t-1}) \inAS 1/k$ according to Lemma~\ref{lem:sum_spacings_strategy},
it follows by Ces\`{a}ro summability that
\begin{align}\label{eqt:asConv_Mt_part2}
    \sum_{t=2}^n \E_\pr(I_t \mid \mathcal{F}_{t-1}) /n\inAS 1/k.
\end{align}
Hence, (\ref{eqt:asConv_Mt}) and (\ref{eqt:asConv_Mt_part2}) imply that
\begin{align}\label{eqt:conv_of_avgI_avg1-I}
    \frac{1}{n} \sum_{t=2}^n I_t \inAS 1/k
    \qquad \text{and} \qquad
    \frac{1}{n} \sum_{t=2}^n (1-I_t) \inAS 1-1/k.
\end{align}
Then, using (\ref{eqt:conv_of_avgI_avg1-I}) and the Ces\`{a}ro means of $I_t(A_{t-1}-A)\inAS 0$,
we have
\begin{align}\label{eqt:decompose_sumIA}
    \frac{1}{n}\sum_{t=2}^n I_t A_{t-1}
    = \left( \frac{1}{n}\sum_{t=2}^n I_t A\right) +
        \left\{ \frac{1}{n}\sum_{t=2}^n I_t (A_{t-1} - A) \right\}
    \inAS \frac{A}{k} + 0.
\end{align}
Similarly, we have that
\begin{align}\label{eqt:decompose_sum(1-I)B}
    \frac{1}{n}\sum_{t=2}^n (1-I_t) B_{t-1} \inAS (1-1/k) B.
\end{align}
Combining (\ref{eqt:decompose_sumIA}) and (\ref{eqt:decompose_sum(1-I)B}), we have
\begin{align}\label{eq:sumSpacing_e_power}
    \frac{1}{n}\log \Lambda_n \inAS \frac{1}{k} \log \frac{1/k}{\pi_k} + \left(1-\frac{1}{k} \right) \log \frac{1-1/k}{1-\pi_k}
    = \KL(\Bern(1/k)\| \Bern(\pi_k)),
\end{align}
hence, the result in \eqref{eq:k_fraction_power} follows.

\subsection{Proof of Proposition \ref{prop:weitghedSum_validity}}
(i) Since $\{w_{t,k}:k\in\mathcal{K}_{t-1}\}  \in \mathcal{F}_{t-1}$,
we know that $\lambda_t^{(m, \mathcal{K})} \in \mathcal{F}_{t}$.
Inductively, we have $\Lambda_n^{(m, \mathcal{K})} \in \mathcal{F}_{n}$.
Using the fact that $\{\Lambda_t^{(m,k)}\}_{t\geq 1}$ is an $e$-process for each $k\in\mathcal{K}_{\infty}$,
we have
\begin{align*}
    \E_{\pr_0}\left( \Lambda_n^{(m, \mathcal{K})} \mid \mathcal{F}_{n-1}\right)
    &= \Lambda_{n-1}^{(m, \mathcal{K})}  \E_{\pr_0}\left(
    \sum_{k\in\mathcal{K}_{n-1}} w_{n,k} \lambda_t^{(m, k)} \mid \mathcal{F}_{n-1}\right) \\
    &=\Lambda_{n-1}^{(m, \mathcal{K})}
    \sum_{k\in\mathcal{K}_{n-1}} w_{n,k} \E_{\pr_0}\left( \lambda_t^{(m, k)} \mid \mathcal{F}_{n-1}\right) \\
    &=\Lambda_{n-1}^{(m, \mathcal{K})}
    \sum_{k\in\mathcal{K}_{n-1}} w_{n,k}  \\
    &=\Lambda_{n-1}^{(m, \mathcal{K})} ,
\end{align*}
where the last line follows from the constraint $\sum_{k\in\mathcal{K}_{n-1}} w_{n,k}=1$.
Similar to Section \ref{sec:proof_prop:eprocess},
since $\Lambda_1^{(m, \mathcal{K})} =\lambda_1^{(m, \mathcal{K})}=1$, then $\{\Lambda_n^{(m,\mathcal{K})}\}_{n\geq 0}$ is
a test martingale, and thus it is
an $e$-process under simple null $\pr_0$.
(ii) Since $\lambda_t^{(m, \mathcal{K})} =\sum_{t\in\mathcal{K}_{t-1}}w_{t, k} \lambda_t^{(m, k)}\geq w_{t, \widehat{k}_{t-1}} \lambda_t^{(m, \widehat{k}_{t-1})}$,
we have
\begin{align*}
    \frac{1}{n}\log \Lambda_n^{(m,\mathcal{K})}
        &\geq \frac{1}{n}\sum_{t=1}^n \log\left\{ w_{t, \widehat{k}_{t-1}} \lambda_t^{(m, \widehat{k}_{t-1})} \right\} \\
        &= \left( \frac{1}{n}\sum_{t=1}^n \log w_{t, \widehat{k}_{t-1}}   \right)
        +\left\{ \frac{1}{n}\sum_{t=1}^n \log \lambda_t^{(m, \widehat{k}_{t-1})} \right\} .
\end{align*}
Taking $\limsup$ on both sides, we obtain
\[
    \liminf_{n\rightarrow\infty}  \frac{1}{n}\log \Lambda_n^{(m,\mathcal{K})} \geq 0+ \KL_{k^{\circ}}.
\]
Thus, the desired result follows.

\subsection{Proof of Theorem \ref{thm:divergent_mathcalK_optimal_softmax}}
\subsubsection{Argmax form}
Consider the argmax form.
We verify the conditions of
Proposition~\ref{prop:weitghedSum_validity} with
$k^{\circ}=k^{\star}$, and set
$\widehat{k}_{t-1} = \widehat{K}_{t-1}$.
Therefore,
$\widehat{k}_{t-1}\in\mathcal{K}_{t-1}$ and
$w_{t,\widehat{k}_{t-1}}=1$.

\noindent\underline{\bf Step 1: Show that $\widehat{k}_{t-1}\inAS k^{\star}$ for $\widehat{k}_{t-1}=\widehat{K}_{t-1}$}\\
We prove it by contradiction.
Fix any $\varepsilon>0$.
Assume $|\widehat{k}_{t-1}-k^{\star}| >\varepsilon$.
By definition of the maximizer $\widehat{k}_{t-1} = \widehat{K}_{t-1}$, and the definitions of $\eta_{t-1}$ and $\Delta_{\varepsilon}$,
we have
\[
    \frac{\log\Lambda_{t-1}^{(m,\widehat{k}_{t-1})}}{t-1}
    \leq \KL_{\widehat{k}_{t-1}} + \eta_{t-1}
    \leq \KL_{k^{\star}} - \Delta_{\varepsilon} + \eta_{t-1}
    < \KL_{k^{\star}} -  \eta_{t-1}
    \leq \frac{\log\Lambda_{t-1}^{(m,k^{\star})}}{t-1},
\]
which shows that
such $\widehat{k}_{t-1}$ will not be selected at time $t-1$.
It is a contradiction.
Hence, $|\widehat{k}_{t-1}-k^{\star}| \leq \varepsilon$
for all $t\geq T$.
Since $\varepsilon>0$ is arbitrary,
we have $\widehat{k}_{t-1} \inAS k^{\star}$.

\noindent\underline{\bf Step 2: Convergence of the $\sum_{t=1}^n\log\lambda_t^{(m, \widehat{k}_{t-1})}$}\\
Now we consider the path $\widehat{k}_{t-1}$.
By the assumption for the argmax form, we know that
\begin{align}\label{eqt:argmax_kpathKL}
    \left\vert\frac{1}{n}\sum_{t=1}^n \left\{ \log \lambda_t^{(m,\widehat{k}_{t-1})} - \log \lambda_t^{(m, k^{\star})}\right\}  \right\vert
    &\leq O_{\AS}\left(\frac1n\right)+ \frac{1}{n}\sum_{t=1}^n
    \sup_{\substack{k\in\mathcal{K}_{t-1}\\ |k-k^{\star}|\leq\varepsilon}}
    \left| \log \lambda_t^{(m,k)} - \log \lambda_t^{(m, k^{\star})}\right|
    \rightarrow 0 ,
\end{align}
which follows by passing $n\rightarrow\infty$ and then $\varepsilon\downarrow 0$.
Using (\ref{eq:k_fraction_power}) and $k^{\star}\in\mathcal{K}_{\infty}$, we know that
\begin{align}\label{eqt:argmax_kstarKL}
    \frac{1}{n} \sum_{t=1}^n \log \lambda_t^{(m,k^{\star})}
    \inAS \KL_{k^{\star}}.
\end{align}
Combining (\ref{eqt:argmax_kpathKL}) and (\ref{eqt:argmax_kstarKL}),
we have
\begin{align}\label{eqt:argmax_kpathKL_final}
    \frac{1}{n}\sum_{t=1}^n \log \lambda_t^{(m,\widehat{t}_{k-1})}
    \inAS \KL_{k^{\star}}.
\end{align}

\noindent\underline{\bf Step 3: Convergence of $\sum_{t=1}^n\log\lambda_t^{(m, \mathcal{K})}$}\\
Since
$w_{t,\widehat{k}_{t-1}} = 1$,
we have
\begin{align}
    \frac{1}{n}\log \Lambda_n^{(m, \mathcal{{K}})}
    &=  \left\{\frac{1}{n}\sum_{t=1}^{T-1}\log\lambda_t^{(m, \mathcal{K})} \right\}
        + \left\{ \frac{1}{n}\sum_{t=T}^n\log\lambda_t^{(m, \mathcal{K})} \right\} \nonumber\\
    &\geq \left\{\frac{1}{n}\sum_{t=1}^{T-1}\log\lambda_t^{(m, \mathcal{K})} \right\}
        + \left\{ \frac{1}{n}\sum_{t=T}^n \log
         \lambda_t^{(m, \widehat{k}_{t-1})} \right\} \label{eqt:decompose_logLambda_argmax}\\
    &\rightarrow 0 + \KL_{k^{\star}} , \nonumber
\end{align}
where the last line follows from
$T<\infty$ almost surely and (\ref{eqt:argmax_kpathKL_final}).

\subsubsection{Softmax form}
Consider the softmax form.

\noindent\underline{\bf Step 1: Decomposing $-\log w_{t,k^{\star}}$ into $G_{t,\varepsilon}$ and $G_{t,\varepsilon}'$}\\
Fix any $\varepsilon>0$.
We decompose $-\log w_{t,k^{\star}}$. By the elementary inequality $\log(1+G+G') \leq \log(1+G) + \log(1+G')$ for $G,G'\geq 0$, we have
\begin{align}\label{eqt:decompose_wtk_softmax}
    -\log w_{t,k^{\star}}
        \le -\log \frac{1}{1+ G_{t, \varepsilon} +  G_{t, \varepsilon}'}
        \leq \log(1+ G_{t, \varepsilon}) + \log(1+ G_{t, \varepsilon}'),
\end{align}
where
\[
     G_{t, \varepsilon} = \sum_{\substack{k\in\mathcal{K}_{t-1}\\ 0<|k-k^{\star}|\leq\varepsilon}} \frac{w_{t,k}}{w_{t,k^{\star}}}
     \qquad \text{and} \qquad
     G_{t, \varepsilon}' = \sum_{\substack{k\in\mathcal{K}_{t-1}\\ |k-k^{\star}|>\varepsilon}} \frac{w_{t,k}}{w_{t,k^{\star}}},
\]

\noindent\underline{\bf Step 2: Bound $G_{t,\varepsilon}'$}\\
There is an almost surely finite $T$
such that for all $t\geq T$,
$\eta_{t-1} < \Delta_{\varepsilon}/4$.
For any $k\in\mathcal{K}_{t-1}$ with $|k-k^{\star}| > \varepsilon$,
the definition of $\Delta_{\varepsilon}$ in (\ref{eqt:condition_validy_weightedVer_softmax}) implies that
\begin{align}\label{eqt:softmax_bound_diff_logLambda}
    \frac{\log \Lambda_{t-1}^{(m,k)}}{t-1} - \frac{\log \Lambda_{t-1}^{(m,k^{\star})}}{t-1}
        \leq (\KL_k + \eta_{t-1})- (\KL_{k^{\star}} - \eta_{t-1})
        \leq -\Delta_{\varepsilon} + 2 \eta_{t-1}
        \leq -\Delta_{\varepsilon}/2.
\end{align}
Then, it follows that $\Lambda_{t-1}^{(m,k)} / \Lambda_{t-1}^{(m,k^*)}= \exp\left(\log \Lambda_{t-1}^{(m,k)} - \log \Lambda_{t-1}^{(m,k^*)}\right)\le \exp(-\Delta_\varepsilon(t-1)/2)\le 1/2$ eventually. Also, by (\ref{eq:k_fraction_power}), $\Lambda_{t-1}^{(m,k^{\star})}\ge \exp((t-1)\KL_{k^*}/2)$ eventually.
According to the definition of softmax weights in (\ref{eqt:argmax_softmax}), for each $k$, we get that
$$
 \frac{w_{t,k}}{w_{t,k^\star}}
 =\exp\!\left[\gamma\left\{
        \Lambda_{t-1}^{(m,k)}-\Lambda_{t-1}^{(m,k^\star)}
                        \right\}\right]
 \leq \exp\!\left\{-\tfrac{\gamma}{2}\Lambda_{t-1}^{(m,k^\star)}\right\}
 \leq \exp\!\left\{-\tfrac{\gamma}{2}
              \exp\{(t-1)\mathrm{KL}_{k^\star}/2\}\right\}.
$$
Then,
$$
\log(1+G_{t, \varepsilon}') \le G_{t, \varepsilon}'
\le |\mathcal{K}_{t-1}| \exp\!\left\{-\tfrac{\gamma}{2}
              \exp\{(t-1)\mathrm{KL}_{k^\star}/2\}\right\}\inAS 0.
$$
By Ces\`{a}ro summability, we also have that
\begin{align}\label{eqt:boundG'_softmax}
    0\leq \frac{1}{n}\sum_{t=1}^{n} \log(1+G_{t, \varepsilon}')
    \inAS 0.
\end{align}

\noindent\underline{\bf Step 3: Convergence of $w_{t,k^{\star}}$}\\
Letting $n\rightarrow\infty$ in (\ref{eqt:decompose_wtk_softmax}) and using (\ref{eqt:boundG'_softmax}),
we have
\[
    \limsup_{n\rightarrow\infty} \frac{1}{n}\sum_{t=1}^n (-\log w_{t,k^{\star}})
        \leq o_{\AS}(1) + \limsup_{n\rightarrow\infty} \frac{1}{n}\sum_{t=1}^n \log(1+G_{t,\varepsilon}).
\]
Since $\varepsilon>0$ is arbitrary, we let $\varepsilon\downarrow 0$ and use the second condition in (\ref{eqt:softmax_cond}) to obtain
\begin{align}\label{eqt:convergence_logw_softmax}
    \frac{1}{n}\sum_{t=1}^n (-\log w_{t,k^{\star}}) \inAS 0.
\end{align}

\noindent\underline{\bf Step 4: Convergence of $\Lambda_n^{(m, \mathcal{K})}$}\\
In particular,
Since $k^{\star}\in\mathcal{K}_{\infty}$,
there is an almost surely finite $T''\ge T$ such that $k^{\star}\in\mathcal{K}_{t-1}$ for all $t\geq T''$.
So, we can take $\widehat{k}_{t-1} = k^{\star} \in\mathcal{K}_{t-1}$ for $t\geq T''$, and
choose any $\widehat{k}_{t-1} = \min \mathcal{K}_{t-1}$ for $t< T''$.
Following similar steps in (\ref{eqt:decompose_logLambda_argmax}) and
using (\ref{eqt:convergence_logw_softmax}) and (\ref{eq:k_fraction_power}), we have
\begin{align}
    \frac{1}{n} \log \Lambda_n^{(m, \mathcal{K})}
        &\geq \left\{\frac{1}{n}\sum_{t=1}^{T''-1}\log\lambda_t^{(m, \mathcal{K})} \right\}
        + \left\{ \frac{1}{n}\sum_{t=T''}^n \log
        w_{t, \widehat{k}_{t-1}} \lambda_t^{(m, \widehat{k}_{t-1})} \right\} \\
        &=\left\{\frac{1}{n}\sum_{t=1}^{T''-1}\log\lambda_t^{(m, \mathcal{K})} \right\}
        + \left\{ \frac{1}{n}\sum_{t=T''}^n \log w_{t, k^{\star}}  \right\}
        + \left\{ \frac{1}{n}\sum_{t=T''}^n \log
        \lambda_t^{(m, k^{\star})} \right\} \\
        &\inAS 0 + 0 + \KL_{k^{\star}}.
\end{align}
This completes the proof of the theorem.

\subsection{Proof of Proposition~\ref{prop:average_rule}}
\subsubsection{Proof of the upper bound of $\E_{\pr}\left(\log\lambda_t^{(m,\mathcal{K})}\mid \mathcal{F}_{t-1}\right)$}
We first prove that
\begin{equation*}
    \limsup_{t\to\infty} \E_{\pr}\left(\log\lambda_t^{(m,\mathcal{K})}\mid \mathcal{F}_{t-1}\right) \le0 \quad \text{almost surely} .
\end{equation*}
Define
\[
    \omega_p(x)=\sup_{E : \int_Ep(u) \dd u \le x}\Leb(E).
\]
Since $\Leb\{u:p(u)=0\}=0$,
we know that $\lim_{x\downarrow0}\omega_p(x) = 0$.
Recall that, by definition,
$$
\widehat{D}_{t-1}^{\agg \star}=\int_{\mathcal U_{t-1}^{(m,k)}}p(u)\,\dd u,
\qquad
\widehat{D}_{t-1}^{(m,k)}=\frac{\lfloor t/k\rfloor}{t}.
$$
By the Dvoretzky--Kiefer--Wolfowitz inequality, we have that
\begin{align}\label{eqt:_uniformBound}
    b_t := \max_{0\le j\le t} \left|F(U_{j,t}) - \frac{j}{t} \right|
        = O_{\AS}\left\{  \left( \frac{\log t}{t} \right)^{1/2} \right\}.
\end{align}
Then for a complete spacing block $\mathcal{B}_{i,t} := [U_{im,t-1}, U_{(i+1)m,t-1}]$, $\pr(\mathcal{B}_{i,t})\le m/t + 2b_t$. Furthermore,
$$
    \widehat{D}_{t-1}^{\agg \star}
    \le \left(\frac{m}{t} + 2b_t\right) \left\lfloor \frac{t}{mk} \right\rfloor\le \frac1k + O_{\AS}(t^{1/2-\phi}\sqrt{\log t}).
$$
There exists an almost surely finite $T$
such that $\widehat{D}_{t-1}^{\agg \star}\le 2/k$ for $t\ge T$.
And also $\widehat{D}_{t-1}^{(m,k)}\le 1/k$ by definition.
So, $D_{t-1}^{(m,k)}\le\omega_p(\widehat{D}_{t-1}^{\agg \star})\le\omega_p(2/k)$ and
\begin{equation}\label{eq:bound_D_y}
    \left|D_{t-1}^{(m,k)}-\widehat{D}_{t-1}^{(m,k)}\right|\le D_{t-1}^{(m,k)}+\widehat{D}_{t-1}^{(m,k)}\le\omega_p\left(\frac2k\right)+\frac1k.
\end{equation}
By the definition of $\lambda_t^{(m,k)}$,
\[
\begin{aligned}
\E_{\pr}\left(\left|\lambda_t^{(m,k)}-1\right|\mid \mathcal{F}_{t-1}\right)
=\int_{\mathcal U_{t-1}^{(m,k)}}
\left|\frac{\widehat{D}_{t-1}^{(m,k)}}{D_{t-1}^{(m,k)}}-1\right|p(u)\,\dd u
+
\int_{[0,1]\setminus\mathcal U_{t-1}^{(m,k)}}
\left|\frac{1-\widehat{D}_{t-1}^{(m,k)}}{1-D_{t-1}^{(m,k)}}-1\right|p(u)\,\dd u.
\end{aligned}
\]
Set \(p_{\max}:= \max_{u\in[0,1]}p(u)\), we obtain
\[
\begin{aligned}
\E_{\pr}\left(\left|\lambda_t^{(m,k)}-1\right|\mid \mathcal{F}_{t-1}\right)
&\le
p_{\max} D_{t-1}^{(m,k)}\left|\frac{\widehat{D}_{t-1}^{(m,k)}}{D_{t-1}^{(m,k)}}-1\right|
+
p_{\max}(1-D_{t-1}^{(m,k)})\left|\frac{1-\widehat{D}_{t-1}^{(m,k)}}{1-D_{t-1}^{(m,k)}}-1\right|\\
&=
p_{\max}|\widehat{D}_{t-1}^{(m,k)}-D_{t-1}^{(m,k)}|+p_{\max}|D_{t-1}^{(m,k)}-\widehat{D}_{t-1}^{(m,k)}|
=2p_{\max}|D_{t-1}^{(m,k)}-\widehat{D}_{t-1}^{(m,k)}|.
\end{aligned}
\]
Using the fact that \(\log x\le|x-1|\) and the triangle inequality, we obtain

\begin{align}
\E_{\pr}\left(\log\lambda_t^{(m,\mathcal{K})}\mid \mathcal{F}_{t-1}\right)
&\le\E_{\pr}\left(\left|\lambda_t^{(m,\mathcal{K})}-1\right|\mid \mathcal{F}_{t-1}\right)
\le
\frac1{|\mathcal K_{t-1}|}
\sum_{k\in\mathcal K_{t-1}}
\E_{\pr}\left(\left|\lambda_t^{(m,k)}-1\right|\mid \mathcal{F}_{t-1}\right) \nonumber \\
&\le
\frac{2p_{\max}}{|\mathcal K_{t-1}|}
\sum_{\substack{k\in\mathcal K_{t-1}\\k\le M_t}}
\left|D_{t-1}^{(m,k)}-\widehat{D}_{t-1}^{(m,k)}\right|,  \label{eqt:Eloglambda_avgRule}
\end{align}
where $M_t:=\min\{1+\Delta_t|\mathcal{K}_{t-1}|, t/m\}$.
Note that it makes sense because, by (\ref{cond:Kt}), we know $k\le 1+\Delta_t|\mathcal{K}_{t-1}|$ and $k\le t/m$ if $\lfloor t/(mk)\rfloor\ge1$.
Now fix \(K>1\) and split the summation in (\ref{eqt:Eloglambda_avgRule})
into two cases: \(k<K\) and \(k\ge K\).
We handle these cases one by one:
\begin{itemize}
    \item For \(k<K\), we use \(|D_{t-1}^{(m,k)}-\widehat{D}_{t-1}^{(m,k)}|\le1\).
    Because the grid spacing is \(\Delta_t\), these candidates have total weight at most
$
   (K+1) / ( \Delta_t |\mathcal{K}_{t-1}| ).
$
    \item For \(k\ge K\), combining (\ref{eq:bound_D_y}) and monotonicity of \(\omega_p\), we have
$
    \left|D_{t-1}^{(m,k)}-\widehat{D}_{t-1}^{(m,k)}\right|
    \le\omega_p\left(2/K\right)+1/K.
$
The total weight of these candidates is at most one.
\end{itemize}
Consequently, using these bounds, (\ref{eqt:Eloglambda_avgRule}) can be bounded as follows:
\[
    \E_{\pr}\left(\log\lambda_t^{(m,\mathcal{K})}\mid \mathcal{F}_{t-1}\right)
\le
2p_{\max}\cdot\frac{K+1}{\Delta_t |\mathcal{K}_{t-1}|}
+
2p_{\max}\cdot\left(\omega_p\left(\frac{2}{K}\right)+\frac1K \right).
\]
Since \(\Delta_t |\mathcal{K}_{t-1}|\asymp t^\nu\to\infty\) by assumption,
letting \(t\to\infty\) with \(K\) fixed yields
\[
\limsup_{t\to\infty} \E_{\pr}\left(\log\lambda_t^{(m,\mathcal{K})}\mid \mathcal{F}_{t-1}\right)
\le
2p_{\max}\left\{\omega_p\left(\frac2K\right)+\frac1K\right\}
\quad \text{almost surely}.
\]
Since $K$ is arbitrary, we let \(K\to\infty\) to obtain
\begin{equation}\label{bound:upper_expect}
    \limsup_{t\to\infty} \E_{\pr}\left(\log\lambda_t^{(m,\mathcal{K})}\mid \mathcal{F}_{t-1}\right) \le0 \quad \text{almost surely},
\end{equation}
where we have used the result \(\lim_{x\downarrow0}\omega_p(x)=0\).

\subsubsection{Proof of the lower bound of $\E_{\pr}\left(\log\lambda_t^{(m,\mathcal{K})}\mid \mathcal{F}_{t-1}\right)$}
It remains to prove
\[
    \liminf_{t\to\infty} \E_{\pr}\left(\log\lambda_t^{(m,\mathcal{K})}\mid \mathcal{F}_{t-1}\right)\ge0 \quad \text{almost surely}.
\]
Since \(\KL\{ \Bern(\widehat{D}_{t-1}^{\agg \star})\| \Bern(D_{t-1}^{(m,k)}) \}\ge0\),
we have
\begin{align}
    &\E_{\pr}\left(\log\lambda_t^{(m,\mathcal{K})}\mid \mathcal{F}_{t-1}\right) \nonumber\\
    &\qquad\ge
    \frac{1}{|\mathcal{K}_{t-1}|}
    \sum_{\substack{k\in\mathcal K_{t-1}\\k\le M_t}}
    \left[
        \KL\{\Bern(\widehat{D}_{t-1}^{\agg \star})\| \Bern(D_{t-1}^{(m,k)})\}
        - \KL\{\Bern(\widehat{D}_{t-1}^{\agg \star})\| \Bern(\widehat{D}_{t-1}^{(m,k)})\}
    \right] \nonumber \\
    &\qquad\ge
    -\frac{1}{|\mathcal{K}_{t-1}|}
    \sum_{\substack{k\in\mathcal K_{t-1}\\k\le M_t}}
    \KL\{\Bern(\widehat{D}_{t-1}^{\agg \star})\| \Bern(\widehat{D}_{t-1}^{(m,k)})\}. \label{eqt:avgRule_lowerBound_step1}
\end{align}
Next, we need the following technical lemma, whose proof is lengthy and
is provided separately in
Section \ref{sec:proof_of_lem:for_proof_in_avg_rule}:
\begin{lemma}\label{lem:for_proof_in_avg_rule}
Under the assumptions in Proposition~\ref{prop:average_rule}, we have
\[
    \Delta_t
\sum_{\substack{k\in\mathcal K_{t-1}\\k\le M_t}}
\KL(\Bern(\widehat{D}_{t-1}^{\agg \star})\| \Bern(\widehat{D}_{t-1}^{(m,k)}))\overset{\AS}{=}O(1).
\]
\end{lemma}
Applying Lemma \ref{lem:for_proof_in_avg_rule} to (\ref{eqt:avgRule_lowerBound_step1}) and using the second condition in (\ref{cond:Kt}),
we have
\begin{align}\label{bound:lower_expect}
\E_{\pr}\left(\log\lambda_t^{(m,\mathcal{K})}\mid \mathcal{F}_{t-1}\right)
&\gtrsim -\frac{1}{|\mathcal{K}_{t-1}|} \frac{1}{\Delta_t}
\asymp t^{-\nu}
\inAS 0.
\end{align}
Combining (\ref{bound:upper_expect}) and (\ref{bound:lower_expect}),
we have
$
\E_{\pr}\left(\log\lambda_t^{(m,\mathcal{K})}\mid \mathcal{F}_{t-1}\right)\inAS 0.
$

\subsubsection{Convergence of $\log\Lambda_n^{(m, \mathcal{K})} /n$}
By the definition of \(\Lambda_n^{(m,\mathcal{K})}\),
\begin{align}
\frac1n\log\Lambda_n^{(m,\mathcal{K})}
=&
\frac1n\sum_{t=1}^n \E_{\pr}\left(\log\lambda_t^{(m,\mathcal{K})}\mid \mathcal{F}_{t-1}\right)
+
\frac1n\sum_{t=1}^n
\left\{\log \lambda_t^{(m,\mathcal{K})} -\E_{\pr}\left(\log\lambda_t^{(m,\mathcal{K})}\mid \mathcal{F}_{t-1}\right)\right\}, \label{eqt:decompose_Lambda_avgForm}
\end{align}
where the first term tends to zero almost surely by Ces\`{a}ro summability. For the second term of (\ref{eqt:decompose_Lambda_avgForm}),
using the martingale convergence theorem and Kronecker's lemma,
we have
\[
\sum_{t=1}^n
\left\{\log \lambda_t^{(m,\mathcal{K})} -\E_{\pr}\left(\log\lambda_t^{(m,\mathcal{K})}\mid \mathcal{F}_{t-1}\right)\right\}
=o_{\AS}(n).
\]
In detail,
since \(p\) is bounded by \(p_{\max}\), the probability that two observations are within distance \(x\) is at most \(2p_{\max}x\). That is, we first compute the conditional probability conditioning on \(U_i=u\). Since \(U_i\) and \(U_j\) are independent, \(U_j\) still has density \(p\). Thus
\[
\pr\bigl(|U_i-U_j|<x\mid U_i=u\bigr)
=\int_{[u-x,u+x]\cap[0,1]}p(v)\,\mathrm dv
\le p_{\max}\cdot 2x.
\]
We then average over \(U_i\) gives
\[
\pr\bigl(|U_i-U_j|<x\bigr)\le 2p_{\max}x.
\]
Also, the probability that an observation is within distance \(x\) of either endpoint is also at most \(2p_{\max}x\). Define $d_{ij}$ as the distance between $U_i$ and $U_j$ with $i\neq j$. A union bound therefore gives
\begin{equation}\label{eq:upper_min_distance}
    \pr\left(\min_{0\le i,j\le t-1}d_{ij}<x\right) \le \binom{t-1}{2}\,2p_{\max}x+(t-1)\,2p_{\max}x
=p_{\max}t(t-1)x
\le 2p_{\max}t^2x.
\end{equation}
For each active factor in (\ref{def:GENe_process_lambda}), both \(\widehat D_{t-1}^{(m,k)}\) and \(1-\widehat D_{t-1}^{(m,k)}\) are at least \(1/t\); $0\le D_{t-1}^{(m,k)}\le 1$. By the definition of $\lambda_t^{(m,k)}$, we have $\lambda_t^{(m,k)}\ge 1/t$. The selected region and its complement each contain at least one elementary spacing, so both \(D_{t-1}^{(m,k)}\) and \(1-D_{t-1}^{(m,k)}\) are at least \(\min_{0\le i,j\le t-1}d_{ij}\); $0\le \widehat D_{t-1}^{(m,k)} \le 1$. The bound holds for each candidate factor \(\lambda_t^{(m,k)}\), so it also holds for their average \(\lambda_t^{(m,\mathcal K)}\). That is,
\[
\frac1t\le\lambda_t^{(m,\mathcal K)}
\le \frac1{\min_{0\le i,j\le t-1}d_{ij}}.
\]
The event $\{-\log(\min_{0\le i,j\le t-1}d_{ij})>x\}$ is exactly the event that the minimum spacing is smaller than \(e^{-x}\). (\ref{eq:upper_min_distance}) implies that \(\pr(-\log(\min_{0\le i,j\le t-1}d_{ij})>x)\le\min\{1,2p_{\max}t^2e^{-x}\}\).

Split the integral $(0,\infty)$ at \(3\log t\). On the first part, bound the probability by \(1\). On the second, use \(2p_{\max}t^2e^{-x}\).
Use the tail-integration identity: for $Y\ge 0$,
$\E_{\pr}[Y^2]=2\int_0^\infty x\,\pr(Y>x)\,\mathrm dx$, we get that
\[
\E_{\pr}\left[
  \left(\log\frac{1}{\min_{0\le i,j\le t-1}d_{ij}}\right)^2
\right]
\le 2\int_0^{3\log t}x\,\mathrm dx
   +4p_{\max}t^2\int_{3\log t}^{\infty}xe^{-x}\,\mathrm dx
=O((\log t)^2).
\]
Consequently, \(\E_{\pr}\left[\left(\log\lambda_t^{(m,\mathcal K)}\right)^2\right]=O\left((\log t)^2\right)\), and
\begin{equation}\label{bound:var}
    \sum_{t=2}^\infty
\frac{\E_{\pr}\left[
\operatorname{Var}_{\pr}\left(\log\lambda_t^{(m,\mathcal K)}\mid\mathcal F_{t-1}\right)\right]}{t^2}
\le
\sum_{t=2}^\infty
\frac{\E_{\pr}\left[\left(\log\lambda_t^{(m,\mathcal K)}\right)^2\right]}{t^2}
<\infty.
\end{equation}
Define
\[
a_t:=\log\lambda_t^{(m,\mathcal K)}
-\E_{\pr}\left(\log\lambda_t^{(m,\mathcal K)}
\mid\mathcal F_{t-1}\right).
\]
Since \(\E_{\pr}(a_t\mid\mathcal F_{t-1})=0\), for \(s<t\), \(a_s\) is \(\mathcal F_{t-1}\)-measurable, so
$\E_{\pr}(a_sa_t)
=\E_{\pr}\!\left[a_s\E_{\pr}(a_t\mid\mathcal F_{t-1})\right]=0$. Thus the cross terms vanish, and then
\[
\E_{\pr}\left[\left(\sum_{t=2}^n\frac{a_t}{t}\right)^2\right]
=\sum_{t=2}^n
\frac{\E_{\pr}\left[
\operatorname{Var}_{\pr}\left(\log\lambda_t^{(m,\mathcal K)}
\mid\mathcal F_{t-1}\right)\right]}{t^2}< \infty,
\]
where the last step follows from (\ref{bound:var}).
Therefore, the \(L^2\) martingale convergence theorem gives almost-sure convergence of \(\sum_{t\ge2}a_t/t\). Kronecker’s lemma then gives
\[
\frac1n\sum_{t=1}^n
\left\{\log\lambda_t^{(m,\mathcal K)}
-\E_{\pr}\!\left(\log\lambda_t^{(m,\mathcal K)}
 \mid\mathcal F_{t-1}\right)\right\}
\inAS 0.
\]
Therefore, we conclude that, if $\Leb(\{u:p(u)=0\})=0$, then
$
n^{-1}\log\Lambda_n^{(m,\mathcal{K})}\inAS 0.
$

\subsection{Proof of Proposition \ref{prop:multi_eprocess}}
The structure of this proof is similar to that of Proposition~\ref{prop:gen_eprocess}.
By definition, $\lambda_1 = 1$. For any time $t \ge 2$, we have
\begin{align}\label{eq:multi_cond_exp}
    \E_{\pr_0} \left( \lambda_t \mid \mathcal{F}_{t-1} \right)
    &= \frac{\widehat{D}_{t-1}}{D_{t-1}} \pr_0 \left(U_t^{[j^{\star}_{t-1}]} \in \mathcal{U}_{t-1}^{[j^{\star}_{t-1}]} \mid \mathcal{F}_{t-1} \right)
    + \frac{1-\widehat{D}_{t-1}}{1-D_{t-1}} \pr_0 \left(U_t^{[j^{\star}_{t-1}]} \notin \mathcal{U}_{t-1}^{[j^{\star}_{t-1}]} \mid \mathcal{F}_{t-1}\right),
\end{align}
where $j^{\star}_{t-1} = \argmax_{j \in \{1, \ldots, d\}} D_{t-1}^{[j]}$ by definition.
Conditioning on $\mathcal{F}_{t-1}$,
both the maximizer $j^{\star}_{t-1}$ and the corresponding interval set $\mathcal{U}_{t-1}^{[j^{\star}_{t-1}]}$ are
deterministic subsets of $[0, 1]$. By Lemma~\ref{lemma:uniform_trans}, under $\mathcal{H}_0$, the components $U_t^{[1]}, \ldots, U_t^{[d]}$ are mutually independent $\Unif[0,1]$ random variables, and the vector $U_t$ is independent of $\mathcal{F}_{t-1}$.

Because $U_t^{[j^{\star}_{t-1}]} \sim \Unif[0,1]$ and is independent of $\mathcal{F}_{t-1}$,
we have
\begin{equation}
    \pr_0 \left(U_t^{[j^{\star}_{t-1}]} \in \mathcal{U}_{t-1}^{[j^{\star}_{t-1}]} \mid \mathcal{F}_{t-1} \right)
    = \textnormal{Leb}\left(\mathcal{U}_{t-1}^{[j^{\star}_{t-1}]}\right)
    = D_{t-1},
\end{equation}
where $\textnormal{Leb}(\cdot)$ denotes
the Lebesgue measure.
Substituting this back into \eqref{eq:multi_cond_exp}, we obtain
\begin{align}
    \E_{\pr_0} \left( \lambda_t \mid \mathcal{F}_{t-1} \right)
    &= \frac{\widehat{D}_{t-1}}{D_{t-1}} D_{t-1}
    + \frac{1-\widehat{D}_{t-1}}{1-D_{t-1}} (1-D_{t-1}) = 1.
\end{align}
Since $\Lambda_n = \prod_{t=1}^n \lambda_t$, this implies that
\begin{equation}
    \E_{\pr_0} \left( \Lambda_n \mid \mathcal{F}_{t-1} \right) = \Lambda_{n-1} \E_{\pr_0} \left( \lambda_n \mid \mathcal{F}_{t-1} \right) = \Lambda_{n-1}.
\end{equation}
By definition, $\Lambda_n$ is constructed from non-negative terms.
It ensures that $\Lambda_n \ge 0$.
Furthermore, $\E_{\pr_0} (\Lambda_n) = \E_{\pr_0}(\Lambda_1) = 1 < \infty$.
Consequently, the stochastic process $\{\Lambda_n\}_{n \ge 1}$ is a test martingale with respect to
$\{\mathcal{F}_n\}_{n \ge 1}$ under $\mathcal{H}_0$.

\subsection{Proof of Theorem \ref{thm:full_general_optimal}}
Note that by the definition of $\lambda_t$ and $\lambda_t^{\star}$, we have
$$
\E_\pr\left(\log \lambda_t \mid \mathcal{F}_{t-1}\right) \le \E_\pr\left(\log \lambda_t^{\star} \mid \mathcal{F}_{t-1}\right)
$$
by following the same procedure as (\ref{align:proof_thm42}) in the proof of Theorem~\ref{thm:general_log_optimal}. Similarly, following the remaining parts of Appendix~\ref{proof:thm42}, the result follows.

\subsection{Proof of Theorem \ref{thm:multi_simple}}
The proof is the same as that of Theorem~\ref{thm:maxSpacing_strategy}.
Recall that we define $\widehat{D}_{t-1}^{(1,\infty)} = \log t / t$ in Theorem~\ref{thm:maxSpacing_strategy}, which is same as here. The difference is that the maximum spacing might occur in different coordinates $j=1,2,\ldots$ at different time $t=1,2,\ldots$.
However, for proving Theorem~\ref{thm:multi_simple}, we do not need to change the proof in Appendix~\ref{proof:thm32} by the corresponding type of definition of $\lambda_t$ and the corresponding Condition~\ref{cond:simple_strategy}.

\subsection{Proof of Theorem \ref{thm:multi_k_spacings}}
The proof is the same as that of Theorem~\ref{thm:sum_spacings_strategy}.
Recall that we define \(\widehat{D}_{t-1}^{(m,k)} := \lfloor t/k \rfloor / t\) in Theorem~\ref{thm:sum_spacings_strategy}, which is same as here. The difference is that the sum of the \(\lfloor t/(mk) \rfloor\) largest $m$th order spacings might occur in different coordinates $j=1,2,\ldots, d$ at different time $t=1,2,\ldots$. According to Lemma~\ref{lem:sum_spacings_strategy}, for each $j\in\{1,\ldots,d\}$ non-uniform transformed marginals, $\mathcal{U}_{t-1}^{(m,k)[j]}\inAS \mathcal{U}^{[j]}  := \{u : p(u)\le p_k^{[j]}\}$. According to Lemma~\ref{lem:Gt_to_1half}, \(\pi_k^{[j]}=1/k\) for uniform transformed marginals. Then the finite number of coordinates and the unique limiting maximizer imply eventual selection of \(j^\star\). It implies that $D_{t-1}^{(m,k)} \inAS D_{t-1}^{(m,k)[j^{\star}]}$.
Following the proof of Theorem~\ref{thm:sum_spacings_strategy} in Appendix~\ref{proof:thm_45},
we can prove Theorem~\ref{thm:multi_k_spacings}
by
rewriting equation (\ref{eq:L1_L0}) as
\begin{align*}
    A_{t-1} &:= \log \frac{\widehat{D}_{t-1}^{(m,k)}}{D_{t-1}^{(m,k)}} \inAS
    \log \frac{\widehat{D}_{t-1}^{(m,k)}}{D_{t-1}^{(m,k)[j^{\star}]}} \inAS \log \frac{1/k}{\pi_k^{[j^{\star}]}} =: A,\\
    B_{t-1} &:= \log \frac{1-\widehat{D}_{t-1}^{(m,k)}}{1-D_{t-1}^{(m,k)}} \inAS
    \log \frac{1-\widehat{D}_{t-1}^{(m,k)}}{1-D_{t-1}^{(m,k)[j^{\star}]}} \inAS
    \log \frac{1-1/k}{1-\pi_k^{[j^{\star}]}} =: B.
\end{align*}
The remaining parts follow by the same proof procedure as in Appendix~\ref{proof:thm_45}.

\section{Proof of auxiliary results}\label{appendix:lemmas}

\subsection{Review of technical results needed}
We need the following lemma retrieved  from Theorem~3,~\cite{deheuvels1984strong}.

\begin{lemma}[Theorem~3 of \cite{deheuvels1984strong}]\label{lem:Gt_locates_low_den}
Suppose the CDF of $\pr$, $F(u)$, has a continuous first derivative $q(u) > 0$ on
$[A, B]$, where
$-\infty < A := \inf\{u: F(u) > 0\} < B := \sup\{u: F(u) < 1\} < \infty$.
Suppose that
there exists $u_0\in [A, B]$ such that $q(u) > q(u_0) > 0$ holds for any $u \in [A, B] \setminus\{u_0\}$.
Then, for any $\delta > 0$,
there exists an almost surely finite $T$ such that, for any $t \geq T$,
$(U_{i,t},U_{i+1,t}) \subseteq (u_0 - \delta, u_0 + \delta)$.
\end{lemma}

\subsection{Proof of Lemma \ref{lem:Gt_to_1half}}
Let
\[
    \Delta_{t-1} := \sup_{u\in [0,1]} | \widehat{F}_{t-1}(u) - u |
    \quad \text{where} \quad
    \widehat{F}_{t-1}(u) = \frac{1}{t-1}\sum_{i=1}^{t-1} \mathbb{1}(U_i\leq u) .
\]
By the Dvoretzky--Kiefer--Wolfowitz (DKW) inequality, we have,
for any $\epsilon>0$, that
$\pr(\Delta_{t-1}> \epsilon) \le 2\exp\{-2(t-1)\epsilon^2\}$.
In particular, taking $\epsilon = \sqrt{(2\log t)/(t-1)}$, we get
\[
    \pr\left(\Delta_{t-1} > \sqrt{(2\log t)/(t-1)}\right)
\le 2t^{-4}.
\]
Because $\sum_{t=1}^{\infty} 2t^{-4} <\infty$, by the Borel–Cantelli lemma, we have that
$$
\Delta_{t-1} = O_{\AS}\left( \sqrt{\frac{2\log t}{t-1}} \right)
= O_{\AS}\left( \sqrt{\frac{\log t}{t}} \right).
$$
By the definition of the empirical CDF $\widehat{F}_{t-1}$ and the order statistics $U_{j,t-1}$,
we have $\widehat{F}_{t-1}(U_{j,t-1}) = j/(t-1)$ for $0\le j\le t$,
where we recall that $U_{0,t-1}=0$ and $U_{t,t-1}=1$.
Then, for $0\le j\le t$,
\[
    \bigg| U_{j,t-1} - \frac{j}{t}\bigg|
    \leq \Delta_{t-1} + \left| \frac{j}{t-1} - \frac{j}{t} \right|
    \leq \Delta_{t-1}  + \frac{1}{t},
\]
which implies
\begin{align}\label{eqt:_uniformBound_U-time}
    a_t := \max_{0\le j\le t} |U_{j,t-1} - j/t |
        = O_{\AS}\{( \log t/t)^{1/2}\}
\end{align}
Therefore, by triangle inequality and (\ref{eqt:_uniformBound_U-time}), we can
bound
$d_{i,t-1}^{(m)} = U_{(i+1)m,t-1} - U_{im,t-1}$ as follows:
\begin{align*}
    \left| d_{i,t-1}^{(m)} - \frac{m}{t} \right|
        &\leq \left| U_{(i+1)m,t-1} - \frac{(i+1)m}{t} \right| + \left| U_{im,t-1} - \frac{im}{t} \right|
        \leq 2a_t  \qquad \text{for $i=0, \ldots, K_{t-1}-1$}; \\
    \left| d_{i,t-1}^{(m)} - \frac{t-K_{t-1}m}{t} \right|
        &= \left| U_{K_{t-1}m, t-1}  - \frac{K_{t-1}m}{t} \right|
        \leq a_t \qquad \text{for $i=K_{t-1}$}.
\end{align*}
In view of (\ref{eq:sum_half_max_spacings}), the preceding displayed inequalities,
and (\ref{eqt:_uniformBound_U-time}),
we have
\begin{align*}
    \left| D_{t-1}^{(m,k)} - \frac{\lfloor t/(mk) \rfloor m}{t}\right|
    &\leq \left\{ \sum_{i\in \mathcal{I}_{t-1}^{(m,k)}} \left| d_{i,t-1}^{(m)} - \frac{m}{t} \right| \right\}
    + \left\{
        \left| d_{K_{t-1},t-1}^{(m)} - \frac{t-K_{t-1}m}{t} \right|
        + \frac{t-K_{t-1}m}{t}
    \right\}
    \mathbb{1}(t/k\not\in\mathbb{Z})  \\
    &\leq |\mathcal{I}_{t-1}^{(m,k)}|(2a_t) + a_t + \frac{m}{t}\\
    &\leq \left(\frac{2t}{\lfloor t^\phi\rfloor k} + 1\right) O_{\AS}\left\{ \left( \frac{\log t}{t}\right)^{1/2} \right\} + \frac{\lfloor t^\phi\rfloor}{t} \\
    &\inAS 0,
\end{align*}
where the last step holds because
$\phi\in (1/2,1)$.
Then, it is trival that
\[
\left| D_{t-1}^{(m,k)} - \frac{1}{k}\right|
\leq \left| D_{t-1}^{(m,k)} - \frac{\lfloor t/(mk) \rfloor m}{t}\right| + \left| \frac{1}{k} - \frac{\lfloor t/(mk) \rfloor m}{t} \right|
\inAS 0.
\]
Thus, the desired result follows.

\subsection{Proof of Lemma~\ref{lem:sum_spacings_strategy}}
\subsubsection{Step 1: Establish $\pr(U_t\in\mathcal{U}_{t-1}^{(m,k)})\inAS 1/k$.}
Let \(p\) and $F$ be the probability density function and CDF of \(U\) under \(\pr\),
respectively.
Define $V_i = F(U_i)$ for $i\in\{1,\ldots, t-1\}$.
Denote the order statistics of $V_1,\ldots, V_{t-1}$
as $0 \equiv V_{0,t-1} \leq V_{1,t-1}\leq \cdots \leq V_{t-1,t-1} \leq V_{t,t-1} \equiv 1$. According to the assumptions in Lemma~\ref{lem:sum_spacings_strategy},
we have that
\[
    \mathcal{U}^{(k)} := \{ u: p(u) \leq p_k\},
    \quad \pr(U_t \in \mathcal{U}^{(k)}) = \frac1k ,
    \quad \text{and} \quad
    \Leb(\{u\in[0,1]: p(u)=p_k\}) = 0.
\]
Since $V_1,\ldots, V_{t-1}\simIID \Unif[0,1]$,
we have, by (\ref{eqt:_uniformBound_U-time}), that
\begin{align}\label{eq:_uniformBound_U-time_V_version}
    b_t := \max_{0\le j\le t} \left|V_{j,t-1} - \frac{j}{t} \right|
        = O_{\AS}\left\{  \left( \frac{\log t}{t} \right)^{1/2} \right\} .
\end{align}
Let $B_{i,t-1}$ be the $i$th spacing block at time $t-1$. For each complete spacing block,
\begin{align}\label{ineq:bound_prob_complete}
    \bigg|\pr(U_t\in B_{i,t-1}\mid \mathcal{F}_{t-1}) - \frac{m}{t}\bigg|=& \bigg|V_{im,t-1} -V_{(i-1)m,t-1} -\frac{m}{t}\bigg|\nonumber\\
    \le& \bigg|V_{im,t-1} -\frac{im}{t}\bigg| + \bigg|V_{(i-1)m,t-1} - \frac{(i-1)m}{t}\bigg|
    \le 2b_t.
\end{align}
For the possibly incomplete spacing block $R_{t-1}^{\mathrm{rem}}$, let $n_t = \lfloor t/m \rfloor$ and then we get
\begin{equation}\label{ineq:prob_remain_term}
    \pr(U_t\in R_{t-1}^{\mathrm{rem}} \mid\mathcal{F}_{t-1})
= 1- F(U_{n_tm,t})
    \leq 1- \left(\frac{n_tm}{t}- b_t\right)
< \frac{m}{t} + b_t.
\end{equation}
Hence, combining (\ref{ineq:bound_prob_complete}) and (\ref{ineq:prob_remain_term}) implies
$$
\left|\pr(\mathcal{U}_{t-1}^{(m,k)})- \left\lfloor \frac{t}{mk} \right\rfloor \frac{m}{t}\right|
\le 2 \left\lfloor \frac{t}{mk} \right\rfloor b_t +\frac{m}{t}+b_t
\inAS 0,
$$
where the last step follows from (\ref{eq:_uniformBound_U-time_V_version}) and the fact that $m=\lfloor t^\phi\rfloor$ with $\phi\in(1/2, 1)$.
Because $\lfloor \frac{t}{mk} \rfloor \frac{m}{t} \rightarrow 1/k$, we have
\begin{equation}\label{eq:prob_1k}
    \pr(\mathcal{U}_{t-1}^{(m,k)})\inAS\frac1k.
\end{equation}

\subsubsection{Step 2: Establish $\liminf_{t\to\infty}\Leb(\mathcal{U}_{t-1}^{(m,k)})\ge\Leb(\mathcal{U}^{(k)})\ \text{almost surely}$.}
Fix any $\varepsilon>0$.
Since $\pr(\mathcal{U}^{(k)}) = 1/k >0$ and $\Leb\{u:p(u)=c\}=0$ for any $c$,
there exists an interval $I_\varepsilon\subseteq\{u:0<p(u)<p_k\}\subseteq \mathcal{U}^{(k)}$ such that $0<\Leb(I_\varepsilon)<\varepsilon$ and $\pr(U_t\in I_\varepsilon)>0$. Set
$$
K_\varepsilon=\mathcal{U}^{(k)}\setminus I_\varepsilon.
$$
Since $p$ is continuous, $\mathcal{U}^{(k)}$ is closed.
Consequently, $K_\varepsilon$ is compact. Furthermore,
\begin{equation}\label{ineq:bound_Ke}
  \Leb(K_\varepsilon)
=\Leb(\mathcal{U}^{(k)})-\Leb(I_\varepsilon) >\Leb(\mathcal{U}^{(k)})-\varepsilon,
\end{equation}
and also
$$
\pr( U_t\in K_\varepsilon)
=\pr(U_t\in \mathcal{U}^{(k)})-\pr(U_t\in I_\varepsilon)
<\frac1k.
$$
By regularity of the probability measure $\pr$, we can find
$O_\varepsilon$ defined as a finite union of $L_{\varepsilon}$ open intervals such that
\begin{equation}
K_\varepsilon\subseteq O_\varepsilon
\quad\text{and}\quad
\pr( U_t\in  O_\varepsilon)<\frac1k.
\end{equation}
Define
the index set of all spacing blocks intersecting $O_\varepsilon$ as
\[
J_t=\{i:B_{i,t-1}\cap O_\varepsilon\ne\varnothing\}.
\]
Then let $N_{t-1}(O_\varepsilon)$ denote the number of $U_{1:t-1}$ belonging to $O_\varepsilon$.
For each interval component of $O_\varepsilon$, at most two spacing blocks $B_{i,t-1}$ intersect its boundary without lying entirely inside it. Every remaining complete $B_{i,t-1}$ contains $m$ observations. Considering one possibly incomplete spacing block $R_t^{\mathrm{rem}}$, we get that
\begin{equation}
|J_t|
\le
\frac{N_{t-1}(O_\varepsilon)}m
+2L_\varepsilon+1,
\end{equation}
where the first term on the right-hand side is an upper bound on the number of blocks $B_{i,t-1}$ entirely contained in $O_\varepsilon$. Then we get
\[
\frac{m}{t}|J_t|
\le
\frac{N_{t-1}(O_\varepsilon)}t
+(2L_\varepsilon+1)\frac{m}{t}.
\]
By the strong law of large numbers,
$
 N_{t-1}(O_\varepsilon) / t
\inAS \pr(U_t \in O_\varepsilon).
$
Recall that $L_\varepsilon$ is finite. It follows that
\[
\limsup_{t\to\infty}\frac{m}{t}|J_t|
\le \pr(U_t \in O_\varepsilon)
<\frac1k
\quad\text{almost surely}.
\]
We also know that $\lfloor \frac{t}{mk} \rfloor \frac{m}{t}\to1/k$. Consequently, for all sufficiently large $t$,
\begin{equation}\label{ineq:Jt_bound}
    |J_t|\le \left\lfloor \frac{t}{mk} \right\rfloor \quad\text{almost surely}.
\end{equation}
Note that, by our strategy,
$\mathcal{U}_{t-1}^{(m,k)}$ contains the $\lfloor \frac{t}{mk} \rfloor$ largest spacing blocks.
Combining this with (\ref{ineq:Jt_bound}),
we almost surely have that, for all sufficiently large $t$,
\begin{align*}
\Leb(\mathcal{U}_{t-1}^{(m,k)})
\ge\sum_{i\in J_t}\Leb(B_{i,t-1})
=\Leb\left(\bigcup_{i\in J_t}B_{i,t-1}\right)
\overset{(a)}{\ge}\Leb(K_\varepsilon)
\overset{(b)}{\ge}\Leb(\mathcal{U}^{(k)})-\varepsilon ,
\end{align*}
where (a) holds since $K_\varepsilon\subseteq O_\varepsilon\subseteq \bigcup_{i\in J_t}B_{i,t-1}$,
while (b) holds according to (\ref{ineq:bound_Ke}).
Since $\varepsilon>0$ is arbitrary,
we can let $\varepsilon\rightarrow 0$.
Then we have
\begin{equation}\label{eq:Uk_upper_bound}
    \liminf_{t\to\infty}\Leb(\mathcal{U}_{t-1}^{(m,k)})\ge\Leb(\mathcal{U}^{(k)})\quad \text{almost surely}.
\end{equation}

\subsubsection{Step 3: Establish $\limsup_{t\to\infty}\Leb(\mathcal{U}_{t-1}^{(m,k)})\le\Leb(\mathcal{U}^{(k)})$ almost surely.}
For any measurable set $S\subseteq[0,1]$, define
\begin{align*}
R(S)
&:=\Leb(\mathcal{U}^{(k)})-\Leb(S)
-\frac{\pr(U_t\in\mathcal{U}^{(k)})-\pr(U_t\in S)}{p_k} \\
&=\int_{\mathcal{U}^{(k)}\setminus S}
\left(1-\frac{p(u)}{p_k}\right) \dd u
+\int_{S\setminus \mathcal{U}^{(k)}}
\left(\frac{p(u)}{p_k} -1\right) \dd u.
\end{align*}
Since $p(u)\le p_k$ for $u \in \mathcal{U}^{(k)}$ and $p(u)>p_k$ for $u\in (\mathcal{U}^{(k)})^{\complement}$, we have $R(S)\ge0$.
Consequently,
\[
\Leb(S)
\le
\Leb(\mathcal{U}^{(k)})
+\frac{\pr(U_t\in S)-\pr(U_t\in \mathcal{U}^{(k)})}{p_k}.
\]
Applying this with $S=\mathcal{U}_{t-1}^{(m,k)}$ and using (\ref{eq:prob_1k}), we have
\begin{equation}\label{eq:Uk_lower_bound}
\limsup_{t\to\infty}\Leb(\mathcal{U}_{t-1}^{(m,k)})
\le\Leb(\mathcal{U}^{(k)})
\quad\text{almost surely}.
\end{equation}
Combining (\ref{eq:Uk_upper_bound}) and (\ref{eq:Uk_lower_bound}), we get that
\begin{equation}\label{eq:Ut_to_Uk}
\Leb(\mathcal{U}_{t-1}^{(m,k)})\inAS\Leb(\mathcal{U}^{(k)}).
\end{equation}

\subsubsection{Step 4: Establish $\Leb(\mathcal{U}_{t-1}^{(m,k)}\triangle \mathcal{U}^{(k)}) \inAS 0$.}
Define $S_\delta = \{u: | p(u) -p_k| \le\delta\}$, where $\delta>0$. We have that
\begin{equation}\label{ineq:symm_diff_bound}
    \Leb(\mathcal{U}_{t-1}^{(m,k)}\triangle \mathcal{U}^{(k)})
\le \Leb(S_\delta)
+\Leb\big\{(\mathcal{U}_{t-1}^{(m,k)}\triangle \mathcal{U}^{(k)})\cap S_\delta^{\complement}\big\}.
\end{equation}
Recall that
\begin{align}\label{eq:R_Ut}
    R(\mathcal{U}_{t-1}^{(m,k)})
    &= \int_{\mathcal{U}_{t-1}^{(m,k)}\triangle\mathcal{U}^{(k)}}
\left|1-\frac{p(u)}{p_k}\right|du
\ge \int_{(\mathcal{U}_{t-1}^{(m,k)}\triangle\mathcal{U}^{(k)}) \cap S_\delta^{\complement}}
\left|1-\frac{p(u)}{p_k}\right|du\nonumber\\
&\overset{(c)}{\ge} \frac{\delta}{p_k}\Leb((\mathcal{U}_{t-1}^{(m,k)}\triangle\mathcal{U}^{(k)}) \cap S_\delta^{\complement}),
\end{align}
where (c) holds since, for any \(u\in S_\delta^{\complement}\), we have $|p(u)-p_k|>\delta$.
Combining (\ref{ineq:symm_diff_bound}) and (\ref{eq:R_Ut}),
we have
\begin{equation}\label{eq:symm_diff_upper_bound}
\Leb(\mathcal{U}_{t-1}^{(m,k)}\triangle \mathcal{U}^{(k)})
\le\Leb(S_\delta)
+\frac{p_k}{\delta} R(\mathcal{U}_{t-1}^{(m,k)}).
\end{equation}
Fix an arbitrary \(\varepsilon>0\).
Because \(\Leb(\{u: p(u)=p_k\})=0\), there exists a fixed \(\delta_\varepsilon>0\) such that
$
\Leb(S_{\delta_\varepsilon})
<{\varepsilon}/{2}.
$
By
(\ref{eq:prob_1k}), (\ref{eq:Ut_to_Uk}), and
the definition of $R(\mathcal{U}_{t-1}^{(m,k)})$,
we get $R(\mathcal{U}_{t-1}^{(m,k)}) \inAS 0$.
Then, there exists an almost surely finite \(T_\varepsilon\) such that, for all \(t\ge T_\varepsilon\),
\[
R(\mathcal{U}_{t-1}^{(m,k)})<\frac{\varepsilon\delta_\varepsilon}{2p_k}\quad \text{almost surely}.
\]
Then, putting it into (\ref{eq:symm_diff_upper_bound}), we obtain
$$
\Leb(\mathcal{U}_{t-1}^{(m,k)}\triangle \mathcal{U}^{(k)})
< \frac{\varepsilon}{2} + \frac{p_k}{\delta_\varepsilon}\frac{\varepsilon\delta_\varepsilon}{2p_k} = \varepsilon
$$
Because \(\varepsilon>0\) is arbitrary, we have
$
\Leb(\mathcal{U}_{t-1}^{(m,k)}\triangle \mathcal{U}^{(k)})
\inAS 0.
$

\subsubsection{Step 5: Properties of $\pi_k$}
Now, we derive the feasible range of $\pi_k$.
Note that
$\bar{p}' \leq p_k \leq \bar{p}''$, where
\[
    \bar{p}' = \frac{1}{\pi_k}\int_{\mathcal{U}^{(k)}} p(u)\, \dd u
        = \frac{1/k}{\pi_k}
    \qquad \text{and} \qquad
    \bar{p}'' = \frac{1}{1-\pi_k}\int_{(\mathcal{U}^{(k)})^{\complement}} p(u)\, \dd u
        = \frac{1-1/k}{1-\pi_k}.
\]
Since $\bar{p}'\le\bar{p}''$,
we have $\bar{p}'< 1$, i.e., $\pi_k > 1/k$.
Thus, $\pi_k \in (1/k, 1)$.

\subsection{Proof of Lemma \ref{lem:for_proof_in_avg_rule}}\label{sec:proof_of_lem:for_proof_in_avg_rule}
\subsubsection{Step 1. Preliminary.}
Define $\varepsilon_t=2b_t t/m$,
where $b_t$ is defined in (\ref{eqt:_uniformBound}).
By the asymptoptic bound in (\ref{eqt:_uniformBound}),
we have that
\begin{equation}\label{eq:epsilon_to0}
    \varepsilon_t
=
O_{\AS}\left(
t^{1/2-\phi}\sqrt{\log t}
\right)
\to0.
\end{equation}
Consider a complete spacing block with endpoints
$U_{im,t-1}$
and
$U_{(i+1)m,t-1}$.
By the definition of $b_t$ in (\ref{eqt:_uniformBound}), we have
\begin{align*}
    \left|F(U_{(i+1)m,t-1})-F(U_{im,t-1})-\frac{m}{t}\right|
    &\le \left|F(U_{(i+1)m,t-1})-\frac{(i+1)m}{t}\right|
        + \left|F(U_{im,t-1})-\frac{im}{t}\right|\\
    &\le 2b_t.
\end{align*}
Thus,
the probability of falling into a complete block, i.e.,
$\pr(U_t\in[U_{im,t-1}, U_{(i+1)m,t-1}] \mid \mathcal{F}_{t-1}) = F(U_{(i+1)m,t-1})-F(U_{im,t-1})$ can be bounded as follows:
\[
\frac{m}{t}(1-\varepsilon_t)
\leq \pr(U_t\in[U_{im,t-1}, U_{(i+1)m,t-1}] \mid \mathcal{F}_{t-1})
\leq
\frac{m}{t}(1+\varepsilon_t).
\]
Similarly, the probability of falling into an
incomplete terminal block is at most
$(1+\varepsilon_t)m/t$ as well.
Since at most one
of the $\lfloor t/(mk) \rfloor$ selected blocks is incomplete,
we have
\begin{equation}\label{bound:a_t}
    \left(\left\lfloor \frac{t}{mk} \right\rfloor-1\right)\frac{m}{t}(1-\varepsilon_t)
\le \widehat{D}_{t-1}^{\agg \star}
\le \left\lfloor \frac{t}{mk} \right\rfloor\frac{m}{t}(1+\varepsilon_t).
\end{equation}
It follows that
\begin{align*}
    \left(\widehat{D}_{t-1}^{\agg \star} - \frac{1}{k} \right)
    \leq \frac{\varepsilon_t}{k}, \quad\text{and}\quad
    -\left(\widehat{D}_{t-1}^{\agg \star} -\frac1k\right)
    \le \frac1k-\left(\left\lfloor \frac{t}{mk} \right\rfloor-1\right)\frac{m}{t}(1-\varepsilon_t)
    \le\frac{2m}{t}+ \frac{\varepsilon_t}{k}.
\end{align*}
Consequently,
\begin{align}\label{eqt:diff_hatD*-1/k}
    \left|\widehat{D}_{t-1}^{\agg \star}-\frac1k\right|\le\frac{\varepsilon_t}{k}+\frac{2m}{t}.
\end{align}
Furthermore, since
$
    \widehat{D}_{t-1}^{(m,k)}= \lfloor t/k\rfloor/t,
$
we have
$
    0\le 1/k-\widehat{D}_{t-1}^{(m,k)}<1/t \le m/t.
$
Together with (\ref{eqt:diff_hatD*-1/k}) and the triangle inequality, we have
\begin{equation}\label{bound:a_t_y}
    \left|\widehat{D}_{t-1}^{\agg \star}-\widehat{D}_{t-1}^{(m,k)}\right|
    \leq \left|\widehat{D}_{t-1}^{\agg \star}-\frac{1}{k}\right| + \left|\widehat{D}_{t-1}^{(m,k)}-\frac{1}{k}\right|
    \leq \frac{\varepsilon_t}{k}+\frac{3m}{t}.
\end{equation}
When $\lfloor t/(mk) \rfloor\ge1$, we have $1/t\le 1/(km)$, and hence
\begin{equation}\label{eq:lowerBound_hatD}
    \widehat{D}_{t-1}^{(m,k)}= \frac{\lfloor t/k\rfloor}{t}\ge \frac{t/k -1}{t}= \frac{1}{k}-\frac1t
\ge \frac{1}{k}\left(1-\frac1{m}\right)
\ge\frac{1}{2k}
\end{equation}
for large $t$.
Also, $\widehat{D}_{t-1}^{(m,k)}\le 1/k$, so
\begin{equation}\label{eq:lowerBound_minushatD}
    1-\widehat{D}_{t-1}^{(m,k)}\ge1-\frac{1}{k}.
\end{equation}

\subsubsection{Step 2. Upper bounds for $\KL(\Bern(\widehat{D}_{t-1}^{\agg \star})\| \Bern(\widehat{D}_{t-1}^{(m,k)}))$.}
Using $\log x\le x-1$ for $x>0$, we obtain
\[
\begin{aligned}
0\leq \KL(\Bern(\widehat{D}_{t-1}^{\agg \star})\| \Bern(\widehat{D}_{t-1}^{(m,k)}))
\le&
\widehat{D}_{t-1}^{\agg \star}\left(\frac{\widehat{D}_{t-1}^{\agg \star}}{\widehat{D}_{t-1}^{(m,k)}}-1\right)
+(1-\widehat{D}_{t-1}^{\agg \star})\left(\frac{1-\widehat{D}_{t-1}^{\agg \star}}{1-\widehat{D}_{t-1}^{(m,k)}}-1\right)\\
=&\frac{(\widehat{D}_{t-1}^{\agg \star}-\widehat{D}_{t-1}^{(m,k)})^2}{\widehat{D}_{t-1}^{(m,k)}(1-\widehat{D}_{t-1}^{(m,k)})}.
\end{aligned}
\]
We consider two cases.
\begin{itemize}
    \item (Case 1: $k\ge2$.)
        Since $1/k\le1/2$, (\ref{eq:lowerBound_hatD}) and (\ref{eq:lowerBound_minushatD}) give
        \[
        \widehat{D}_{t-1}^{(m,k)}(1-\widehat{D}_{t-1}^{(m,k)})
        \ge\frac1{2k}\left(1-\frac1k\right)
        \ge\frac1{4k}.
        \]
        Combining this with (\ref{bound:a_t_y}),
        \begin{align}\label{bound:KL_upper_Kge2}
        \KL(\Bern(\widehat{D}_{t-1}^{\agg \star})\| \Bern(\widehat{D}_{t-1}^{(m,k)}))
        \le 4k\left(\frac{\varepsilon_t}{k}+\frac{3m}{t}\right)^2
        \le \frac{8\varepsilon_t^2}{k} +\frac{72m^2k}{t^2}.
        \end{align}
    \item (Case 2: $1<k<2$.)
        By (\ref{eq:lowerBound_hatD}) and (\ref{eq:lowerBound_minushatD}),
        \[
        \widehat{D}_{t-1}^{(m,k)}(1-\widehat{D}_{t-1}^{(m,k)})
        \ge\frac{k-1}{2k^2}
        >\frac{k-1}{8}.
        \]
        Combining this with (\ref{bound:a_t_y}),
        \begin{align}\label{bound:KL_upper_Kle2}
        \KL(\Bern(\widehat{D}_{t-1}^{\agg \star})\| \Bern(\widehat{D}_{t-1}^{(m,k)}))
        \le\frac8{k-1}
        \left(\frac{\varepsilon_t}{k}+\frac{3m}{t}\right)^2
        <\frac{8(\varepsilon_t+3m/t)^2}{k-1}
        <\frac{72(\varepsilon_t+m/t)^2}{k-1}.
        \end{align}
\end{itemize}

\subsubsection{Step 3. Sum over the grid.}
Recall that
$
\mathcal K_{t-1}
=
\{1+j\Delta_t:1\le j\le |\mathcal K_{t-1}|\}.
$
We consider the two cases again:
\begin{itemize}
    \item (Case 1: $k\geq 2$) In this case,
    the intervals $[k-\Delta_t,k]$ have disjoint interiors and are contained in $[1,\infty)$.
Since $x\mapsto1/x$ is decreasing,
$
 \Delta_t / k \le\int_{k-\Delta_t}^{k} 1/x \dd x.
$
Therefore,
\begin{align}\label{bound:Delta_1k}
\Delta_t
\sum_{\substack{k\in\mathcal K_{t-1}\\2\le k\le M_t}}
\frac1k
\le
\int_1^{M_t}\frac{1}{x} \dd x
=\log M_t
\le\log t.
\end{align}
Next, the number of grid points satisfying $k\le M_t$ is
at most $(M_t-1)/\Delta_t$. Thus
\begin{align}\label{bound:Deltak}
\Delta_t
\sum_{\substack{k\in\mathcal K_{t-1}\\2\le k\le M_t}}
k
\le
\Delta_t M_t
\frac{M_t-1}{\Delta_t}
\le M_t^2.
\end{align}
Using (\ref{bound:KL_upper_Kge2}), (\ref{bound:Delta_1k}) and (\ref{bound:Deltak}), we obtain
\begin{align}\label{bound:final_kge2}
\Delta_t
\sum_{\substack{k\in\mathcal K_{t-1}\\2\le k\le M_t}}
\KL(\Bern(\widehat{D}_{t-1}^{\agg \star})\| \Bern(\widehat{D}_{t-1}^{(m,k)}))
&\le
8\varepsilon_t^2 \Delta_t
\sum_{\substack{k\in\mathcal K_{t-1}\\2\le k\le M_t}}
\frac1k
+\frac{72m^2}{t^2}
\Delta_t
\sum_{\substack{k\in\mathcal K_{t-1}\\2\le k\le M_t}}
k\nonumber\\
&\le C\left(
\varepsilon_t^2\log t+\frac{m^2}{t^2}M_t^2
\right),
\end{align}
where $C$ is a constant.

\item (Case 2: $1<k<2$). Write $k=1+j\Delta_t$.
Then
\begin{align}\label{bound:1k_1}
\Delta_t \sum_{\substack{k\in\mathcal K_{t-1}\\1<k<2}}
\frac1{k-1}
\le
\sum_{1\le j<1/\Delta_t}\frac1j
\le1+\log\left(\frac{1}{\Delta_t}\right),
\end{align}
where the last inequality follows from the elementary harmonic-sum bound $\sum_{j=1}^J 1/j\le1+\log J$.
Combining (\ref{bound:KL_upper_Kle2}) and (\ref{bound:1k_1}),
we have
\begin{align}\label{bound:final_kle2}
\Delta_t
\sum_{\substack{k\in\mathcal K_{t-1}\\1<k<2}}
\KL(\Bern(\widehat{D}_{t-1}^{\agg \star})\| \Bern(\widehat{D}_{t-1}^{(m,k)}))
\le
72\left(\varepsilon_t+\frac{m}{t}\right)^2
\left(1+\log\left(\frac{1}{\Delta_t}\right)\right).
\end{align}
Combining (\ref{bound:final_kge2}) and (\ref{bound:final_kle2}), we conclude that
\begin{align}\label{bound:KL_finally}
&\Delta_t \sum_{\substack{k\in\mathcal K_{t-1}\\k\le M_t}}
\KL(\Bern(\widehat{D}_{t-1}^{\agg \star})\| \Bern(\widehat{D}_{t-1}^{(m,k)})) \nonumber \\
&\qquad\le C'\left\{
\varepsilon_t^2\log t
+\frac{m^2}{t^2} M_t^2
+\left(\varepsilon_t+\frac{m}{t}\right)^2
\left(1+\log\left(\frac{1}{\Delta_t}\right)\right)
\right\},
\end{align}
where $C'$ is a constant.
\end{itemize}

\subsubsection{Step 4. Show that the right-hand side is bounded.}
By (\ref{eq:epsilon_to0}),
$\varepsilon_t^2\log t
=
O_{\AS}\left(
t^{1-2\phi}(\log t)^2
\right)
\to0$.
Moreover, we have
$$
\frac{m^2}{t^2}(\log t)^2 \le t^{2\phi-2}(\log t)^2
\to0.
$$
By the assumption that
$\log(1/\Delta_t)=O(\log t)$,
it follows that
\begin{align*}
\left(\varepsilon_t+\frac{m}{t}\right)^2
\left(1+\log\left(\frac{1}{\Delta_t}\right)\right)
\le C\left(\varepsilon_t+\frac{m}{t}\right)^2\log t
\inAS 0.
\end{align*}
Finally, $M_t\le t/m$ from definition,
and therefore $(mM_t/t)^2\to c\le1$.
Substituting these bounds into (\ref{bound:KL_finally}), we obtain
\[
\begin{aligned}
0
\le
\Delta_t
\sum_{\substack{k\in\mathcal K_{t-1}\\k\le M_t}}
\KL(\Bern(\widehat{D}_{t-1}^{\agg \star})\| \Bern(\widehat{D}_{t-1}^{(m,k)}))
\le O(1)
\qquad\text{almost surely.}
\end{aligned}
\]
It completes the proof.

\section{Additional results and discussions}

\subsection{Supplement to Section \ref{sec:maxSpacing_mainResult}: Examples satisfying Case 2 in Condition~\ref{cond:simple_strategy}}\label{proof:examples}

Example~\ref{example:beta} verifies (\ref{eqt:cond_D=0}) and (\ref{eq:cond_Q=0}) for $\operatorname{Beta}(\alpha,\beta)$ when $\max(\alpha,\beta)>1$. Example~\ref{example:continuous_beta} extends the analysis to beta mixtures and identifies the applicable case of Condition~\ref{cond:simple_strategy}. Example~\ref{example:holderClass} shows that H\"older densities satisfying the interior positivity and endpoint conditions of Example~\ref{example:holderClass} can be approximated in Kullback--Leibler divergence by distributions satisfying Case 2 in Condition~\ref{cond:simple_strategy}.

\begin{example}\label{example:beta}
Let $U_1, U_2, \ldots\simIID\Beta(\alpha, \beta)$,
where $\alpha, \beta>0$ such that $\max(\alpha,\beta)>1$.
Then $\delta_t \inAS \infty$. Since any Beta distribution is atomless, Case 2 in Condition~\ref{cond:simple_strategy} is satisfied for $\Beta(\alpha, \beta)$ with $\max(\alpha,\beta)>1$.
See Appendix~\ref{app:proof_example31} for the proof. \qed
\end{example}

\begin{example}[Continuation of Example~\ref{example:beta}]\label{example:continuous_beta}
Let $\delta\in(0, \infty)$ be fixed.
Denote $\mathcal{P}_\delta$ as the class of all distributions $\Beta(\alpha, \beta)$ with $\alpha \geq 1+\delta$ and $\beta>0$.
Define $\mathrm{conv}(\mathcal{P}_\delta)$ as the
convex Choquet hull of $\mathcal{P}_\delta$ \citep{lardy2024reverse}.
Denote $\pr\in \mathrm{conv}(\mathcal{P}_\delta)$ with mixing distribution $W$. If $W$ satisfies that $\int \frac{1}{B(\alpha,\beta)}\, \dd W(\alpha,\beta)
<\infty$,
the distribution $\pr$ satisfies Condition~\ref{cond:simple_strategy}.
Similarly, let $\mathcal{P}'_{\delta}$ be the class of
all distribution $\Beta(\alpha, \beta)$ with $\alpha>0$ and $\beta \geq 1+\delta$.
Then $\pr\in \mathrm{conv}(\mathcal{P}_\delta')$ with such $W$ also
satisfies Condition~\ref{cond:simple_strategy}.
See Appendix~\ref{proof:exampleC2} for the proof. \qed
\end{example}

\begin{example}[Continuation of Example~\ref{example:continuous_beta}]\label{example:holderClass}
We show that
a subset of $\mathrm{conv}(\mathcal{P}_\delta)$,
denoted as $\mathcal{P}_a^W$ in (\ref{eqt:PaW}),
approximates the class of all Hölder continuous distributions on $[0, 1]$.

Let \(\pr_{a,\epsilon} = \Beta\left(a/(1-\epsilon),\, a/\epsilon\right)\), with density \(p_{a,\epsilon}(\cdot)\), where \(\epsilon \in (0,1)\) and \(a>1\) is fixed.
Let \(W\) be a distribution supported on \((0,1)\) with density $w(\cdot)$.
We then define the distribution \(\pr_a^W\) as a $W$-mixture of $\pr_{a,\cdot}$ so that
its density \(p_a^W : [0,1] \to [0,\infty)\) is given by
$
    p_a^W(u) := \int_0^1 p_{a,\epsilon}(u)\, \dd W(\epsilon).
$
Then the family of distributions of such $\pr_a^W$ is denoted as
\begin{align}\label{eqt:PaW}
\mathcal P_a^W :=\left\{\pr_a^W: W\text{ is a probability distribution on }(0,1),\
\int_0^1
\frac{1}{B\!\left(a/(1-\epsilon),a/\epsilon\right)}
\dd W(\epsilon)<\infty
\right\}.
\end{align}
Since \(\mathcal{P}_a^W\) is a class of mixture distributions with a fixed \(a>1\),
it is a subset of the mixture class \(\mathrm{conv}(\mathcal{P}_\delta)\) and \(\mathrm{conv}(\mathcal{P}'_\delta)\) in Example~\ref{example:continuous_beta} whenever $0<\delta\le a-1$.
Corollary 3.1 of \cite{10.1214/09-AOS703} shows that distributions in \(\mathcal{P}_a^W\) can approximate Hölder-continuous distributions on \([0,1]\).
Specifically, let
$\mathcal{H}(L,k) = \left\{ p : [0,1] \to \mathbb{R} \,\bigm|\, \lvert p^{(r)}(x) - p^{(r)}(y) \rvert \le L \lvert x-y \rvert^{k-r} \right\}$
be the class of Hölder functions with regularity \(k>0\),
where \(r = \max\{r'\in\mathbb{Z}:r'<k\}\),
and \(p^{(r)}\) denotes the \(r\)th derivative of \(p\).
Suppose \(U_1,U_2,\ldots\simIID\pr\), where
$\pr$ has a density \(p\in \mathcal{H}(L,k)\) such that
$
p(u)>0,\ u\in (0,1),
$
and that there exist integers $r_0,r_1<k$ such that
$
|p^{(r_0)}(0)|>0,\
|p^{(r_1)}(1)|>0
$.
Then, for all sufficiently large
$a$, there exists a probability distribution $W^{\star}$ with finite
support in $(0,1)$ such that
$$
\KL(\pr \,\|\, \pr_a^{W^{\star}}) \le C a^{-k},
$$
where $C>0$ is a constant independent of $a$. Because $W^{\star}$ has finite support in $(0,1)$, i.e.,
\[
W^{\star}=\sum_{j=1}^J w^{\star}_j\delta_{\epsilon_j},
\qquad
0<\epsilon_j<1,\quad
w^{\star}_j>0,\quad
\sum_{j=1}^J w^{\star}_j=1.
\]
For each fixed $a$, it follows that
\[
\int_0^1
\frac{1}{B\!\left(a/(1-\epsilon),a/\epsilon\right)}
\,dW^{\star}(\epsilon)
=\sum_{j=1}^J
\frac{w^{\star}_j}
{B\!\left(a/(1-\epsilon_j),a/\epsilon_j\right)}
<\infty,
\]
where the last inequality holds because the sum has finitely many terms
and every $\epsilon_j$ lies strictly between zero and one.
Therefore, $\pr_a^{W^{\star}}$ satisfies the integrability condition in
Example C.2.

Hence $\pr_a^{W^{\star}}\in\mathcal P_a^W$, and Example~\ref{example:continuous_beta} implies that
$\pr_a^{W^{\star}}$ satisfies Case 2 of Condition~\ref{cond:simple_strategy}. Therefore, the KL approximation rate $Ca^{-k}$ is achieved by a
sequence of distributions satisfying Case 2 of Condition~\ref{cond:simple_strategy}.
In a nutshell,
\(
    \mathcal{H}(L,k) \lesssim \mathcal{P}_a^W \subseteq \mathrm{conv}(\mathcal{P}_\delta)
\) when $1+\delta \leq a$.
\qed
\end{example}

\subsubsection{Proof of Example \ref{example:beta}}\label{app:proof_example31}
Without loss of generality, assume $\alpha = \max(\alpha, \beta) > 1$. The density of the beta distribution is
$$
p_{\alpha,\beta}(u) = \frac{1}{B(\alpha, \beta)} u^{\alpha - 1} (1 - u)^{\beta - 1}.
$$
Because $\alpha > 1$, the density vanishes at the left boundary, i.e., $p_{\alpha,\beta}(0) = 0$.

We just need to prove \eqref{eqt:cond_D=0}, i.e., $\lim_{t \to \infty} \delta_t = \infty$ almost surely. Let $F$ be the CDF. For any arbitrarily large constant $M > 0$, choose $u_0 > 0$ small enough such that $p_{\alpha,\beta}(u) \le 1/M$ for all $u \in [0, u_0]$. Let $v_0 = F(u_0)$.
By the probability integral transform, the transformed points $V_i = F(U_i)$ are i.i.d. $\Unif[0,1]$. Let $N_t$ be the number of $V_i$ falling in the interval $[0, v_0]$ at time $t$ (i.e., we have $t$ samples). By the strong law of large numbers, $N_t / t \inAS v_0$. Conditioned on $N_t$, the points in $[0, v_0]$ act as $N_t$ uniform random variables scaled by $v_0$. The maximum uniform spacing restricted to this interval, denoted $D_{t, \text{sub}}^V$, obeys standard extreme value theory for uniform spacings, yielding
$$
D_{t, \text{sub}}^V \sim v_0 \frac{\log N_t}{N_t}\inAS \frac{\log N_t}{t} \sim \frac{\log t}{t}.
$$

Let $D_t$ be the maximal spacing of $U_{1:t}$ on $[0,1]$. Let this maximal sub-spacing occur between adjacent uniform points $V_{k,t}$ and $V_{k+1,t}$, corresponding to $U_{k,t}$ and $U_{k+1,t}$ strictly within $[0, u_0]$. By the Mean Value Theorem,
\begin{equation*}
    V_{k+1,t} - V_{k,t} = \int_{U_{k,t}}^{U_{k+1,t}} p_{\alpha,\beta}(u) \dd u \le (U_{k+1,t} - U_{k,t}) \sup_{u \in [0, u_0]} p_{\alpha,\beta}(u) \le D_t \frac{1}{M}.
\end{equation*}
Therefore,
$$
D_t \ge M (V_{k+1,t} - V_{k,t}) \sim M \frac{\log t}{t}\qquad \text{almost surely.}
$$
Since $M$ is arbitrary, $\delta_t = D_t / (\log t / t) \inAS \infty$, fulfilling \eqref{eqt:cond_D=0}.

\subsubsection{Proof of Example \ref{example:continuous_beta}}\label{proof:exampleC2}
We first consider $\mathrm{conv}(\mathcal{P}_\delta)$ with $W$ such that $K_W := \int \frac{1}{B(\alpha,\beta)}\,dW(\alpha,\beta)
<\infty$. Because \(\alpha\ge 1+\delta\) for every component and write
\begin{equation}\label{eq:mixture_beta}
    p(u)=\int
\frac{u^{\alpha-1}(1-u)^{\beta-1}}{B(\alpha,\beta)}
\,dW(\alpha,\beta).
\end{equation}
For \(0<u\le 1/2\), we have \(u^{\alpha-1}\le u^\delta\). Also, for each \(\beta>0\), we have that
\[
(1-u)^{\beta-1}\le (1-u)^{-1}\le 2.
\]
Denote $F$ as the CDF of $\pr$. Applying both inequalities into (\ref{eq:mixture_beta}) gives
\[
p(u)\le 2K_Wu^\delta\qquad\text{and}\qquad
F(u)\le \frac{2K_W}{1+\delta}u^{1+\delta}.
\]
Choose $\gamma$ such that $1 /(1+\delta)<\gamma<1$.
Then, for all sufficiently large \(i\),
\[
\pr(U_i\le i^{-\gamma})
=F(i^{-\gamma})
\le \frac{2K_W}{1+\delta}\,i^{-\gamma(1+\delta)}.
\]
Since \(\gamma(1+\delta)>1\), these probabilities are summable. Borel–Cantelli lemma indicates a finite \(N\) such that \(U_i>i^{-\gamma}\) for every \(i\ge N\) almost surely.

Now consider the minimum of the first \(t-1\) observations. For \(N\le i<t\), we have \(U_i>i^{-\gamma}\ge t^{-\gamma}\) almost surely. The finitely many earlier observations \(U_1,\ldots,U_{N-1}\) are all strictly positive, so they also exceed \(t^{-\gamma}\) once \(t\) is sufficiently large. Consequently,
\[
D_{t-1}\ge U_{1,t-1}-0 \ge \min_{1\le i<t}U_i\ge t^{-\gamma}
\qquad\text{a.s.}
\]
By the definition of \(\delta_t\), we obtain
\[
\delta_t=\frac{D_{t-1}}{\log t/t}
\ge \frac{t^{1-\gamma}}{\log t}
\inAS\infty.
\]
If \(\beta>1\), apply the same argument to \(1-U_i\sim\operatorname{Beta}(\beta,\alpha)\). Its first spacing is \(1-\max_{i<t}U_i\), the right boundary spacing of the original observations. This proves (\ref{eqt:cond_D=0}) whenever \(\alpha>1+\delta\) or \(\beta>1+\delta\). Because each $\pr\in\mathrm{conv}(\mathcal{P}_\delta)$ and $\pr\in\mathrm{conv}(\mathcal{P}'_\delta)$ are atomless, using the result in Example~\ref{example:delta_t_to_infty}, (\ref{eq:cond_Q=0}) also holds.

\subsection{Supplement to Section~\ref{sec:oracle_log_optimal}: Pilot estimator for $\widehat{D}_{t-1}^{\star}$}
We provide an alternative, simple algorithm (Example~\ref{example:LLN}) that uses a pilot sample to approximate \(\widehat{D}_{t-1}^{\star}\) from Section~\ref{sec:oracle_log_optimal},
while the construction in Example~\ref{example:LLN} is non-seqeuntial in nature.

\begin{algorithm}

\caption{Estimated log-optimal $e$-process based on a training sample}
\label{algo:LLN}
\SetKwInOut{KIN}{Input}
\KIN{testing dataset: $X_1, X_2, \ldots$\;
training dataset: $X_1', \ldots, X_{n'}'$\;
Stopping time: $\tau$.
}
\SetKwInOut{KOUT}{Output}
\KOUT{$e$-value sequence $\Lambda_1, \Lambda_2, \ldots, \Lambda_\tau$, where $\tau$ is stopping time.}

Initialize $D_0 = 1$ and $\Lambda_0 = 1$; transform $U_i' = F^X_0(X_i')$ for $i=1,\ldots, n'$.

\For{$t = 1,2,\ldots$} {
Transform $U_t = F^X_0(X_t)$\;
Sort $U_1, \ldots, U_{t-1}$ in an ascending order, i.e.,
$0\equiv U_{0,t-1} \leq U_{1,t-1} \leq \cdots \leq U_{t-1,t-1} \leq U_{t,t-1} \equiv 1$\;
Define spacings $d_i = U_{i+1,t-1} - U_{i,t-1}$ for $i = 0,\ldots, t-1$\;
Compute the max spacing $D_{t-1} = \max\limits_{i\in \{1,\ldots, t-1\}} d_i$\;
Define $\mathcal{U}_{t-1} = [U_{i_{t-1},t-1}, U_{i_{t-1}+1,t-1}]$\;
Compute $\widehat{D}_{t-1}^{\pilot} = \sum_{i=1}^{n'} \mathbb{1}(U'_i \in \mathcal{U}_{t-1}) /n'$\;
    \eIf{$U_t \in \mathcal{U}_{t-1}$}{
        $\Lambda_t := \Lambda_{t-1}  {\widehat{D}_{t-1}^{\pilot}}/{D_{t-1}}$;
    }{
        $\Lambda_t := \Lambda_{t-1}\ {(1-\widehat{D}_{t-1}^{\pilot})}/{(1-D_{t-1})}$;
    }

    \If{$\tau = t$}{
    Stop and return $\Lambda_1, \ldots, \Lambda_{\tau}$,
    }
}
\end{algorithm}

\begin{example}[Pilot estimator]\label{example:LLN}
Direct sequential estimation of $\pr$ is typically infeasible
because $\mathcal{H}_1$ is composite and highly complex.
A natural approach is to estimate $\widehat{D}_{t-1}^{\star} = \pr\left(D_t < D_{t-1} \mid \mathcal{F}_{t-1} \right)$
by a pilot estimator based on a training sample.

Let $X'_1, \ldots, X'_{n'}\simIID \pr$ a prior training sample that is independent of $X_1, \ldots, X_n$.
Define $U_i' = F^X_0(X_i')$ for $i=1, \ldots, n'$.
Let $i_t$
be the index at time $t$ such that $D_t = U_{i_t + 1} - U_{i_t}$.
We approximate $\widehat{D}_{t-1}^{\star}$ by
\begin{align}\label{eqt:pilotDhat}
    \widehat{D}_{t-1}^{\pilot}
        = \frac{1}{n'}\sum_{i=1}^{n'} \mathbb{1}( U_i' \in \mathcal{U}_{t-1}),
\end{align}
where $\mathcal{U}_{t-1} := [U_{i_t}, U_{i_t+1}]$.
For each fixed $t$,
conditioning on $U_{1:t-1}$, we have
$\widehat{D}_{t-1}^{\pilot} \inP \widehat{D}_{t-1}^{\star}$
as $n'\rightarrow\infty$ by the law of large numbers.
The procedure is detailed in Algorithm~\ref{algo:LLN}. \qed
\end{example}

\subsection{Supplement to Section \ref{sec:general_framework}: Additional examples of $\widehat{D}_{t-1}$}\label{sec:additionalExample_hatD}
Other choices of the generalized $\widehat{D}_{t-1}^{\agg}$ defined
in Section \ref{sec:general_framework} are provided below:

\begin{example}[Parametric plugin]\label{eg:param}
Suppose $p_{\theta}(\cdot)$ is the density of
a continuous distribution supported on $[0, 1]$ with a parameter $\theta\in \Theta$.
Let $\widehat{\theta}_{t-1}:= \argmax_{\theta \in \Theta} \prod_{i=1}^{t-1} p_{\theta}(U_i)$ be the MLE of $\theta$
based on the sample $U_{1:t-1}$.
Then we can set $\widehat{D}_{t-1}$ as
$
   \widehat{D}_{t-1}^{\para}
   = \int_{\mathcal{U}_{t-1}} p_{\widehat{\theta}_{t-1}}(u) \dd u.
$
\end{example}

\begin{example}[Nonparametric plugin]\label{sec:Glivenko-Cantelli}
Denote the empirical distribution of $U_0 , U_1, \ldots, U_{t-1}$ by
$
    \widehat{F}_{t-1}(u) = \sum_{i=0}^{t-1} \mathbb{1}(U_i \leq u)/t,
$
where $U_0\equiv 0$ is included to avoid degeneration.
Note that
$[ U_t \mid U_1, \ldots, U_{t-1}] \sim F \approx \widehat{F}_{t-1}$.
Then
we can set $\widehat{D}_{t-1}$ as
$
    \widehat{D}_{t-1}^{\npara}
    	= \int_{\mathcal{U}_{t-1}} \dd \widehat{F}_{t-1}
        = {|\mathcal{I}_{t-1}|}/{t},
$
where $|\cdot|$ is the cardinality of a finite set.
\end{example}

The estimators $\widehat{D}_{t-1}^{\para}$ and $\widehat{D}_{t-1}^\npara$
approximate
$\widehat{D}_{t-1}^{\agg \star} = \pr\left(U_t^{\agg} \in \mathcal{U}_{t-1} \mid \mathcal{F}_{t-1} \right)$ when
$|\mathcal{I}_{t-1}| \asymp n$ and
$\mathcal{I}_{t-1}$ is independent of $U_{1:t-1}$;
see Remark \ref{rem:max_aggregation} for more discussion.
Although $\widehat{D}_{t-1}$ in Examples \ref{sec:Glivenko-Cantelli} may not satisfy
$\widehat{D}^{\star}_{t-1}/\widehat{D}_{t-1}^\npara \inAS 1$,
we show that this relation holds under the maximum $1/k$ fraction $m$th order spacings strategy $\mathcal{I}_{t-1}:= \mathcal{I}^{(m,k)}_{t-1}$, which is developed in Section~\ref{sec:max_sum_half_spacings}.
We prove that the $e$-process with this aggregation strategy is powerful and consistency.

\begin{remark}[Warnings about using aggregation strategy $\mathcal{I}_{t-1}$]\label{rem:max_aggregation}
    First, the plugin estimators $\mathcal{I}_{t-1}$ in Examples \ref{eg:param}--\ref{sec:Glivenko-Cantelli}
    may lead to the problems of model misspecification or
    ``using the data twice''.
    In particular, to see the problem in Example \ref{sec:Glivenko-Cantelli},
    we consider $\mathcal{I}_{t-1} = \{i_{t-1}\}$,
    where $d_{i_{t-1},t-1}=D_{t-1}^{(1,\infty)}$.
    It leads to the maximum spacing strategy.
    In this case, $\mathcal{I}_{t-1}$ is dependent on $U_{1:t-1}$.
    Consequently, it requires us to use the extreme value theory of spacing
    in Section~\ref{sec:maxSpacing_mainResult} to handle it, and leads to
    \(\widehat{D}_{t-1}^{\md} = (\log t)/t\) in (\ref{eqt:Dtildestar}) rather than $\widehat{D}_{t-1}^{\npara} = 1/t$ in Example \ref{sec:Glivenko-Cantelli}.
    This ensures that $\widehat{D}^{\star}_{t-1}/\widehat{D}_{t-1}^{\md} \inAS 1$.
    Besides, a degenerate $\mathcal{I}_{t-1}$ leads to
    diminishing $\widehat{D}_{t-1}$.
    Hence, the uniform consistency of $\widehat{F}_{t-1}$ ensured by
    Glivenko--Cantelli theorem is insufficient to
    ensure that $\widehat{D}^{\agg \star}_{t-1}/\widehat{D}_{t-1}^{\npara} \inAS 1$.

\end{remark}

\subsection{Supplement to Section \ref{sec:finite_mathcalK}: Finite and fixed $\mathcal{K}_{t}$}\label{sec:armax}

As a corollary of Theorem \ref{thm:divergent_mathcalK_optimal_softmax},
if $\mathcal{K}_t$ is a fixed finite set for all $t\geq 0$,
then the asymptotic optimality can be proven without
the technical conditions in
(\ref{eqt:argmax_cond})--(\ref{eqt:softmax_cond}) and (\ref{eqt:condition_validy_weightedVer_softmax}).

\begin{corollary}\label{coro:combined_eProcess_same_power_as_largest}
Let $m = \lfloor t^\phi \rfloor$, where $\phi \in (0.5,1)$.
Let $\mathcal{K}_t \equiv \mathcal{K} = \{k_1, \ldots, k_K\}$ for all $t=0, 1, \ldots$, where $0<K<\infty$.
Assume that
(\ref{eq:k_fraction_power}) holds for each $k\in\mathcal{K}$,
$k^{\star}$ in (\ref{eq:def_k*}) is unique, and $\KL_{k^{\star}} >0$.
Then
$
\liminf_{n\rightarrow\infty} n^{-1}\log \Lambda_n^{(m,\mathcal{K})}
\geq \KL_{k^{\star}}.
$
\end{corollary}

\begin{proof}[Proof of Corollary \ref{coro:combined_eProcess_same_power_as_largest}]
We consider
the argmax and softmax forms one by one.

\noindent\underline{\textbf{Part 1: Argmax form}}\\
Consider the argmax form.
From Theorem~\ref{thm:sum_spacings_strategy},
we know that
$
    n^{-1}\log \Lambda_n^{(m,k)} \inAS
    \KL_k
$
for each $k \in \mathcal{K}$.
By the uniqueness of $k^{\star}$, for each fixed $k\neq k^{\star}$,
\[
    \frac{\log \Lambda_n^{(m,k^{\star})} - \log \Lambda_n^{(m,k)}}{n} \inAS \KL_{k^{\star}} -\KL_k >0.
\]
Since $\mathcal{K}$ is a finite set,
there exists an almost surely finite $T$ such that for all $n\geq T$,
\[
    \max_{k\in\mathcal{K}\backslash \{k^{\star}\}}\left(\frac{\log \Lambda_n^{(m,k^{\star})} - \log \Lambda_n^{(m,k)}}{n}\right)  >0    .
\]
So, for $n\geq T$, we have $\lambda_t^{(m,\mathcal{K})} = \lambda_t^{(m,k^{\star})}$, then
\begin{align*}
    \frac{1}{n}\log \Lambda_n^{(m,\mathcal{K})}
    &= \frac{1}{n}\log \Lambda_T^{(m,\mathcal{K})} +
        \frac{1}{n}\sum_{t=T+1}^n \log \lambda_t^{(m,k^{\star})}   \\
    &= \frac{1}{n}\log \Lambda_n^{(m,k^{\star})} +
        \frac{1}{n}\left( \log \Lambda_T^{(m,\mathcal{K})} - \log \Lambda_T^{(m,k^{\star})} \right)  \\
    &\inAS \KL_{k^{\star}} + 0.
\end{align*}

\noindent\underline{\textbf{Part 2: Softmax form}}\\
Consider the softmax form.
Since $|\mathcal{K}|<\infty$ and $k^{\star}$
is the unique maximizer, we know that
\[
    \Delta := \KL_{k^{\star}} - \max_{k\in\mathcal{K}\setminus\{k^{\star}\}} \KL_k >0.
\]
There is an almost surely finite $T$ such that
for all $t\geq T$ and all $k\neq k^{\star}$,
\begin{align}\label{eqt:finiteFixed_mathcalK_diff_Lambda}
    \log \Lambda_t^{(m,k^{\star})}
    \geq \frac{t}{2} \KL_{k^{\star}}
    \qquad \text{and} \qquad
    \log \Lambda_t^{(m,k^{\star})} - \log \Lambda_t^{(m,k)}
    \geq \frac{t\Delta}{2} .
\end{align}
There is an almost surely finite $T'\geq T$ such that $e^{-t\Delta/2} \leq 1/2$ for all $t\ge T'$. So, together with (\ref{eqt:finiteFixed_mathcalK_diff_Lambda}),
\begin{align*}
    \Lambda_{t}^{(m,k^{\star})} - \Lambda_{t}^{(m,k)}
     = \Lambda_{t}^{(m,k^{\star})} \left\{ 1 - \frac{\Lambda_t^{(m,k)}}{\Lambda_t^{(m,k^{\star})}}\right\}
     &\geq e^{t \KL_{k^{\star}}/2} (1-e^{-t\Delta/2})
     \geq e^{t \KL_{k^{\star}}/2}  (1-1/2).
\end{align*}
Using this and the inequality $\log(1+x)\leq x$ for $x\geq 0$, we can bound the weight as follows:
\begin{align*}
    0\leq -\log w_{t,k^{\star}}
    &= \log\left(1+\sum_{k\neq k^{\star}} \exp[\gamma\{ \Lambda_t^{(m,k)} - \Lambda_t^{(m,k^{\star})} \} ] \right) \\
    &\leq \sum_{k\neq k^{\star}} \exp\left\{-
        \frac{\gamma}{2} e^{t \KL_{k^{\star}}/2}
    \right\}
    =(K-1) \exp\left\{-
        \frac{\gamma}{2} e^{t \KL_{k^{\star}}/2}
    \right\} \inAS 0.
\end{align*}
By Ces\`{a}ro summability,
$
  n^{-1} \sum_{t=1}^n \log w_{t,k^{\star}} \inAS 0
$.
So,
\begin{align}
    \frac{1}{n}\log \Lambda_{n}^{(m,\mathcal{K})}
        \ge \frac{1}{n} \sum_{t=1}^n \log w_{t, k^{\star}} \lambda_t^{(m, k^{\star})}
        \inAS 0 + \KL_{k^{\star}}.
\end{align}
Thus, the desired result follows.
\end{proof}

\section{Auxiliary simulations}
\subsection{Supplement to Section \ref{sec:comparison_other_methods}: Comparison with other methods}\label{sec:table}
The type-I error probabilities of all methods that are mentioned in Section~\ref{sec:comparison_other_methods} are reported in Table \ref{tab:null-rejection-probabilities_full}.

\begin{table}[!t]
\centering
\caption{The type-I error probability (\%) under each $\mathcal{H}_0$ and $n\in\{2500,5000\}$
        at the nominal level of $5\%$, based on 2,000
            replications each.
  }
\label{tab:null-rejection-probabilities_full}
\small
\setlength{\tabcolsep}{4pt}
\begin{tabular}{ll rr rr rr}
\toprule
& & \multicolumn{2}{c}{Normal}
& \multicolumn{2}{c}{Beta}
& \multicolumn{2}{c}{Notch}
\\
\cmidrule(lr){3-4}\cmidrule(lr){5-6}
\cmidrule(lr){7-8}
Type & Method
& \(n=2500\) & \(n=5000\)
& \(n=2500\) & \(n=5000\)
& \(n=2500\) & \(n=5000\)
\\
\midrule
$e$-spacing & Maximum spacing & 2.30 & 2.35 & 2.10 & 2.10 & 2.45 & 2.50 \\
(proposal) &Finite argmax   & 3.30 & 3.65 & 2.80 & 3.30 & 2.95 & 3.40 \\
&Dyadic argmax   & 3.15 & 3.20 & 3.05 & 3.10 & 4.00 & 4.00 \\
&Dyadic softmax  & 3.60 & 3.70 & 3.05 & 3.10 & 4.15 & 4.15 \\
&Dyadic average  & 3.95 & 4.05 & 3.35 & 3.35 & 4.65 & 4.75 \\
&Switch          & 2.35 & 2.45 & 3.10 & 3.15 & 3.65 & 3.65 \\
\cmidrule(lr){1-8}
KS-based &\textsf{KS-unadjusted} & 61.40 & 65.95 & 60.35 & 64.30 & 62.55 & 65.80 \\
&\textsf{KS-Bonf}       &  8.20 &  8.25 &  7.35 &  7.35 &  7.05 &  7.05 \\
&\textsf{KS-PAS}        &  1.50 &  1.50 &  1.50 &  1.50 &  1.40 &  1.40 \\
&\textsf{KS-DPAS}       &  4.00 &  4.00 &  3.95 &  3.95 &  3.55 &  3.55 \\
\cmidrule(lr){1-8}
KSD-based &\textsf{Gaussian-ONS}    & 0.00 & 0.00 & 0.00 & 0.00 & 0.00 & 0.00 \\
&\textsf{Gaussian-LBOW}   & 0.00 & 0.00 & 0.00 & 0.00 & 0.00 & 0.05 \\
&\textsf{Gaussian-aGRAPA} & 0.10 & 0.15 & 0.10 & 0.20 & 0.20 & 0.20 \\
&\textsf{IMQ-ONS}         & 0.00 & 0.00 & 0.00 & 0.00 & 0.00 & 0.00 \\
&\textsf{IMQ-LBOW}        & 0.00 & 0.00 & 0.00 & 0.00 & 0.10 & 0.10 \\
&\textsf{IMQ-aGRAPA}      & 0.20 & 0.25 & 0.30 & 0.35 & 0.25 & 0.25 \\
\bottomrule
\end{tabular}
\end{table}

\subsection{Supplement to Section \ref{sec:powerMCexp}: Multivariate data}\label{sec:multi_MC}
We investigate the finite-sample performance of the proposed test
with multivariate data
in Example \ref{eg:multiMC}.
The testing setups are the same as Section \ref{sec:powerMCexp}.

\begin{example}[Multivariate data]\label{eg:multiMC}
We study the proposed $\Lambda_n$ under the
following three possible strategies:
\begin{itemize}
    \item Maximum spacing strategy $\Lambda_n^{(1,\infty)}$;
        see (\ref{eqt:Dtildestar}).
    \item Maximum spacing set strategy $\Lambda_n^{(m, \mathcal{K})}$ in (\ref{eq:cond_e_R}) with
    $m=\lfloor t^{0.51}\rfloor$, where
    $\{\mathcal{K}_t\}$ is selected as
    a dyadic set $\mathcal{K}_t$ in Example \ref{eg:infinite_K} under
    argmax rule;
    see Definition \ref{def:argmax_softmax}.
    \item Automatic switch strategy $\Lambda_n^{\text{switch}}$,
            which uses the argmax rule to switch
            between the above two strategies; see Example \ref{eg:switch}.
\end{itemize}
We consider two data-generating processes $\pr$:
\begin{enumerate}
    \item[(a)] Let data be generated from a bivariate normal distribution:
            \[
                X_1, X_2, \ldots \simIID \Normal_2
                \left(
                    \begin{bmatrix}
                        0 \\ \mu
                    \end{bmatrix},
                    \begin{bmatrix}
                        1 & \rho \\
                        \rho & 1
                    \end{bmatrix}
                \right),
            \]
            where $\mu\in\mathbb{R}$ and $\rho\in(-1,1)$.
            We test $\mathcal{H}_0:(\mu, \rho) = (0,0.6)$ against
            $\mathcal{H}_1:(\mu, \rho) \neq (0,0.6)$.
      \item[(b)]
      Let $X_1,X_2, \ldots $ be i.i.d.~sample from
                a distribution with density
                $p(x) = a \{ 1 - 0.95 \mathbb{1}(x\in G_1 \cup G_2) \}$
                for $x\in[0,1]^2$, where
                \[
                    a=\frac{1}{1-0.0095\{ 9 (|\theta_1|+|\theta_2|)-|\theta_1\theta_2| \}},
                    \quad
                    G_1 = W(\theta_1)\times[0.1, 1], \quad
                    G_2=[0.1,1]\times W(\theta_2),
                \]
                and $W(\theta) = [0.25+0.1\min(0,\theta), 0.25+0.1\max(0,\theta)]$.
                The parameters $|\theta_1|$ and $|\theta_2|$
                control the widths of the notches $G_1$ and $G_2$,
                respectively.
                Similar to the univariate case,
                the signs of $\theta_1$ and $\theta_2$
                control the orientation of the notches. We test $\mathcal{H}_0:(\theta_1,\theta_2) = (0,0)$ against
            $\mathcal{H}_1:(\theta_1,\theta_2) \neq (0,0)$.

\end{enumerate}
The results are shown in Figure \ref{fig:fixed_variance}.
We found that the tests remain powerful when the data are multivariate.
We observe similar findings as in the univariate case:
The maximum spacing strategy is more powerful
for detecting local discrepancies from the null (see the notch case),
whereas
the maximum spacing set strategy is more powerful
for detecting more global discrepancies from the null (see the normal case).

\begin{figure}[t]
\centering
\includegraphics[width=.7\textwidth]
{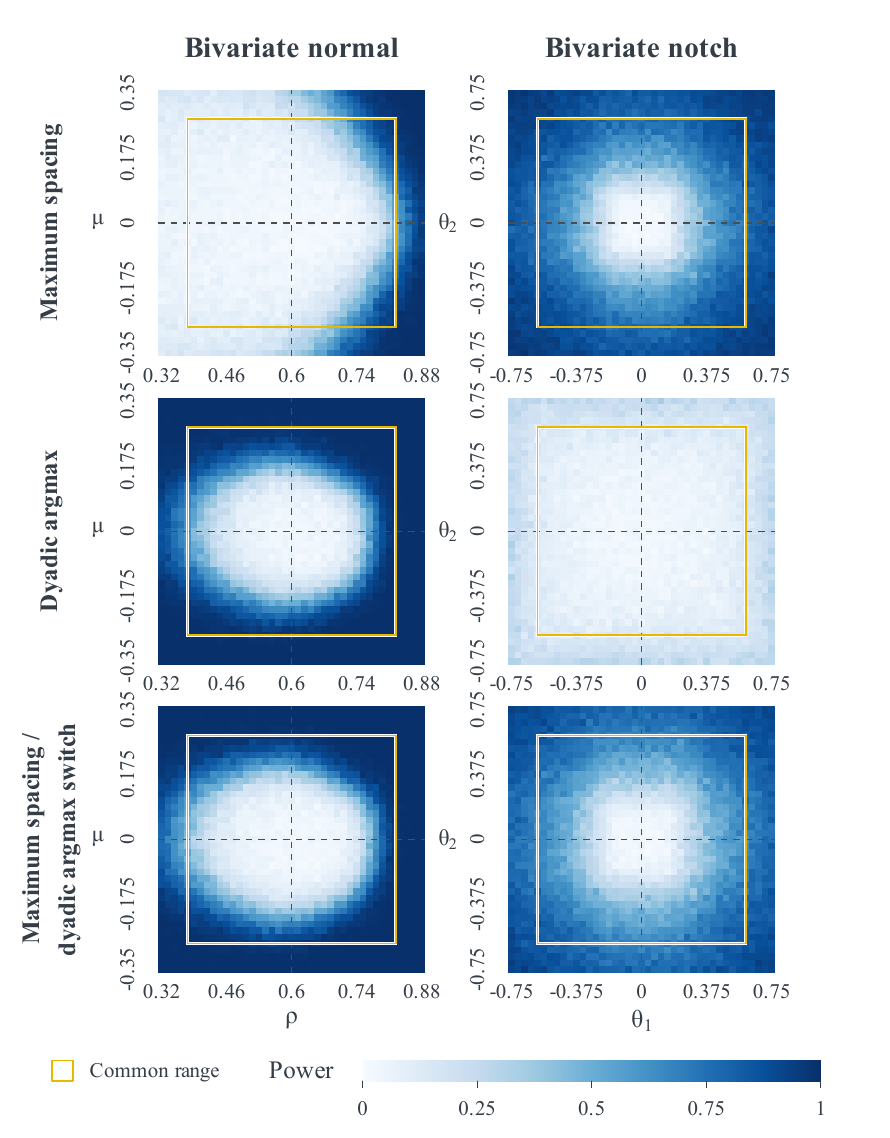}
    \caption{
    The power of the proposed test $\Lambda_n$
    with three strategies under two data-generating processes
    in Example \ref{eg:multiMC}.
    The null case is indicated by a small circle.
    A darker cell means higher power.
    Rectangles indicate the common range of parameters within each data-generating process.
    }
    \label{fig:fixed_variance}
\end{figure}

\end{example}

\subsection{Supplement to Section \ref{sec:realData}: Sensitivity analysis for the real-data application}
\begin{figure}
    \centering
    \includegraphics[width=\linewidth]
    {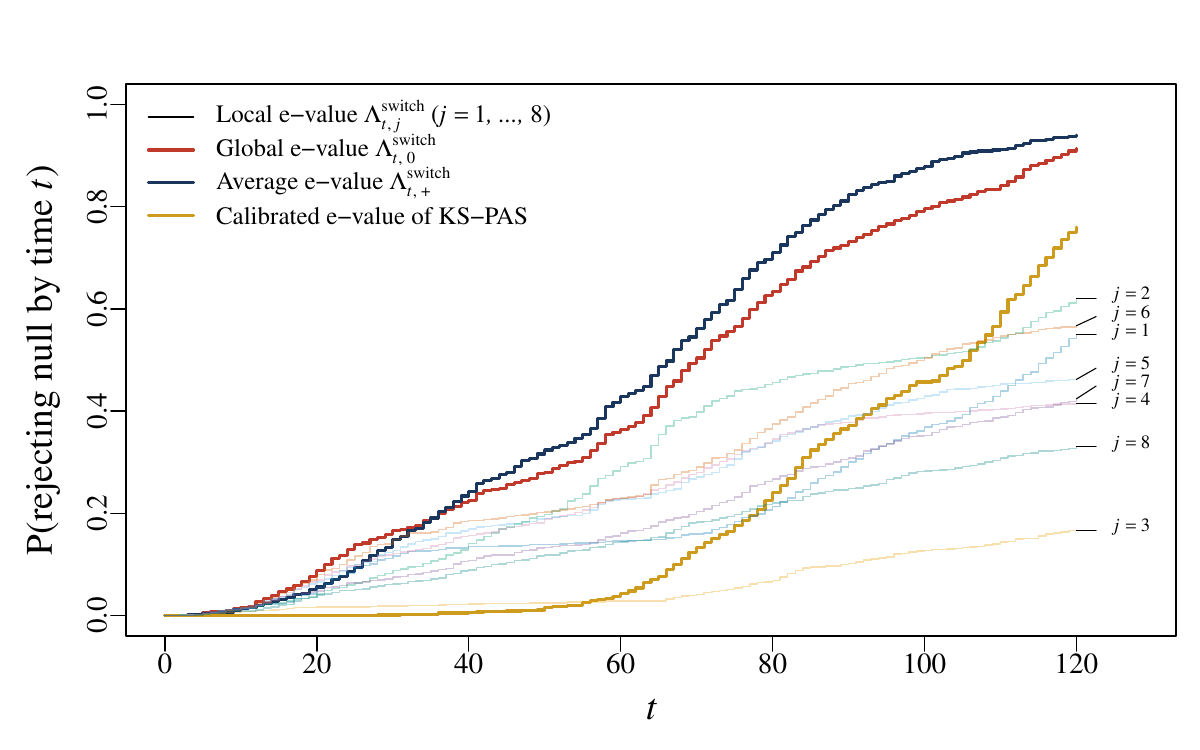}
    \caption{The estimated probabilities of rejecting $\mathcal{H}_0$ on or before $t$
    based on $\Lambda_{t,0}^{\text{switch}}$, $\Lambda_{t,+}^{\text{switch}}$, and \textsf{KS-PAS} are shown. The estimates are produced based on the
    Kaplan--Meier estimator over 1000 replications.
    }
    \label{fig:A05_realData_rejTime}
\end{figure}

In Section \ref{sec:realData}, we replicate the real-data analysis
for 1,000 times as a sensitivity analysis.
The estimated probabilities of rejecting $\mathcal{H}_0$ on or before $t$
based on $\Lambda_{t,0}^{\text{switch}}$, $\Lambda_{t,+}^{\text{switch}}$, and \textsf{KS-PAS} are shown in Figure \ref{fig:A05_realData_rejTime}.
We found that the proposed tests
$\Lambda_{t,0}^{\text{switch}}$ and $\Lambda_{t,+}^{\text{switch}}$
the most powerful test as they can reject $\mathcal{H}_0$ earlier.

\section{Discussion and future work
}\label{sec:discussion}

\subsection{Discussion}\label{sec:discussion}
This article proposes several spacing-based \(e\)-processes for
GoF test.
They are proven to be power-consistent under mild conditions,
and can readily be applied to
multivariate data.
However, the multivariate test has a limitation.
To see it, define
$U_i= (U_i^{[1]},\ldots,U_i^{[d]} )^\T$
as in (\ref{eqt:def_U_multi}).
According to Lemma \ref{lemma:uniform_trans},
$U_i^{[1]},\ldots,U_i^{[d]}\simIID \Unif[0,1]$ if and only if $\mathcal{H}_0$ is true.
Hence, it is possible that
each $U_i^{[1]}\sim \Unif[0,1]$ marginally but are dependent on the other under $\mathcal{H}_1$.
Under such $\mathcal{H}_1$, the proposed tests are not power consistent, motivating further research.
One possible approach is
to sort the \(nd\) transformed observations
\(
U_1^{[1]},\ldots,U_1^{[d]},\ldots,U_n^{[1]},\ldots,U_n^{[d]}
\)
to define the spacings.
It is not hard to see that, by Lemma~\ref{lemma:uniform_trans}, these observations are i.i.d. if and only if \(\mathcal{H}_0\) holds.

It is important to emphasize, however, that the multivariate method loses power only when the transformed sample \(U\) has the same marginal distributions under \(\mathcal{H}_0\) and \(\mathcal{H}_1\), not necessary when the original random vector \(X\) itself does. That is, the method can remain powerful even when \(X\sim\pr\) and \(X\sim\pr_0\) have identical marginal distributions.
Consider the following simple example. Let
$
X\sim\pr=\Normal_2(\mu,\Sigma),
$
and suppose that we test
$
\cH_0:X\sim\pr_0=\Normal_2(\mu,\Sigma_0),
$
where \(\Sigma_0\) is non-diagonal, so that the components of \(X\) are dependent under \(\cH_0\);  \(\Sigma_0\) and \(\Sigma\) have the same main diagonal elements. In this setting, $U$ has different marginal distributions under \(\mathcal{H}_0\) and \(\mathcal{H}_1\). That is, the multivariate method is power-consistent under the mild conditions specified in the consistency theorems. Further technical details are provided in Example~\ref{app:test_multiGaussian}.

\begin{example}{Covariance test for bivariate normal data}\label{app:test_multiGaussian}
Let the data be generated as follows:
\[
    X=\left(X^{[1]},X^{[2]}\right)^{\T}
    \sim \pr=
    \Normal_2\left(
        \begin{pmatrix}
            \mu_1\\
            \mu_2
        \end{pmatrix},
        \Sigma
    \right),
    \qquad \text{where} \qquad
    \Sigma =
    \begin{pmatrix}
        \sigma_{11} & \sigma_{12}\\
        \sigma_{12} & \sigma_{22}
    \end{pmatrix}>0,
\]
Let $c\neq 0$ be fixed.
Suppose that the null distribution of the data is
\[
    \pr_0 =
    \Normal_2\left(
        \begin{pmatrix}
            \mu_1\\
            \mu_2
        \end{pmatrix},
        \Sigma_0
    \right),
    \qquad \text{where} \qquad
    \Sigma_0=
    \begin{pmatrix}
        \sigma_{11} & c\\
        c & \sigma_{22}
    \end{pmatrix} > 0.
\]
The Rosenblatt transform based on $\pr_0$ is
\begin{align*}
    U^{[1]}
    &= F_{0}\left(X^{[1]}\right)
    = \Phi\left(
            \frac{X^{[1]}-\mu_1}{\sqrt{\sigma_{11}}}
        \right), \\
    U^{[2]}
    &= F_{0}\left(X^{[2]}\mid X^{[1]}\right)
    = \Phi\left( Z_c\right),
    \quad \text{where} \quad
    Z_c = \left\{ X^{[2]}-\mu_2-\frac{c}{\sigma_{11}}\left(X^{[1]}-\mu_1\right) \right\}
        \left(\sigma_{22}-\dfrac{c^2}{\sigma_{11}} \right)^{-1/2}
\end{align*}
where $\Phi(\cdot)$ is the CDF of $\Normal(0,1)$.
Since $X^{[1]}\sim \Normal(\mu_1,\sigma_{11})$,
we have $U^{[1]}\sim \Unif[0,1]$ trivially.
Simple algebras give
\[
    Z_c \sim \Normal\left( 0,
                    \left\{
                        \sigma_{22}
                        -\frac{2c\sigma_{12}}{\sigma_{11}}
                        +\frac{c^2}{\sigma_{11}}
                    \right\}\left(\sigma_{22}-\dfrac{c^2}{\sigma_{11}} \right)^{-1}
                \right).
\]
Hence, $U^{[2]}\sim\Unif[0,1]$ if and only if $Z_c\sim \Normal(0,1)$,
which is equivalent to
\[
    \sigma_{22}
    -\frac{2c\sigma_{12}}{\sigma_{11}}
    +\frac{c^2}{\sigma_{11}}
    =
    \sigma_{22}-\frac{c^2}{\sigma_{11}}
    \qquad \text{or} \qquad
    2c(c-\sigma_{12})=0.
\]
By our assumption $c\neq 0$, we therefore conclude that
\[
    U^{[2]}\sim\Unif[0,1]
    \quad\Longleftrightarrow\quad
    c=\sigma_{12}.
\]
In other words, $\pr=\pr_0$ if and only if $c=\sigma_{12}$.
\end{example}

\subsection{Another future research direction}
Another possible future research direction is to extend the proposed methods to composite null hypotheses. A feasible approach, following \cite{martinez2024sequential}, is as follows:
For a composite null \(\mathcal{H}_0\), construct an $e$‑process (may not $e$-process for whole $\cH_0$) \(\Lambda_t^{\pr_0}\) for each \(\pr_0 \in \mathcal{H}_0\) as in the simple‑null case, and then define
$
\Lambda_t^{\pr_0^{\star}} := \min_{\pr_0\in\mathcal{H}_0} \Lambda_t^{\pr_0}.
$
By definition, $\Lambda_t^{\pr_0^{\star}}$ forms an $e$-process under the composite null $\cH_0$. This construction is essentially the same as the universal inference (UI)–type $e$‑values of \cite{wasserman2020universal}.
However, UI‑type $e$‑values can lose power compared with growth‑rate optimal (GRO) $e$‑values in some settings; see \cite{hao2024values}.
Motivated by this, one may adapt the GRO $e$‑value framework to develop more powerful $e$‑processes for GoF testing under composite nulls, thereby improving over UI‑based constructions.

\bibliographystyle{rss}
\bibliography{references_01, references_02_SAVI}

\end{document}